\documentclass{JHEP3}

\usepackage{amsmath, amssymb}
\usepackage{graphicx}
\usepackage[mathscr]{eucal}

\newcommand{\rf}[1]{(\ref{#1})}
\newcommand{\ba}{\begin{array}}
\newcommand{\ea}{\end{array}}
\newcommand{\dis}{\displaystyle}
\newcommand{\bracket}[2]{\bra{#1}\,#2\rangle}
\newcommand{\bra}[1]{\langle\,#1\,|}
\newcommand{\ket}[1]{|\,#1\,\rangle}
\newcommand{\p}{\partial}
\newcommand{\ud}{\mathrm{d}}
\newcommand{\pathD}{\!\mathscr{D}}
\newcommand{\Det}{\text{Det}\,}

\newcommand{\G}{\Gamma}
\renewcommand{\d}{\delta}
\newcommand{\ep}{\epsilon}

\newcommand{\f}{\varphi}

\newcommand{\x}{{\bf x}}
\newcommand{\y}{{\bf y}}
\newcommand{\z}{{\bf z}}
\newcommand{\q}{{\bf q}}
\newcommand{\pp}{{\bf p}}
\newcommand{\kk}{{\bf k}}
\newcommand{\X}{\mathbf{X}}

\newcommand{\BB}{\mathbf{B}}

\renewcommand{\k}{\mathcal{K}}

\newcommand{\dirac}[1]{\delta\left(#1\right)}
\newcommand{\til}{\widetilde}
\def\Journal#1#2#3#4{{#1} {\bf #2} (#3) #4}

\title{Timelike T-duality in the string field Schr\"odinger functional}
\author{A. Ilderton\\Centre for Particle Theory, University of Durham, Durham DH1 3LE, UK\\ E-mail: \email{anton@cantab.net}}
\author{P. Mansfield\\Centre for Particle Theory, University of Durham, Durham DH1 3LE, UK\\ E-mail: \email{p.r.w.mansfield@dur.ac.uk}}
\abstract{Timelike T-duality of string theory appears as a symmetry of time evolution in string field theory, exchanging evolution through times $t$ and $1/t$, and exchanging boundary states with backgrounds. This is demonstrated by constructing the string field Schr\"odinger functional, the generator of time evolution, based on Feynman diagram arguments and in analogy with quantum field theory. There the functional can be described using only properties of first quantised particles on a timelike orbifold. Using new sewing rules applicable to both open and closed strings we generalise this approach to bosonic string theory and express the string field Schr\"odinger functional in terms of strings on $\mathbb{S}^1/\mathbb{Z}_2$.}
\begin{document}
\section{Introduction}
The natural picture in which to investigate time evolution in quantum field theory is the Schr\"odinger representation. The field operator is diagonalised on the initial quantisation surface and arbitrary data on this surface is then evolved through time by the action of the Schr\"odinger functional.

It is not clear how to realise this in Witten's string field theory \cite{Witten} where the time variable is associated with the midpoint of the timelike extension of the string $X^0(\pi/2)$ \cite{Siopsis}, \cite{Maeno}, rather than being a global time for the whole string. Due to the difficulties of working with a Lagrangian formalism, in this paper we construct the string field Schr\"odinger functional using graphical arguments based on the connections between first and second quantised particles and strings.

We begin by summarising the Schr\"odinger picture approach to quantum field theory, using a scalar field in $D+1$ dimensions to illustrate. We describe the functional prescription for the Schr\"odinger functional as given by Symanzik \cite{Symanzik} and its Feynman diagram expansion. We then show that the Schr\"odinger functional can be expressed in terms of particles moving on $\mathbb{R}^D$ times a compactification of the time direction using the sum over paths representation of the field propagator. Since the sum over paths has a natural generalisation to the Polyakov integral for strings, it is suggested that we can construct the string field Schr\"odinger functional from first quantised strings by analogy.

To realise this we begin by deriving the `gluing property'. This is a property of the free space propagator fundamental to second quantisation, but derived in first quantisation. It is a method for gluing together reparametrisation invariant propagators and diagrams, including their moduli spaces, appropriate to the Schr\"odinger representation. Using this property we show that time evolution defined through the action of the Schr\"odinger functional can be described graphically. The Schr\"odinger functional itself is characterised solely by the gluing property and its Feynman diagram expansion.

In section four we give an application of our arguments to interacting theories. We demonstrate how the vacuum wave functional of $\phi^4$ theory can be constructed as the generator of vacuum expectation values and that our construction agrees with the conventional description of the vacuum wave functional in terms of a large time functional integral. Our arguments are shown to hold in Euclidean space and in the representation in which the field momentum is diagonal. In this representation the Schr\"odinger functional is described by particles moving on $\mathbb{R}^D\times\mathbb{S}^1/\mathbb{Z}_2$.

Our construction of the Schr\"odinger functional is invertible. If we 
know the n-point functions and have a gluing property then the 
Schr\"odinger functional is determined by the Feynman diagram expansion. 
The gluing property, although a part of the field theory, is derivable 
as a property of the sum over paths defining the propagator in first 
quantisation. This is of course the language in which most progress in 
string theory has been made. In section five we review the properties of 
the string field propagator constructed from the Polyakov integral. We 
then generalise the gluing property to a method of sewing worldsheets, 
appropriate to a Schr\"odinger representation.  We stress that our goal 
is to construct the objects of second quantised string theory such as 
time evolution operators and wave-functionals using the functional 
integrals of first quantisation. This is a much more modest programme 
than the normal approach of postulating an action and using that to 
derive physical quantities. However it means that the usual problems
of string field theory, such as how to construct an action that ensures 
that when world-sheets are sewn together to generate Feynman diagrams a 
single cover of the moduli space of surfaces results, need not be 
addressed. Thus in section 5.5 we construct the lowest order part of the 
vacuum functional that includes a string loop. We require that 
expectation values computed using this wave-functional yield the 
expressions familiar in first quantisation.
The wave-functional thus consists of two pieces, one that gives rise 
directly to the required vacuum expectation value
(and so, by construction, is an integral over a single cover of moduli 
space),
and another which cancels unwanted terms coming from the sewing of 
diagrams.

In section six we discuss the properties of the string field Schr\"odinger functional in the field momentum representation. The orbifold naturally leads to T-duality. We show that timelike T-duality of string theory becomes a large/small time duality of string field theory, where evolution through time $t$ is exchanged with evolution through time $1/t$ and an interchange of string fields and backgrounds. Our results apply to both the open and closed string. 

Finally, we describe the role of BRST symmetry in our formalism in section seven and give our conclusions. An outline of the results of this paper can be found in \cite{us1}.

\section{The Schr\"odinger functional}\label{Schro-sect}

We begin by briefly reviewing the Schr\"odinger representation in quantum field theory. The scalar field action with metric $\eta_{\mu\nu}=\text{diag}(1,-1\ldots)$ is
\begin{equation*}
  S[\phi] = \int\!\ud^{D}\x\,\ud t\, \frac{1}{2}\big(\partial_t \phi\partial_t \phi - \nabla\phi\cdot\nabla\phi\big) - V(\phi).
\end{equation*}
To quantise, the field $\phi$ and its conjugate momentum $\pi$ are promoted to operators obeying the equal time commutation relation $[\hat\phi(\x), \hat\pi(\y)] = i\hbar\,\dirac{\x-\y}$.
The common approach to canonical quantisation is then to build a Fock space on which the Fourier modes of these operators act as creation and annihilation of particles. In the Schr\"odinger representation we instead diagonalise the field operator restricted to a quantisation surface (which we take to be time $t=0$). That is, a basis for the state space is given by
\begin{equation*}
  \bra{\phi}\hat\phi(\x,0) = \phi(\x)\bra{\phi}.
\end{equation*}
Using the canonical commutation relations the dependence on the field is made explicit by writing
\begin{equation*}
\bra{\phi}=\bra{D}\exp\bigg(\frac{i}{\hbar}\int d\x\,\,\phi(\x)\hat{\pi}(\x,0)\bigg)
\label{eq2} 
\end{equation*}
where the Dirichlet state $\langle\,D\,|$ is annihilated by $\hat{\phi}(\x,0)$. We also have
\begin{equation*}
  \bra{\phi} \hat{\pi}(\x) = -i\hbar\frac{\d}{\d\phi(\x)}\bra{\phi}.
\end{equation*}
A quantum state $\Psi$ is a functional of the field and depends explicitly on the time,
\begin{equation*}
  \bracket{\phi}{\Psi} = \Psi\left[ \phi(\x), t\right].
\end{equation*}
The Schr\"odinger equation for time evolution becomes a functional differential equation
\begin{eqnarray*}
  i\hbar\frac{\partial}{\partial t}\bracket{\phi}{\Psi} &=& \bra{\phi}\hat H\ket{\Psi} \\
  \rightarrow i\hbar\frac{\partial}{\partial t}\Psi[\phi(x),t] &=& \int\!\ud^D\x\, \left( -\frac{\hbar^2}{2}\frac{\delta^2}{\delta \phi^2} + \frac{1}{2}\nabla\phi\cdot\nabla\phi + V(\phi)\right)\Psi[\phi(x),t].
\end{eqnarray*}
The Schr\"odinger functional is defined as
\begin{equation}\label{schr-def}
\begin{split}
	\mathscr{S}[\phi_2,t_2;\phi_1,t_1] &=\bra{\phi_2}e^{-i\hat H (t_2 - t_1)/\hbar}\ket{\phi_1} \\
	  &= \bra{D}e^{i\int\phi_2\hat\pi/\hbar}e^{-i\hat H t/\hbar}e^{-i\int\phi_1\hat\pi/\hbar}\ket{D}
\end{split}
\end{equation}
where the states $\bra{\phi_i}$ are again eigenvectors of the field operator, and describes time evolution as follows. Suppose we have some state $\Psi[\phi,t_1]$. Then this state at later time $t_2$ is, inserting a complete set,
\begin{eqnarray}\label{S-evolves}
	\nonumber \Psi[\phi,t_2] &\equiv& \bra{\phi}e^{-i\hat H (t_2-t_1)/\hbar}\ket{\Psi} \\
	\nonumber &=& \bra{\phi}e^{-i\hat H (t_2-t_1)/\hbar} \left[\int\pathD \varphi\, \ket{\varphi}\bra{\varphi}\right] \ket{\Psi} \\
	&=& \int\pathD \varphi\,\, \mathscr{S}[\phi,t_2;\varphi,t_1]\,\Psi[\varphi,t_1].
\end{eqnarray}
Our goal is to construct this functional for string field theory -- of course we do not know the Hamiltonian so the above definition does not seem immediately applicable. However, we will find that we can express the Schr\"odinger functional in terms of a Feynman diagram expansion generalisable to the string.

The Feynman prescription for the Schr\"odinger functional is
\begin{equation}\label{s1}
	\mathscr{S}[\phi_2, t_2; \phi_1, t_1] = \int\pathD \varphi\,\, e^{iS[\varphi]/\hbar}\bigg|^{\varphi(t_2) = \phi_2}_{\varphi(t_1) = \phi_1.}
\end{equation}
The change of variable
\begin{equation}\label{s2}
  \tilde\varphi = \varphi + \theta(t-t_2)\phi_2(\x) + \theta(-t+t_1)\phi_1(\x)
\end{equation}
where $\theta$ is the step function ($\theta(0)=1$), leaves the measure invariant and will move the $\phi$--dependence in the integration limits to boundary terms in the action. Naively the $\theta$ terms do not contribute to the potential, since the action is evaluated between times $t_1$ and $t_2$.  Our path integral becomes, dropping the tilde,
\begin{equation}\label{co-ord rep}\begin{split}	
	 \int\pathD \varphi\, \exp\bigg(\frac{i}{\hbar}S[\varphi] &+ \frac{i}{\hbar}\int\!\ud^D\mathbf{x}\, \phi_2(\x)\,\dot\varphi|_{t=t_2}  -\frac{i}{\hbar}\int\!\ud^D\mathbf{x}\,\phi_1(\x)\,\dot\varphi|_{t=t_1} \\
	 &+\frac{i}{2\hbar}\int\!\ud^D\mathbf{x}\,\phi_2(\x)^2\delta(0) - \frac{i}{2\hbar}\int\!\ud^D\mathbf{x}\,\phi_1(\x)^2\delta(0)\bigg)
\end{split}
\end{equation}
and the integration variable obeys $\varphi=0$ on the boundaries $t=t_1$ and $t=t_2$. The first three terms in this expression relate directly to the canonical expression in Eq.(\ref{schr-def}). The final two terms require regularisation. As Symanzik has discussed \cite{Symanzik}, placing source terms on the boundary leads to divergences, and the Schr\"odinger functional is an example. The divergences appear in perturbation theory because the field is placed at the same point as the image charges which enforce boundary conditions, and are in addition to the usual free space UV divergences. In order to regulate the image divergences we should split in time the fields-- thus the fields in \rf{sf-c-m rep},  \rf{corr}, \rf{sf-mom rep} will be defined at different ordered times, with the difference acting as regulator. 

Interactions will be dealt with in due course but for now let us carry out the free field integral in \rf{co-ord rep} and introduce some notation. If we take the time splitting regularisation to be understood then we can drop the delta functions, and with the usual $i\varepsilon$ prescription the Gaussian converges to
\begin{equation}\label{sf-c-m rep}
\begin{split}
	\mathscr{S}[\phi_2,t_2;\phi_1,t_1] = N_\mathscr{S}\exp\bigg( -\frac{1}{\hbar}\iint\ud^D(\x,\y)\,\, &\frac{1}{2}\phi_2(\y) \ddot{G}_D(\y, t_2, \x, t_2) \phi_2(\x) \\
	&- \phi_2(\y) \ddot{G}_D(\y, t_2, \x, t_1)\phi_1(\x) \\
	&+  \frac{1}{2}\phi_1(\y)\ddot{G}_D(\y, t_1,\x,t_1)\phi_1(\x)\bigg)
\end{split}
\end{equation}
where the propagator $G_D$ obeys Dirichlet boundary conditions on the surfaces $t=t_1$ and $t=t_2$, reflecting the fact that $\langle\,D\,|\hat{\phi}(\x)=0$,  and the dots denote the differentiation with respect to time which results from $\phi$ being coupled to $\dot\varphi$,
\begin{equation}
  \ddot{G}_D(\x_2,t_2,\x_1,t_1)\equiv \frac{\partial^2}{\partial t\, \partial t'}\,G_D(\mathbf{x}_2,t,\mathbf{x}_1,t')\big|_{t=t_2,\,\, t'=t_1}. 
\end{equation}
The normalisation constant $N_\mathscr{S}$ will be discussed below. The free field Schr\"odinger functional can be written
\begin{equation}\label{sf-c-m-pic}
    \includegraphics[width=0.9\textwidth]{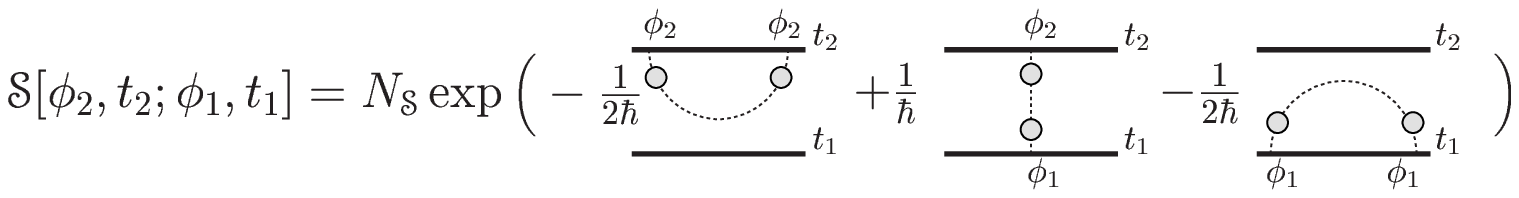}
\end{equation}
The heavy lines denote the boundaries at $t=t_1, t_2$. In our diagrams a dotted line will denote a propagator with Dirichlet conditions on all boundaries shown (whether or not the propagator ends on them all). The grey dot is the time derivative.

\section{Time evolution from a sum over paths}

In this section we give a graphical description of time evolution in field theories which we will later generalise to the string field. Our approach is based on interpreting sums over field histories in terms of sums over particle histories. The Schr\"odinger functional between times $0$ and $t$ (without loss of generality) is built from the Green's function $G_D$ which vanishes on the hypersurfaces at times $0$ and $t$. Beginning in free space, we identify free space points with their images under an $\mathbb{S}^1/\mathbb{Z}_2$ (orbifold) compactification of the time direction, radius $t/\pi$. In the sum over paths from points $(\x,t_i)$ to $(\y,t_f)$ on this spacetime we must include the paths to the image points since they are considered equivalent. If we attach a minus sign each time a path crosses a reflection of the quantisation surfaces at times $nt$ for $n\in\mathbb{Z}$, 
\begin{equation*}
  \includegraphics[width=0.3\textwidth]{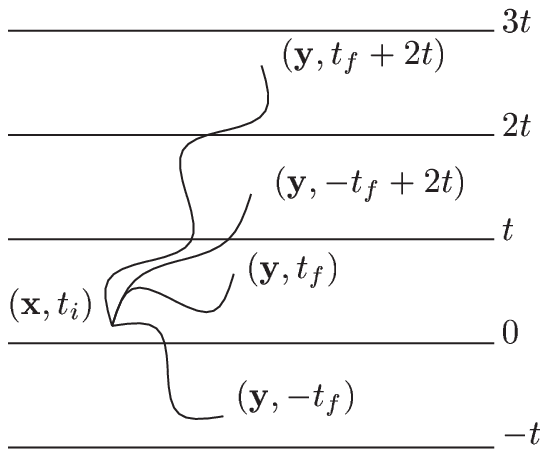}
\end{equation*}
then the sum over paths to each image gives a free space propagator weighted with a sign. The sign is positive if the image is $t_f + 2nt$ and negative if $-t+2nt$ for $n\in\mathbb{Z}$. All together, we have
\begin{equation}
  \sum\limits_{n\in\mathbb{Z}} G_0(\x_1,t_1;\x_2, t_2 + 2nt) - G_0(\x_1,t_1;\x_2, -t_2 + 2nt) \equiv G_D(\x_1,t_1;\x_2, t_2)
\end{equation}
which is equal to the desired propagator $G_D$ because the sum on the left hand side is the method of images imposition of the boundary conditions on the right hand side.

In this section we will use this interpretation to show that time evolution can be described using the gluing property and the diagram expansion. It is this structure which we will later generalise to the string field.

\subsection{The gluing property}\label{glue-prop}

It is well known that the off-shell free space propagator from $x_i \equiv(\x_i, t_i)$ to $x_f$ may be written as a sum over all paths from $x_i$ to $x_f$ with an action involving an intrinsic metric $g$
\cite{Brink} \cite{PolyakovBook}. Integrating out $g$ gives a Boltzmann weight equal to the exponential of the length of the path,
\begin{equation}\label{prop-path}\begin{split}
  G_0(x_f; x_i) &= \int \frac{\pathD (x,\,g)}{\text{Vol Diff}}\,\, e^{i\int_0^1 d\xi\,\, ({\dot x\cdot\dot x}/(2g)+m^2g/2)}\bigg|_{x(0)=x_i}^{x(1)=x_f} \\
  & =\int \pathD x\,\, e^{im\int_0^1 d\xi\,\, \sqrt{\dot x\cdot\dot x}}\bigg|_{x(0)=x_i}^{x(1)=x_f}.
\end{split}\end{equation}
The path is parameterised by $\xi$ and $x(0)$ is the point $x_i$,
$x(1)$ the point $x_f$. 
We have corrected for the over counting of equivalent paths resulting from reparametrisation invariance of the action by dividing out the volume of the space of reparametrisations $f(\tau)$. This is the particle analogue of the Polyakov integral in string theory. Any metric can be reduced, by a suitable gauge transformation, to
\begin{equation*}
  g(\tau) = T^2 \bigg(\frac{\ud f(\tau)}{\ud \tau}\bigg)^2
\end{equation*}
for some $f$ and $T$, the latter of which is the analogue of a string modular parameter. The Jacobian for the change of variables is
\begin{equation*}
  \pathD g = \frac{\ud T}{\sqrt{T}}\,\Det^{1/2}\bigg(-\frac{1}{T^2}\frac{\ud^2}{\ud \tau^2}\bigg)\,\pathD f.
\end{equation*}
We can now evaluate the path integral (\ref{prop-path}) by using the reparametrisation invariance to set $g=T^2$, constant, and then integrate over $T$,
\begin{equation*}
  G_0(x_f;x_i) = \int\limits_0^\infty\!\frac{\ud T}{\sqrt{T}}\Det^{1/2}\left(-\frac{1}{T^2}\frac{\ud^2}{\ud \tau^2}\right)\int\pathD x^\mu\,\,e^{iS(x,T)}
\end{equation*}
The above determinant is computed with Dirichlet boundary conditions and can be zeta function regulated to give
\begin{equation*}
  \Det^{1/2}\left(-\frac{1}{T^2}\frac{\ud^2}{\ud \tau^2}\right) = \bigg(\prod\limits_{n=1}\frac{\pi^2 n^2}{T^2}\bigg)^{1/2} \rightarrow \text{const.}\sqrt{T}.
\end{equation*}
Splitting $x^\mu$ into classical and quantum pieces, the integral over $x$ is over the quantum piece which has the Fourier expansion
\begin{equation*}
  x = \sum\limits_{m=1}x_m\sin(m\pi\tau)\sqrt{\frac{2}{T}}.
\end{equation*}
Each $x$ integration results in the inverse of the previous determinant. To do the final integration we can rotate the contour $T\rightarrow -iT$ and obtain the integral of the heat kernel $\k$ of the Laplacian with the normalisation $\k(T=0)=\delta^{D+1}(x_f-x_i)$,
\begin{equation*}
  \begin{split}  
G_0(x_f; x_i)&= i\displaystyle{\int\limits_0^\infty\!\frac{\ud T}{(4\pi T)^{(D+1)/2}}\, e^{ -\frac{1}{4T}(x_f-x_i)^2 + m^2 T}} \\
&=\displaystyle{ i\int\limits\!\frac{\ud^{D+1}k}{(2\pi)^{D+1}}\, \frac{e^{-ik\cdot(x_f - x_i)}}{k_\mu k^\mu - m^2}}.
  \end{split}
\end{equation*}
The derivation of the gluing property begins with the simple observation that paths from $(\x_1,t_1)$ to $(\x_2,t_2)$ must cross the plane at time $t$ at least once if $t_2>t>t_1$. This implies the sum over paths defining the propagator can factorised so that formally
\begin{equation}\label{fact}
\sum_{\textrm{paths AB}} e^{-\textrm{length(AB)}}= \sum\limits_\textrm{C}  \bigg( \sum_{\textrm{paths AC}}
e^{-\textrm{length(AC)}}\bigg) \bigg( \sum_{\textrm{paths CB}} e^{-\textrm{length(CB)}}\bigg) 
\end{equation}
where $C$ lies in the plane at time $t$. To make this factorisation explicit, insert into (\ref{prop-path}) a resolution of the identity,
\begin{equation}\label{prop-trick}
	G_0=\int\pathD x\,\, \left[ \int\!\ud\tau'\, J(x^0)\dirac{x^0(\tau')-t}\right]\,\, e^{iS[x]}.
\end{equation}
For $t_2 > t > t_1$ the delta function always has support on the worldline. The Jacobian $J$ which makes the insertion unity is easily found to be $J=\dot x^0(\tau')$. Taking the integral over $\tau'$ outside and distinguishing between worldline times earlier and later than $\tau'$, the path integral is
\begin{equation*}
	\int\!\ud\tau'\, \int\left[\prod_{\tau<\tau'}\ud x^\mu(\tau)\right]\ud^D \x(\tau')\dot{x}^0(\tau')\left[ \prod_{\tau>\tau'} \ud x^\mu(\tau)\right]\, \exp\left(i\sum\limits_{\tau<\tau'} S[x(\tau)] + \sum\limits_{\tau>\tau'} S[x(\tau)]\right).
\end{equation*}
We can write $\dot{x}^0(\tau')$ as a two sided derivative
\begin{equation*}
  \dot{x}(\tau')^0 = \lim_{h\rightarrow 0} \frac{x^0(\tau'+h)- x^0(\tau'-h)}{2h}
\end{equation*}
which splits the path integration into a pair of terms each with an insertion. The integrals are invariant under reparametrisations of the worldline and so have no explicit $\tau'$ dependence, the integral over which gives a finite volume, leaving
\begin{equation}
\begin{split}
	G_0=\int\!\ud^D\mathbf{y}\int\pathD x^\mu\, \dot{x}^0(\tau_\text{final})&e^{iS[x]} \int\pathD x^\mu\,e^{iS[x]} \\
	&+ \int\!\ud^D\y\int\pathD x^\mu\, e^{iS[x]}\int\pathD x^\mu\, \dot{x}(\tau_\text{initial})e^{iS(x)},
\end{split}
\end{equation}
where the integral over $\y$ is over the boundary spatial co-ordinate data. The path integrals can be done in the Polyakov approach, fixing the metric $g=T^2$. The insertion can be taken outside the integral as a derivative w.r.t. boundary data, giving us the factorisation of the propagator
\begin{equation}\label{factorisation}
	G_0(\x_f, t_f, \x_i, t_i) = -i\int\!\ud^D\y\, G_0(\x_f, t_f, \y, t)\overleftrightarrow{\frac{\partial}{\partial t}}G_0(\y, t, \x_i, t_i), \qquad t_2 > t > t_1.
\end{equation}
The explicit calculation is easiest using the Fourier representation. Consider
\begin{align*}
 &\int \ud^D\y\,\, G_0(\x_2,t_2,\y,t)\frac{\p}{\p t}G_0(\y,t,\x_1,t_1)\\
\\
= \int&\!\frac{\ud^D(\pp\,\q,\y)\,\ud(p_0,q_0)}{(2\pi)^{2(D+1)}}\,\, (-iq_0)\,
\frac{e^{-i[\pp(\x_2-\y)+p_0(t_2-t)+\q(\y-\x_1)+q_0(t-t_1)]}}
{(p_0^2-E^2(\pp))(q_0^2-E^2(\q))}\\
\\
=\int&\frac{\ud^D\pp\,\ud(p_0,q_0)}{(2\pi)^{(D+2)}}\,\,(-iq_0)\,
\frac{e^{i[\pp(\x_2-\x_1)+p_0(t_2-t)+q_0(t-t_1)]}}
{(p_0^2-E^2(\pp))(q_0^2-E^2(\pp))}
\end{align*}
which follows from doing the $\y$ integration giving $(2\pi)^D\dirac{\pp-\q}$ (all with the $i\varepsilon$ prescription). The $q_0$ integration depends on the sign of $t-t_1$, $\ep(t-t_1)$, giving
\begin{equation*}
  \frac{i}{2} \ep(t-t_1)\int\frac{\ud^D\pp\, \ud p_0}{(2\pi)^{D+1}}\,\,\frac{e^{-ip_0(t_2-t)-i\pp(\x_2-\x_1)-iE(\pp)|t-t_1|}}{p_0^2-E^2(\pp)}.
\end{equation*}
The $p_0$ integration gives
\begin{equation*}
  \frac{i}{2} \ep(t-t_1)\int\frac{\ud^D \pp}{(2\pi)^{D}}\,\,\frac{e^{-i\pp(\x_2-\x_1)-iE(\pp)(|t_2-t|+|t-t_1|)}}{2E(\pp)}.
\end{equation*}
The leading factors are taken care of if we include the second term in \rf{factorisation}, the result of which follows. We can write this with the insertion acting on a single propagator by replacing
\begin{equation*}
  \overleftrightarrow{\frac{\partial}{\partial t}}\longrightarrow -2\frac{\partial}{\partial t}  
\end{equation*}
and the following additional cases are immediate,
\begin{equation}
\int\!\ud^D \y \,\, G_0(\x_2,t_2;\y,t)\left(-2\frac{\p}{\p t}\right)G_0(\y,t;\x_1,t_1)=
\left\{ \ba{cr}
{\dis  iG_0(\x_2,t_2;\x_1,t_1)} & t_2> t>t_1
\\
{\dis  -iG_0(\x_2,t_2;\x_1,t_1)} & t_1>t>t_2
\\
{\dis  iG_I(\x_2,t_2;\x_1,t_1)} & t>t_1,t_2
\\
{\dis  -iG_I(\x_2,t_2;\x_1,t_1)} & t<t_1,t_2
\ea
\right.
\label{prop-cases}
\end{equation}
where $G_I$ is the ``image propagator'' equal to the free space propagator for the points $(\x_2,t_2)$ and the reflection of $(\x_1,t_1)$ in the plane at time $=t$. In short, if the two points are on opposite sides of the plane at time $=t$, the two propagators are glued
to form the usual propagator, if they are on the same side gluing produces the image propagator.

This result should not be confused with the self-reproducing property of heat-kernels, (this will be apparent when we discuss strings) but plays a nonetheless fundamental role in field theory. For example, applying it twice gives
\begin{equation}
\ba{c}
{\dis \int\ud^D\x_3\, \ud^D\x_2 \,\, G_0(\x_4,t_4,\x_3,t_3)
\bigg( 4\frac{\p^2}{\p t_3\, \p t_2 }\,G_0(\x_3,t_3,\x_2,t_2)\bigg)
G_0(\x_2,t_2,\x_1,t_1)}\\
\\
{\dis = G_0(\x_4,t_4,\x_1,t_1)\quad\text{for}\quad t_4>t_3>t_2>t_1}.
\ea
\end{equation}
Taking all the $t_i$ to zero gives a useful relation which may be expressed as 
\begin{equation*}
  \includegraphics[width=0.5\textwidth]{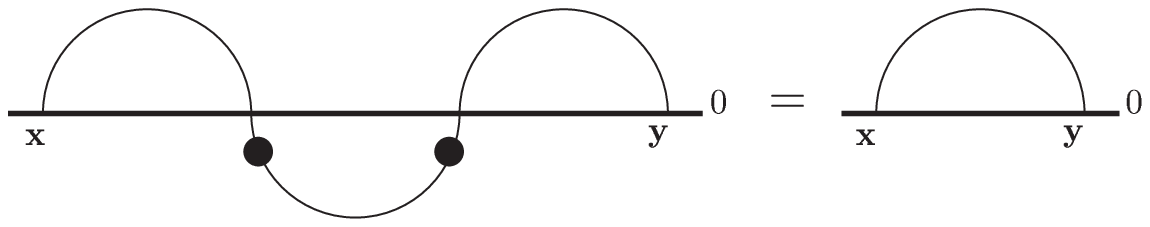}
\end{equation*}
where the heavy line is the plane at time $t=0$, the unbroken line is the free space propagator and a black dot is $-2$ times a time derivative\footnote{Note that the orientation of this diagram is unimportant and can be chosen for convenience.}. Thus
\begin{equation}
  \includegraphics[width=0.35\textwidth]{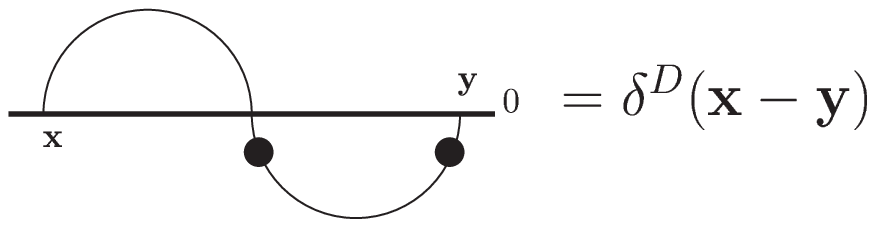}
\end{equation}
From this we deduce that the inverse of the free space propagator at equal time is
\begin{equation}
  \includegraphics[width=0.13\textwidth]{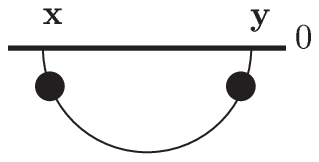}.
\end{equation}

\subsection{Time evolution of the vacuum}\label{vac-sect}
The vacuum wave functional (VWF) is the generator of vacuum expectation values (equal time Green's functions). To lowest order the VWF is that of a free theory with form
\begin{equation}\label{g20-def}
  \Psi_0[\phi] =\exp\left( -\frac{1}{2\hbar}\int\!\ud^D(\x,\y)\, \phi(\x)\,\Gamma^0_2(\x,\y)\phi(\y)\right)
\end{equation}
and must generate the free space propagator restricted to the boundary $t=0$ via
\begin{equation}\label{inverse}
  \hbar\, G_0(\x,0,\y,0) = \langle\phi(\x,0)\phi(\y,0)\,\rangle = \int\pathD \phi\, \phi(\x)\phi(\y) e^{-\int\phi\,\Gamma^0_2\phi/\hbar} =\hbar\,\big(2\,\Gamma^0_2(\x,\y)\big)^{-1}.
\end{equation}
The required inverse was calculated in the previous section. It follows that to lowest order the VWF can be written
\begin{equation}\label{vwf-first}
   \includegraphics[width=0.4\textwidth]{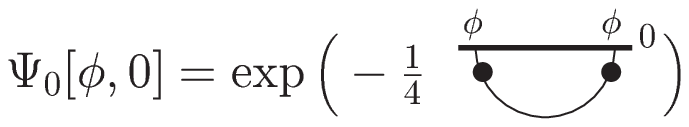}.
\end{equation}
The VWF is also an eigenstate of the Hamiltonian. Using this and the Schr\"odinger functional we have two expressions for the state at time $t$ (we will set $\hbar=1$ in the remainder of this section),
\begin{equation}\label{vac-evol}
  \Psi_0[\f_2,t] = e^{-iE_0 t}\Psi_0[\f,0] = \int\pathD \f_1\, \mathscr{S}[\f_2,t;\f_1,0]\Psi_0[\f_1,0].
\end{equation}
The integral is Gaussian to lowest order and can be carried out directly, but we will represent it using the diagram expansion.  The integral is
\begin{equation}\label{evol-c-m1}
  \includegraphics[width=\textwidth]{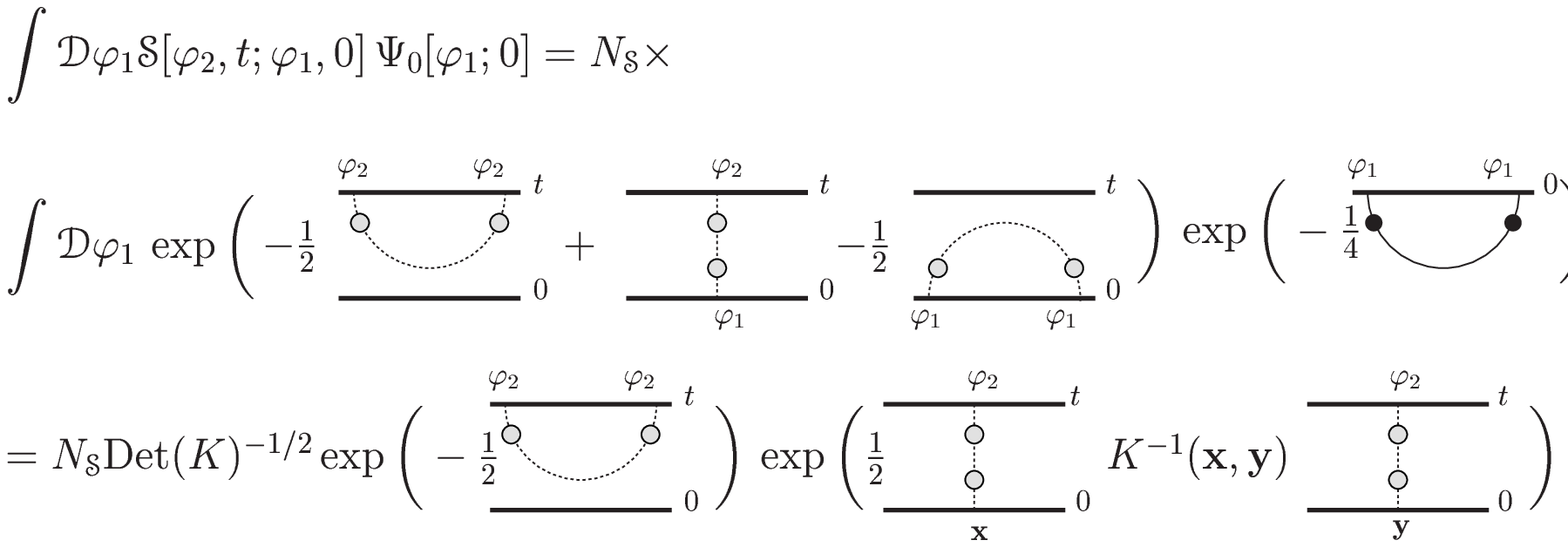}
\end{equation}
The first three terms are the Schr\"odinger functional, the final term is the vacuum. The Gaussian integral in $\varphi_1$ gives the third line, where the symmetric operator $K$ and its inverse are defined by
\begin{equation}\label{evol-c-m2}
  \includegraphics[width=0.8\textwidth]{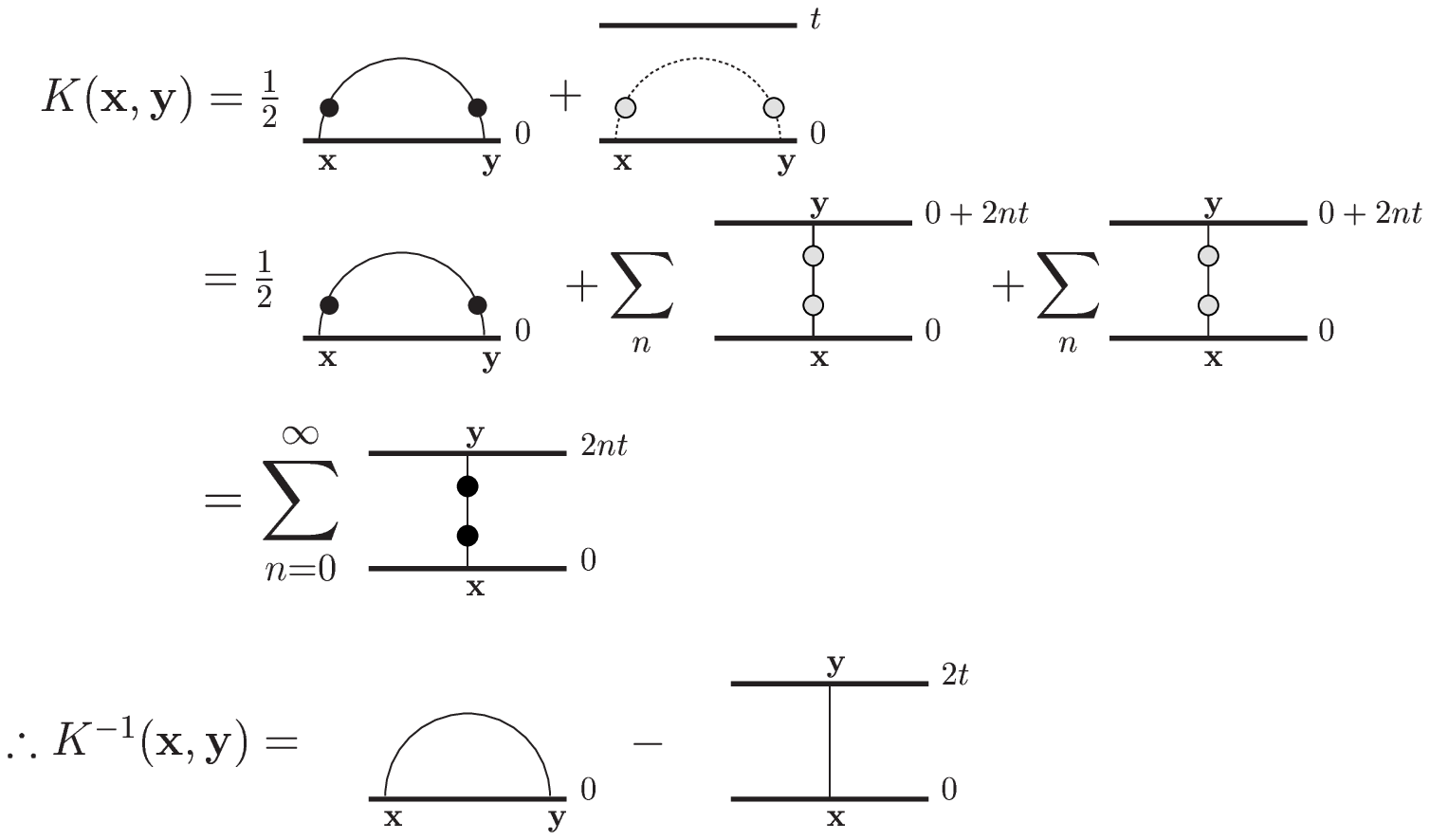}
\end{equation}
In the definition of $K$, the derivatives on the propagator lead to all the images entering with the same sign (plus). The indices in the diagrams can be checked using the finite dimensional case,
\begin{equation}
  \int\!\prod\limits_k\ud u_k\,\, e^{-\frac{1}{2}u_i A_{ij}u_j + v_j B_{ji} u_i} = \Det(A)^{-1/2} e^{\frac{1}{2}v_k B_{ki}A^{-1}_{ij} v_m B_{mj}}.
\end{equation}
We can check explicitly that the inverse is correct, using the gluing properties (\ref{prop-cases}).
\begin{equation}\label{evol-c-m3}
    \includegraphics[width=0.9\textwidth]{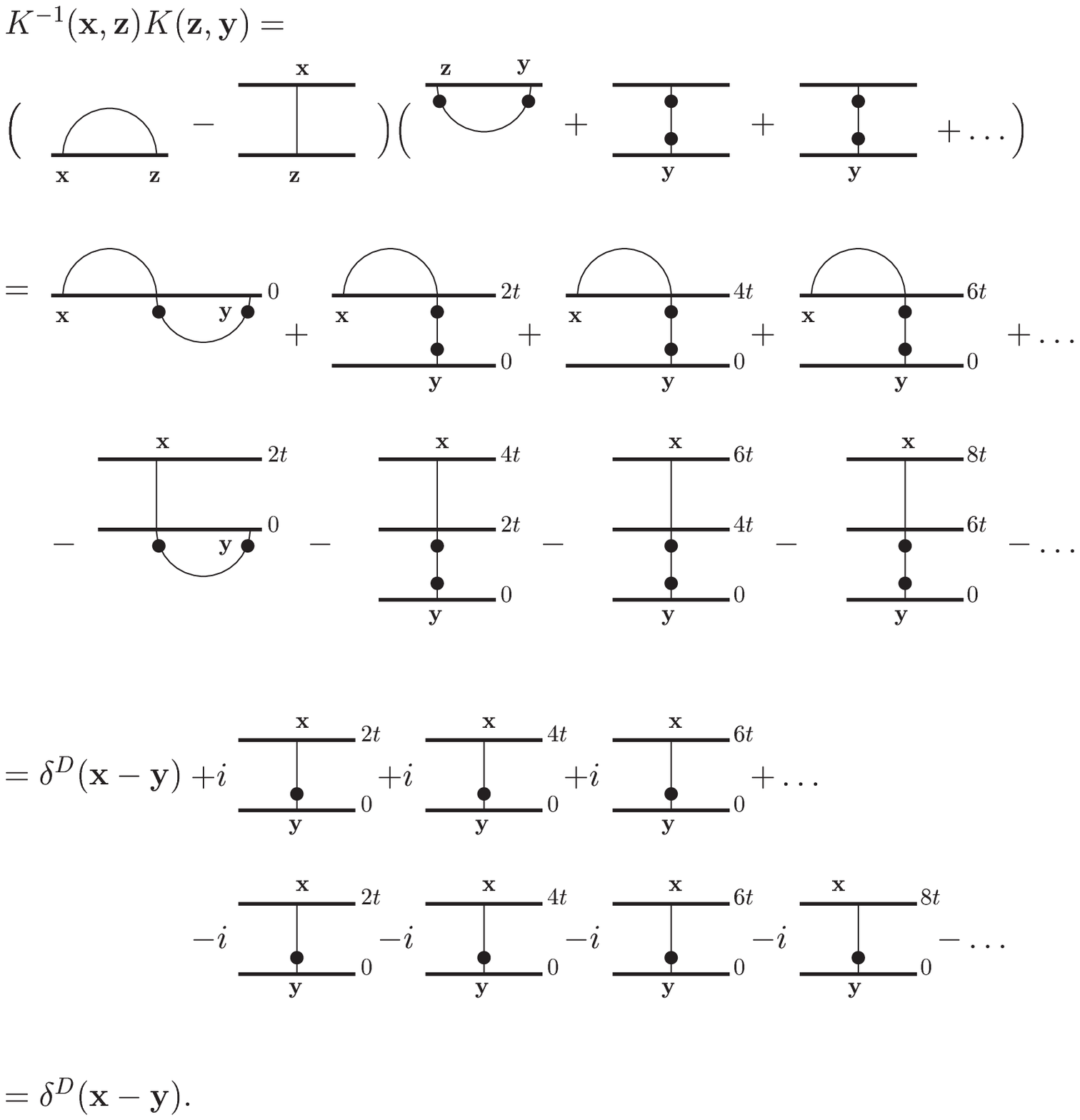}
\end{equation}
To understand the terms in the third line, either recall that gluing propagators which end on the same side of the boundary produces an image propagator, or, as we have illustrated, use the time dependence of the propagator to translate the diagrams. For example,
\begin{equation*}
    \includegraphics[width=0.3\textwidth]{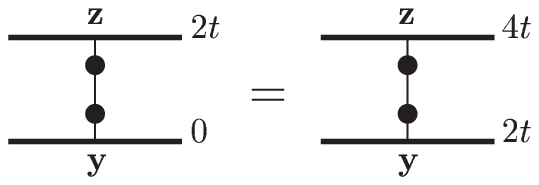}.
\end{equation*}
The Schr\"odinger functional term to be contracted with $K^{-1}$ is
\begin{equation}\label{evol-c-m4}
    \includegraphics[width=0.8\textwidth]{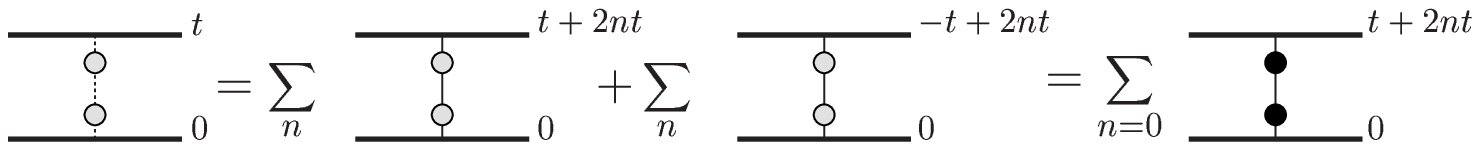}
\end{equation}
Carrying this out gives the final result of the Gaussian integration,
\begin{equation}\label{evol-c-m5}
    \includegraphics[width=\textwidth]{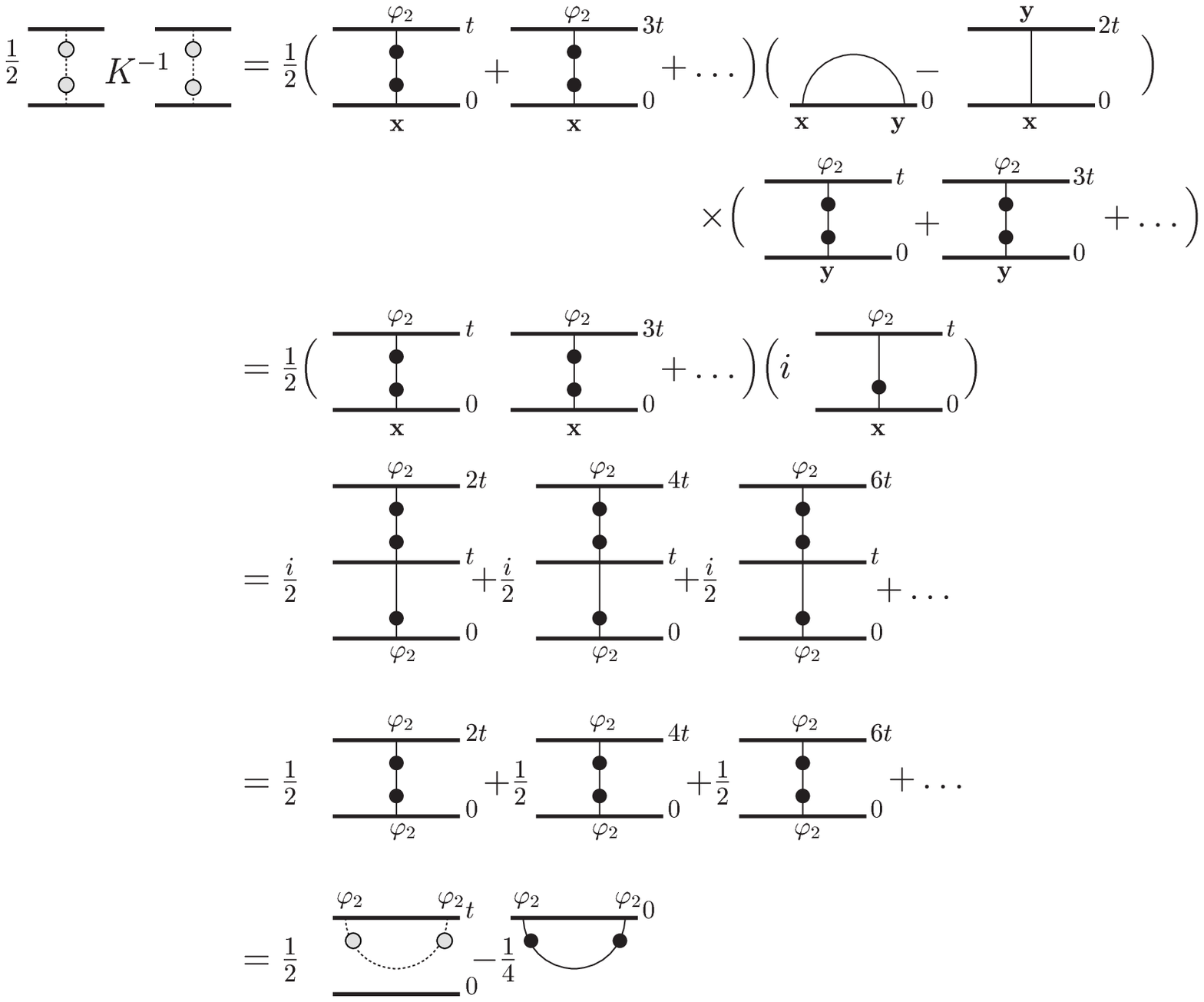}
\end{equation}
This removes the remaining term in (\ref{evol-c-m1}) and implies
\begin{equation}\label{evol-c-m6}
    \includegraphics[width=0.8\textwidth]{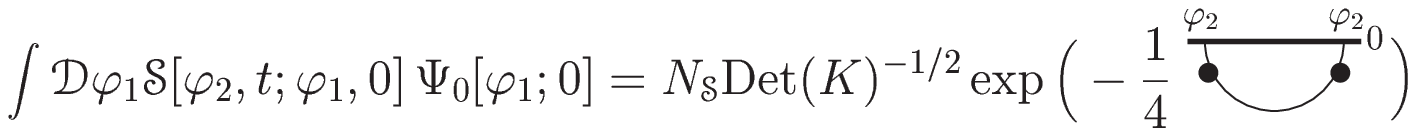}
\end{equation}
which is the correct diagram since the exponent should be independent of time. From (\ref{co-ord rep}) we see that the normalisation of the Schr\"odinger functional is the determinant of the Laplacian with Dirichlet conditions at times $0$ and $t$, to the power minus one half. We can write this as the determinant of the propagator $G_D$ to the power plus one half. Comparing (\ref{vac-evol}) and (\ref{evol-c-m6}) we have an unregulated expression for the free vacuum energy,
\begin{equation}
e^{-iE_0 t} = \frac{\Det^{1/2}(G_D)}{\Det^{1/2}(K)}.
\end{equation}

\subsection{Time dependence of the two-point function}\label{2-sect}

We will now apply our methods to an explicitly time dependent object. The two point function can be written in terms of the Schr\"odinger functional and the VWF,
\begin{equation}
  \langle \phi(\x,t)\phi(\y, 0)\rangle = \int\pathD (\varphi_2, \varphi_1)\Psi_0[\varphi_2]\, \varphi_2(\x)\,\, \mathscr{S}[\varphi_2, t; \varphi_1, 0]\,\varphi_1(\y)\,\Psi_0[\varphi_1].
\end{equation}
In the free theory the $\varphi_1$ integral is
\begin{equation}\label{2point1cm}
    \includegraphics[width=0.8\textwidth]{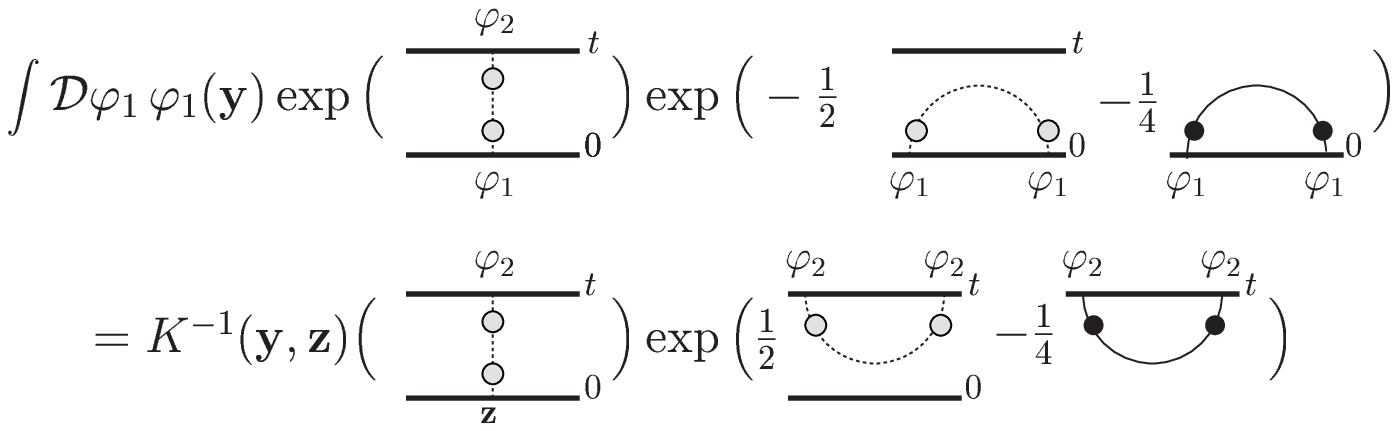}
\end{equation}
The exponential terms are the same as for the evolution of the vacuum, but the insertion of $\varphi_1(\y)$ brings down the leading factor. Again the indices can be checked by comparison with the finite dimensional case. The remaining integral over $\varphi_2$ ties together $\x$ and $\y$ giving us

\begin{equation*}
    \includegraphics[width=0.7\textwidth]{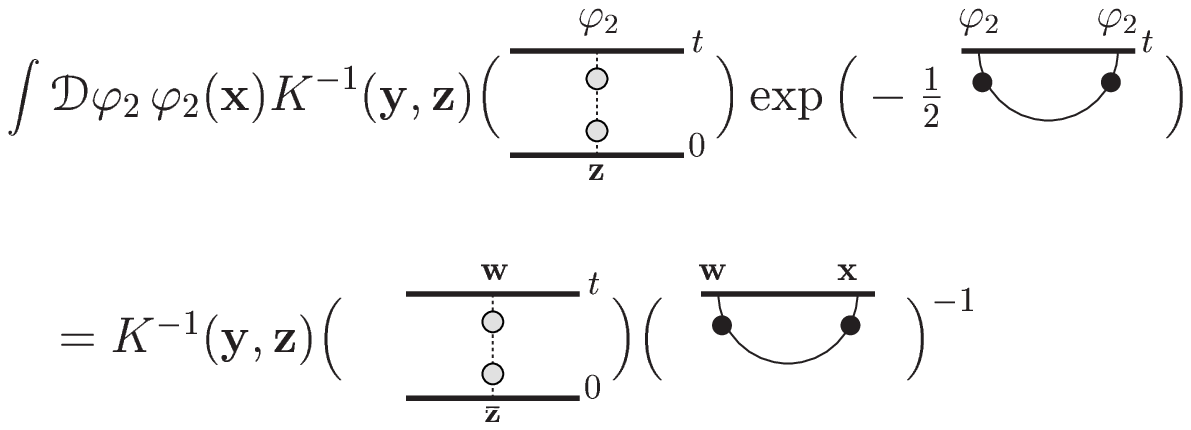}
\end{equation*}
\begin{equation}\label{2point2cm}
   \hspace{2.33cm}\includegraphics[width=0.7\textwidth]{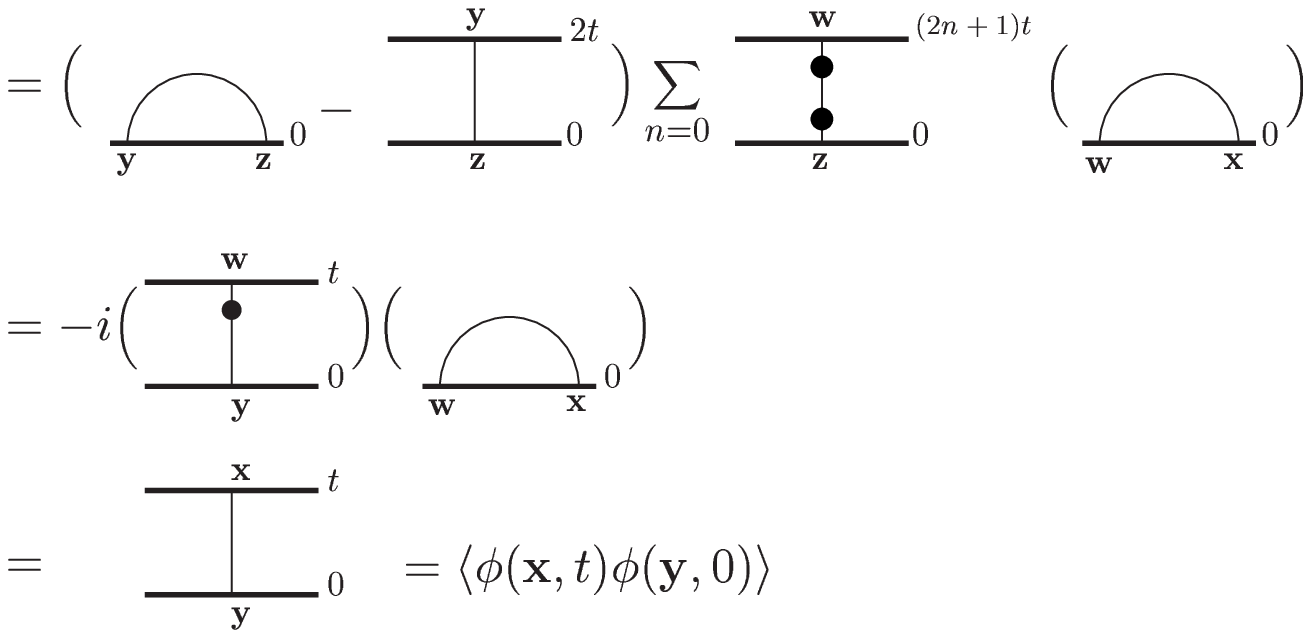}
\end{equation}
Using the gluing property alone we have shown that equation (\ref{schro-diags}) for the Schr\"odinger functional leads to the correct result for the two point function at unequal times. This argument is invertible; If we know the two point function we can construct the Schr\"odinger functional as in (\ref{schro-diags}) provided the gluing property holds. If we can generalise the gluing property to string theory we can repeat the diagrammatic arguments and construct the second quantised string Schr\"odinger functional.

\section{Interacting, Euclidean, field momentum extensions}

We begin this section by describing application of our graphical techniques to interacting theories, using the example of constructing the vacuum state in $\phi^4$ theory.

We then show that our results hold equally well in Euclidean space, in preparation for string theory where the Euclidean Polyakov integrals are better defined. We will also examine the properties of the Schr\"odinger functional in the field momentum representation, which will be later related to T-duality.

\subsection{Reconstructing the vacuum functional}

The aim of this section is to use diagrammatic methods to show that the vacuum wave functional, $\Psi_0[\phi(\x)] \equiv\bracket{\phi(\x)}{0}$, can be constructed from the requirement that it must generate known results for vacuum expectation values (equal time correlation functions) through
\begin{equation}
\begin{split}
\langle\,\phi(\x_1,0)\cdot\cdot\cdot\phi(\x_n,0)\,\rangle &\equiv \bra{0}\hat\phi(\x_1,0)\ldots\hat\phi(\x_n,0)\ket{0} \\
&= \int \pathD\phi(\x)\,\, \phi(\x_1)\cdot\cdot\cdot \phi(\x_n)\,\, |\Psi_0[\phi]|^2.
\label{corr}
\end{split}
\end{equation}
We will focus on massive $\lambda\phi^4$ theory and build the VWF perturbatively by order in $\hbar$ and $\lambda$, using the known diagram expansion of $n$--point functions, where the gluing property will play a central role (the diagram expansion of the  VWF has been considered previously, \cite{Paul-Vevs}, but we are taking a different approach).

We determined the free field contribution earlier using precisely this approach for $n=2$. If we consider interactions then the logarithm of the VWF has an expansion not only in powers of $\lambda$ but in $\hbar$ also. We expand the logarithm, $W[\phi]$, as
\begin{equation*}
W[\phi]=\frac{1}{\hbar}\sum_{n=1}^\infty \frac{i^{2n}}{(2n)!}\int\!\ud^D(\x_1\ldots\x_{2n})\,\,
\phi(\x_1) \cdots\phi(\x_{2n})\, \G_{2n}(\x_1,\ldots,\x_{2n})\,
\end{equation*}
where each of the $\Gamma_{2n}$ has an expansion in powers of $\hbar$,
\begin{equation}
  \G_{2n} = \G^0_{2n} + \G^\hbar_{2n} + \G^{\hbar^2}_{2n} + \cdots
\end{equation}
with their dependence given by the superscript. There can be no correction to $\Gamma_2^0$ (defined in \ref{g20-def}) from the addition of interactions since this would give a leading order correction to the two point function, but we know that the first corrections are of order $\lambda\hbar^2$,
\begin{equation}
\includegraphics[width=0.6\textwidth]{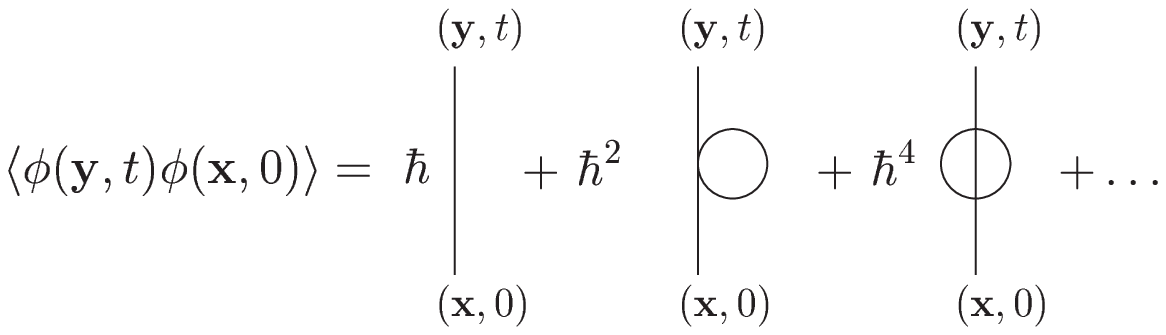}
\label{2ptnorm}
\end{equation}
Therefore (\ref{inverse}) still holds.

Keeping only $\G^0_2$ in the exponent and expanding the other contributions to $W[\phi]$, call this
$\overline{W}[\phi]$, yields vacuum expectation values
\begin{equation}
\langle\,\phi(\x_1,0)\ldots\phi(\x_{2n},0)\,\rangle = \int\pathD\phi \,\, e^{-\int\!\phi\,\G^0_2\,\phi} \sum\limits_{m=0}\frac{(2\overline{W}[\phi])^{2m}}{(2m)!}\, \phi(\x_1)\cdots\phi(\x_{2n}),
\end{equation}
so in general we have to contract $\phi(\x_1)\ldots\phi(\x_{2n})$ with the $\phi$ in the diagrams contributing to $\overline{W}$ using the inverse of $2\G^0_2$ which is the equal time propagator in free space.

To identify the first interaction vertex $\G_4^0$ consider the (connected) four point function
\begin{equation}\label{4point1}
  \includegraphics[width=0.9\textwidth]{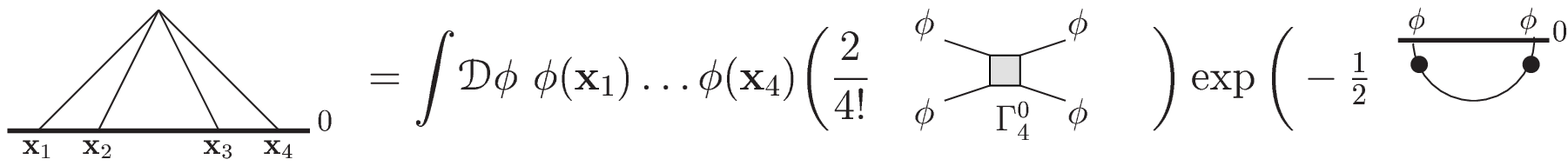}
\end{equation}
The four external legs are sewn to the four fields from $\Gamma^0_4$ using the equal time propagator, at a single vertex , which must give us the usual four point function restricted to the boundary. We can invert the propagators attached to $\G^0_4$ using $2\G^0_2$ and we find
\begin{equation}\label{4point2}
  \includegraphics[width=0.55\textwidth]{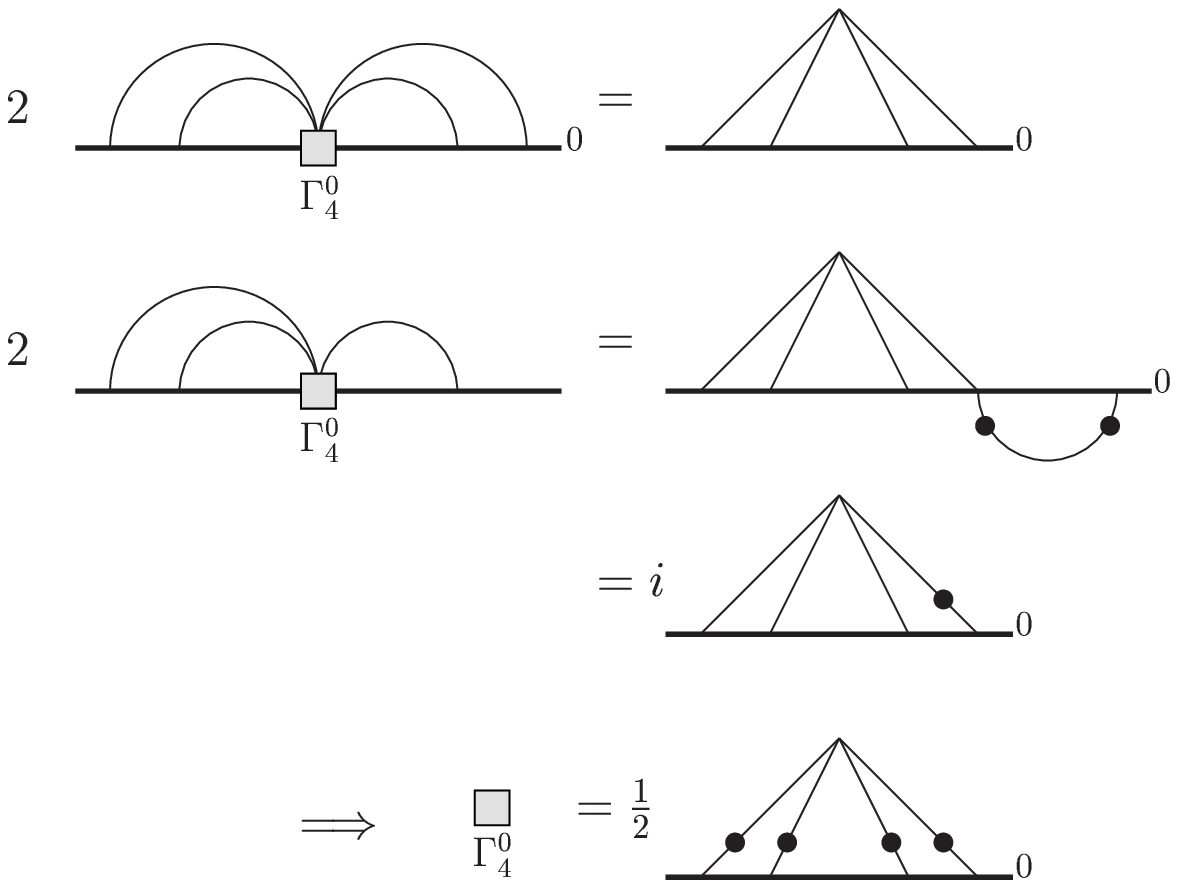}
\end{equation}
Keeping track of the combinatoric factors the vertex is
\begin{equation}
  \G_4^0(\x_1,\ldots,\x_4) = \frac{\lambda}{2}\int\!\bigg[\prod\limits_{i=1}^4 \frac{\ud\pp_i}{(2\pi)^D} e^{-i\pp.\x_i}\bigg]\,\frac{\dirac{\pp_1+\ldots+\pp_4}}{(E(\pp_1)+\ldots + E(\pp_4))}.
\end{equation}
All the tree level contributions $\G^0_{2n}$ can be similarly derived from considering the tree level $2n$-point function at equal time, as we have done for $n=1, 2$ above.

Using $\G^0_4$ we can determine the one loop correction $\G^\hbar_2$. The one loop self energy graph in the two point function is of order $\lambda\hbar^2$, so the only terms contributing are $\Gamma^0_4$ and $\Gamma^\hbar_2$. Expanding to first order in $\lambda$ we calculate
\begin{equation}
  \includegraphics[width=0.9\textwidth]{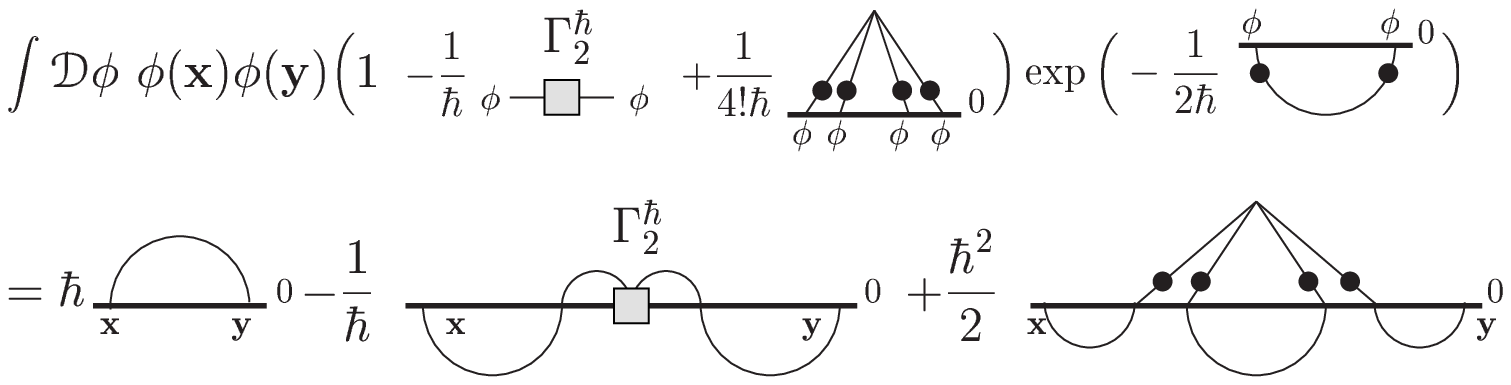}
\end{equation}
To make sense of the final diagram we appeal to the gluing property with appropriate conditions $t_1, t_2\lessgtr 0$,
\begin{equation}\label{image}
\begin{split}
  \int\!\ud^D(\x,\y)\bigg(2\frac{\partial}{\partial t}G_0(\x, x_1; \y, t)\bigg)&G_0(\y, 0; \z, 0)\bigg(2\frac{\partial}{\partial t'}G_0(\z, t'; \x_2, t_2)\bigg)\bigg|_{t'=t=0} \\
  &=-G_I(\x_1, x_1; \x_2, t_2)
\end{split}
\end{equation}
so the effect of gluing two free propagators with the inverse of $2\G^2_0$ on the boundary is to produce the image propagator. The external legs become propagators as in (\ref{4point2}), and the internal lines become an image propagator loop, denoted by an $I$,
\begin{equation}
  \includegraphics[width=0.55\textwidth]{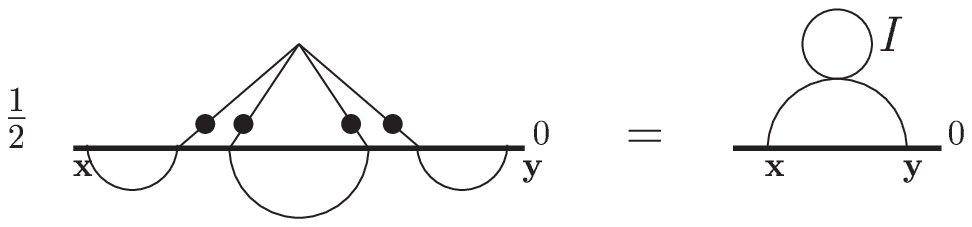}.
\end{equation}
The factor of $1/2$ on the left is the symmetry factor implicit in the diagram on the right. Loops of image and free propagators are unequal, so this term must be removed by $\Gamma^\hbar_2$. We are led to the result
\begin{equation}\label{vac-quad}
  \includegraphics[width=0.9\textwidth]{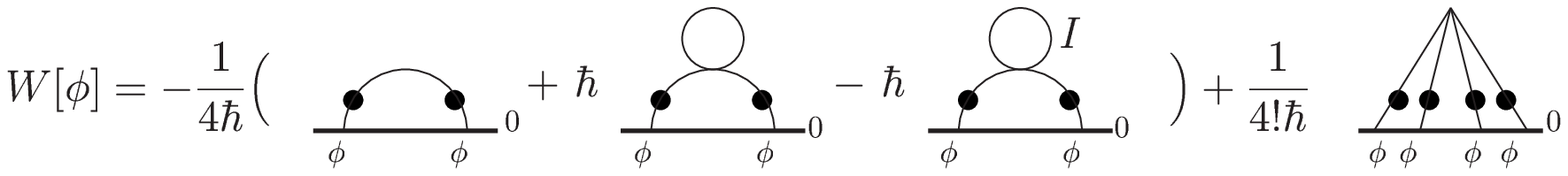}
\end{equation}

\subsection{The VWF from a sum over paths}

The conventional method of constructing the VWF \cite{Symanzik} is via a large time path integral. If we apply the time evolution operator $\exp(-i\hat{H}t/\hbar)$ to any state $|\,\upsilon\,\rangle$ not orthogonal to the vacuum, then for large times
\begin{equation}
\exp(-i\hat{H}t/\hbar)|\,\upsilon\,\rangle\sim |\,0\,\rangle e^{-iE_0t/\hbar}\langle\,0\,|\,\upsilon\,\rangle\,\,\,\,(t\rightarrow \infty)
\end{equation}
where $E_0$ is the energy of the vacuum, and the larger energy eigenvalues cause rapid oscillations which do not contribute. Thus
\begin{equation}
\Psi[\phi]= \lim_{t\rightarrow \infty} N e^{iE_0 t/\hbar}\langle\,D\,|e^{i\int d\x\,\, \phi(\x)\hat{\pi}
(\x,0)/\hbar}e^{-i\hat{H}t/\hbar}|\,\upsilon\,\rangle
\end{equation}
(the overall normalisation of the vacuum will not concern us here, so we set it equal to one). We obtain the path integral representation following the same arguments as in section \ref{Schro-sect},
\begin{equation}\label{VWF-a rep}
\Psi_0[\phi] = \int \pathD \f(\x,t) \, e^{iS[\f]/\hbar}{\Big |}_{\f(\x,0)=0}
\end{equation}
with action
\begin{equation}
S[\f]= \int\limits_{-\infty}^0\!\ud t\!\int\! \ud^D\x\, \frac{1}{2}\bigg(\dot\varphi^2 - (\nabla\varphi)^2 - m^2\varphi^2 \bigg)- \frac{\lambda}{4!}\varphi^4
+\int\!\ud^D\x\,\dot{\f}(\x,0)\phi.
\label{newact}
\end{equation}
The boundary condition on the field at $t=-\infty$ is that it must be regular. Standard field theory results then imply that the logarithm $W[\phi]$ of the VWF is given by the sum of connected diagrams constructed from the new action \rf{newact}. There is a boundary at $t=0$ where the field vanishes; the propagators $G_{d(0)}$ in the Feynman diagrams satisfy Dirichlet conditions on the boundary, where all external legs must end with a time derivative. Using the notation we introduced in section \ref{Schro-sect} we can write $W[\phi]$ generated from \rf{newact} as
\begin{equation}  \label{logpsi-m}
  \includegraphics[width=0.9\textwidth]{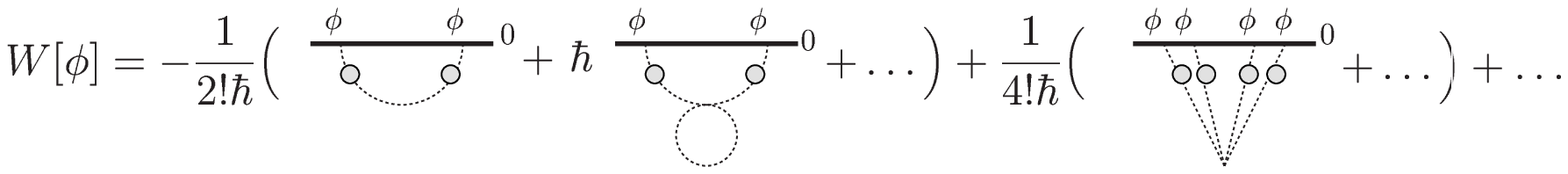}
\end{equation}
Let us compare this with our expression (\ref{vac-quad}). The lowest order contribution comes from turning off the interaction and doing the Gaussian integration in \rf{VWF-a rep}, which implies
\begin{equation}\label{compare}
  \frac{1}{2}\Gamma^0_2(\x, \y) = \frac{1}{2}\frac{\partial^2}{\partial t\partial t'} G_{d(0)}(\x,t,\y,t')\bigg|_{t=t'=0}
\end{equation}
By the method of images the propagator is
\begin{equation}
  \begin{split}
  G_{d(0)}(\x,t,\y,t')&=G_0(\x,t,\y,t')-G_0(\x,t,\y,-t') \\
  &= G_0(\x,t,\y,t')-G_I(\x,t,\y,t').
  \end{split}
\label{D-prop}
\end{equation}
Using this we can verify the time independence of the vacuum wave functional; for propagators $G_{d(t)}$ which obey Dirichlet conditions on the plane at time $t$ we have
\begin{equation}
  \ddot{G}_{d(t_1)}(\x,t_1,\y,t_1) = \ddot{G}_{d(t_2)}(\x,t_2,\y,t_2).
\end{equation}
We can now verify the consistency of both (\ref{inverse}) and (\ref{compare}). In the latter, the differentiation leads to $G_0$ and its image contributing equally and implies
\begin{equation}
  \Gamma_2^0(\x,\y) = 2i\int\!\frac{\ud^D\kk\,\ud k^0}{(2\pi)^{D+1}}\frac{k_0^2}{k_0^2 - E(\kk)^2 +i\varepsilon}e^{-i\kk.(\x-\y) - ik_0\epsilon}.
\end{equation}
The $k^0$ integral becomes divergent, so we have introduced the regulator. Carrying out the $k_0$ integral and taking the regulator to zero we find
\begin{equation}
  \Gamma_2^0(\x,\y) =2\lim_{\epsilon=0} \int\!\frac{\ud^D\kk}{(2\pi)^{D+1}}\,\pi\,E(\kk)\,e^{i\kk.(\x-\y) - i\epsilon E(\kk)} = \int\frac{\ud^D\mathbf{k}}{(2\pi)^D}\,E(\kk)\,\, e^{-i\mathbf{k}\cdot(\x-\y)}.
\end{equation}
recovering \rf{inverse}. The interaction part of the Lagrangian is only integrated over half of all space. To interpret \rf{logpsi-m} in terms of ordinary Feynman diagrams, we can use the symmetries of the propagators attached to the boundary to write the interaction vertex integrated over all space by inserting a factor of $1/2$ in front of the diagram. Using (\ref{D-prop}) the one loop term is
\begin{equation}
  \includegraphics[width=0.55\textwidth]{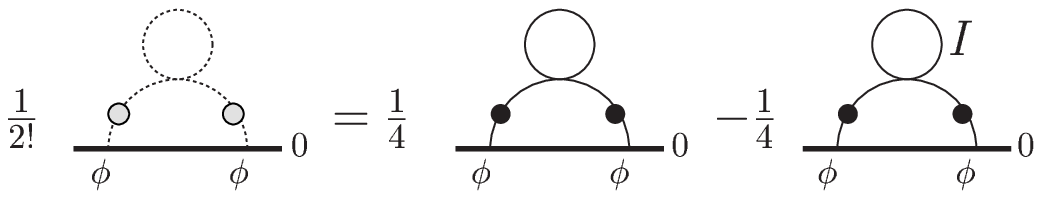}
\end{equation}
in agreement with our result \rf{vac-quad}. 

In summary, we have seen that the VWF has a diagram expansion in a propagator obeying Dirichlet boundary conditions, and this can be constructed using the gluing properties. To complete the discussion of the VWF we describe how the propagator is given in first quantisation. If we identify spacetime points with their reflection in $t=0$, the quantisation surface, then when we sum over paths from $(\mathbf{x},t_1)$ to $(\mathbf{y},t_2)$ we have to include paths from $(\mathbf{x},t_1)$ to $(\mathbf{y},-t_2)$, the reflection of $(\mathbf{y},t_2)$,
\begin{equation*}
\includegraphics[width=0.25\textwidth]{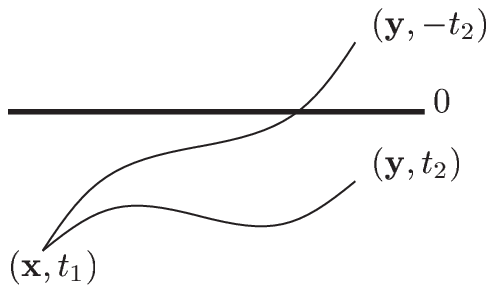}
\end{equation*}
Now weight paths with a minus sign each time they cross the surface $t=0$. Paths from $(\mathbf{x},t_1)$ to $(\mathbf{y},-t_2)$ must cross the quantisation surface an odd number of times and acquire an overall minus sign, whereas paths directly from $(\mathbf{x},t_1)$ to $(\mathbf{y},t_2)$ cross an even number of times and are weighted with an overall plus sign. The contribution from the latter paths gives $G_0$ and from the former $- G_I$ in expression \rf{D-prop} for the Dirichlet propagator $G_{d(0)}$.

\subsection{Euclidean extension}

Rotating $x^0\rightarrow -ix^0$ the Euclidean action is
\begin{equation}
  S_E[\phi] = \int\!\ud^{D+1}x\,\,\frac{1}{2}\big(\dot{\phi}^2 + \nabla\phi.\nabla\phi + m^2\phi^2\big) + \frac{\lambda}{4!}\phi^4.
\end{equation}
and the Euclidean propagator is
\begin{equation}\label{E-prop}
\begin{split}
  G(\x_2,t_2;\x_1,t_1) &= \int\limits_0^\infty\!\ud T\,\,\frac{1}{(4\pi T)^{(D+1)/2}}e^{-(x_f-x_i)^2/4T-m^2T} \\
  &= \int\!\frac{\ud^{D+1}k}{(2\pi)^{D+1}}\, \frac{e^{-ik\cdot(x_f - x_i)}}{k_ik_i + m^2} =\int\!\frac{\ud^{D}\kk}{(2\pi)^D}\, \frac{e^{-i\kk\cdot(\x_f - \x_i)}}{2E(\kk)}e^{ - E(\kk)|t_f-t_i|}.
\end{split}
\end{equation}
In Euclidean space the propagator factorises as
\begin{equation}\label{E-glue}
  G_0(\x_2,t_2;\x_1,t_1) = \int\!\ud^D\x\,G_0(\x_2,t_2;\x,t)\overleftrightarrow{\frac{\partial}{\partial t}}G_0(\x,t;\x_1,t_1)\quad\ t_2>t>t_1.
\end{equation}
The gluing properties are
\begin{equation}
\int\! \ud^D\y \,\, G_0(\x_2,t_2;\y,t)\left(-2\frac{\p}{\p t}\right)G_0(\y,t;\x_1,t_1)=
\left\{ \ba{cr}
{\dis  G_0(\x_2,t_2;\x_1,t_1)} & t_2> t>t_1
\\
{\dis  -G_0(\x_2,t_2;\x_1,t_1)} & t_1>t>t_2
\\
{\dis  G_I(\x_2,t_2;\x_1,t_1)} & t>t_1,t_2
\\
{\dis  -G_I(\x_2,t_2;\x_1,t_1)} & t<t_1,t_2
\ea
\right.
\label{prop-cases-E}
\end{equation}
where the only difference is the now missing factor of $i$ and the equal time inverse picks up a minus sign, 
\begin{equation}\label{inversions-E}
  \includegraphics[width=0.45\textwidth]{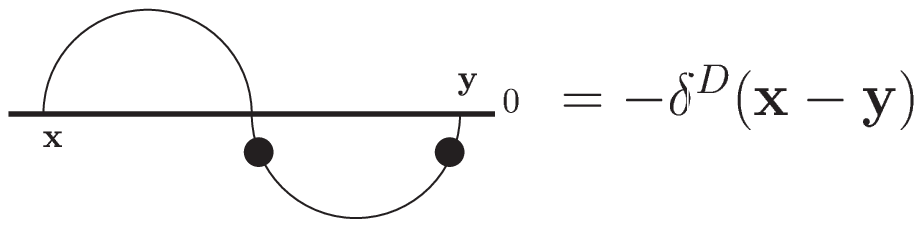}.
\end{equation}
In Euclidean space the vacuum is given by applying the imaginary time evolution operator $\exp(-\hat{H}t)$ to any state, $|\,\upsilon\,\rangle$, not orthogonal to the vacuum, then for large times
\begin{equation*}
\exp(-\hat{H}t)|\,\upsilon\,\rangle\sim |\,0\,\rangle e^{-E_0t}\langle\,0\,|\,\upsilon\,\rangle\,\,\,\,(t\rightarrow \infty)
\end{equation*}
where $E_0$ is the energy of the vacuum, and the larger energy eigenvalues are exponentially damped. The path integral representation now follows as before,
\begin{equation*}
  \Psi_0[\phi] = \int\pathD\varphi\, \exp\bigg(-S[\varphi] - \int\!\ud^D\x\,\dot\varphi(\x,0)\phi(\x)\bigg)\bigg|_{\varphi(\x,0)=0}
\end{equation*}
with the action evaluated on the half space $t<0$. The Schr\"odinger functional has a similar expression,
\begin{equation*}
  \mathscr{S}[\phi_2,t_2;\phi_1,t_1] = \int\pathD\varphi\,\exp\bigg(-S_E[\varphi] - \int\!\ud^D\x\, \dot\varphi(\x,t_2)\phi_2(\x) + \int\!\ud^D\x\, \dot\varphi(\x,t_1)\phi_1(\x)\bigg)\bigg|^{\varphi(t_2)=0}_{\varphi(t_1)=0}.
\end{equation*}
All of the calculations we have presented can be repeated in Euclidean space, but the only differences to keep track of are minus signs. Instead we will remain in Euclidean space but use a representation of the state space which will, in Section 5, link us to T-duality.

\subsection{Field momentum representation}

In this section we work in a basis in which the field momentum is diagonal,
\begin{equation}
  \bra{\pi}\hat\pi(\x) = \pi(\x)\bra{\pi}, \quad i\frac{\delta}{\delta \pi(\x)}\bra{\pi} = \bra{\pi}\hat{\phi}(\x).
\end{equation}
This representation can be viewed simply as a functional Fourier transform on the space of state functionals,
\begin{equation}
  \Psi[\pi,t]:=\int\pathD\phi\,\exp\bigg(i\int\!\ud^D\x\, \pi(\x)\phi(\x)\bigg)\Psi[\phi,t]
\end{equation}
but there is more to learn by working from first principals. The free field VWF is
\begin{equation*}
\begin{split}
  \Psi_0[\pi] \equiv \bracket{\pi}{\Psi_0} &= \int\pathD\phi\, \exp\bigg( -S_E[\phi] -i\int\!\ud^D\phi(\x)\pi(\x)\bigg)\bigg|_{\dot\phi(\x,0)=0} \\
  &= \exp\bigg(-\frac{1}{2}\int\!\ud^D(\x,\y)\pi(\x)G_{n(0)}(\x,0,\y,0)\pi(\y)\bigg)
\end{split}
\end{equation*}
where the propagator $G_{n(0)}$ obeys Neumann conditions on the boundary at $t=0$. A dashed line will represent a propagator with Neumann conditions as the dotted line did for Dirichlet so we write
\begin{equation}
  \includegraphics[width=0.37\textwidth]{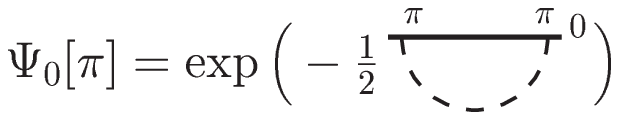}
\end{equation}
As in earlier sections, a variety of methods can be used to reconstruct the VWF without the definition above. By defining $G_0$ at equal times as a vacuum expectation value we find
\begin{equation}
\includegraphics[width=0.8\textwidth]{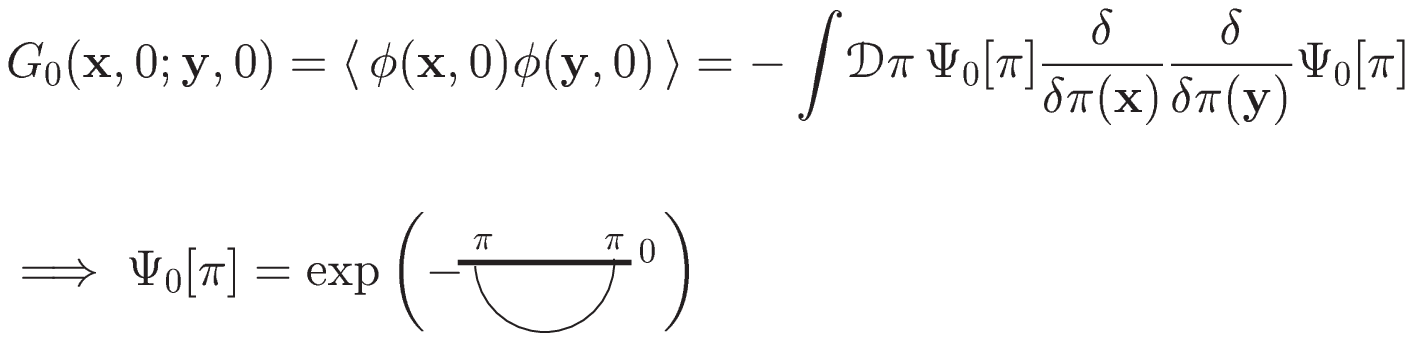}
\end{equation}
The two above expressions for the vacuum are equivalent, as can be seen from the method of images,
\begin{equation}
  G_{n(0)}(\x,t_f,\y, t_i) = G_0(\x,t_f,\y, t_i) + G_0(\x,t_f,\y, -t_i).
\end{equation}
The Schr\"odinger functional now describes imaginary time evolution. It is defined by an expression analogous to (\ref{schr-def}),
\begin{equation}
\begin{split}
  \mathscr{S}[\pi_2, t; \pi_1, 0] &= \bra{\pi_2}e^{-H(t)}\ket{\pi_1} \\
  &= \bra{N}e^{-i\int\pi_2\hat\phi}e^{-Ht}e^{i\int\pi_1\hat\phi}\ket{N} \\
  &= \int\pathD\phi\, \exp\bigg( -S_E[\phi] -i\int\!\ud^D\phi(\x,t)\pi_2(\x)+ i\int\!\ud^D\phi(\x,0)\pi_1(\x)\bigg)\bigg|^{\dot\phi(\x,t)=0}_{\dot\phi(\x,0)=0}
\end{split}
\end{equation}
where $\bra{N}$ is the Neumann state $\bra{N}\hat\pi=0$. The result of the free field integral is
\begin{equation}\label{sf-mom rep}
\begin{split}
	\mathscr{S}[\pi_2, t; \pi_2, 0] = \exp\bigg( -\iint\ud^D\x\ud^D\y&\,\, \frac{1}{2}\pi_1(\y)G_\text{orb}(\y, 0,\x,0))\pi_1(\x) \\
	&- \pi_2(\y) G_\text{orb}(\y, t, \x, 0)\pi_1(\x) \\
	&+ \frac{1}{2}\pi_2(\y) G_\text{orb}(\y, t, \x, t) \pi_2(\x) \bigg)
\end{split}
\end{equation}
where $G_\text{orb}$ obeys Neumann, rather than Dirichlet, boundary conditions on $x^0=t$ and $x^0=0$, and there is no longer a time derivative on the propagators since this has been moved into the fields,  $\pi=\dot\phi$. In terms of Feynman diagrams
\begin{equation}\label{schro-diags}
    \includegraphics[width=0.9\textwidth]{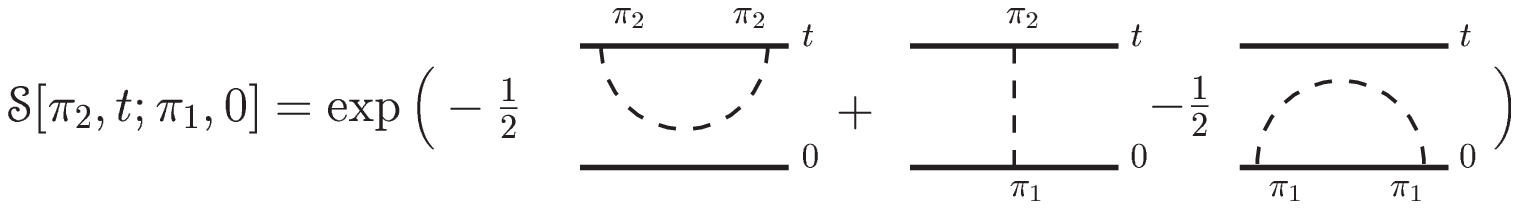}
\end{equation}
The method of images with the free space propagator gives us the required boundary conditions,
\begin{equation}\label{Orb-prop def}
  G_orb(\x_f, t_f, \x_i, t_i) = \sum\limits_{n\in\mathbb{Z}}\,\, G_0(\x_f, t_f, \x_i, t_i + 2nt) + G_0(\x_f, t_f, \x_i, -t_i + 2 nt).
\end{equation}
The sum over images is written as a sum over paths by again opacifying the time direction on the $\mathbb{S}^1/\mathbb{Z}_2$ orbifold of radius $t/\pi$ but without the minus sign weightings of the representation in which the field was diagonal. Therefore the Schr\"odinger functional is constructed from a sum over paths of particles moving on $\mathbb{S}^1/\mathbb{Z}_2\times\mathbb{R}^D$.

Although the following calculations are similar to those in sections (\ref{vac-sect}) and (\ref{2-sect}), we present them for completeness. We begin with the time dependence of the vacuum. The free field integral evolving the vacuum between times $0$ and $t$ is
\begin{equation}\label{sh-evol1}
    \includegraphics[width=\textwidth]{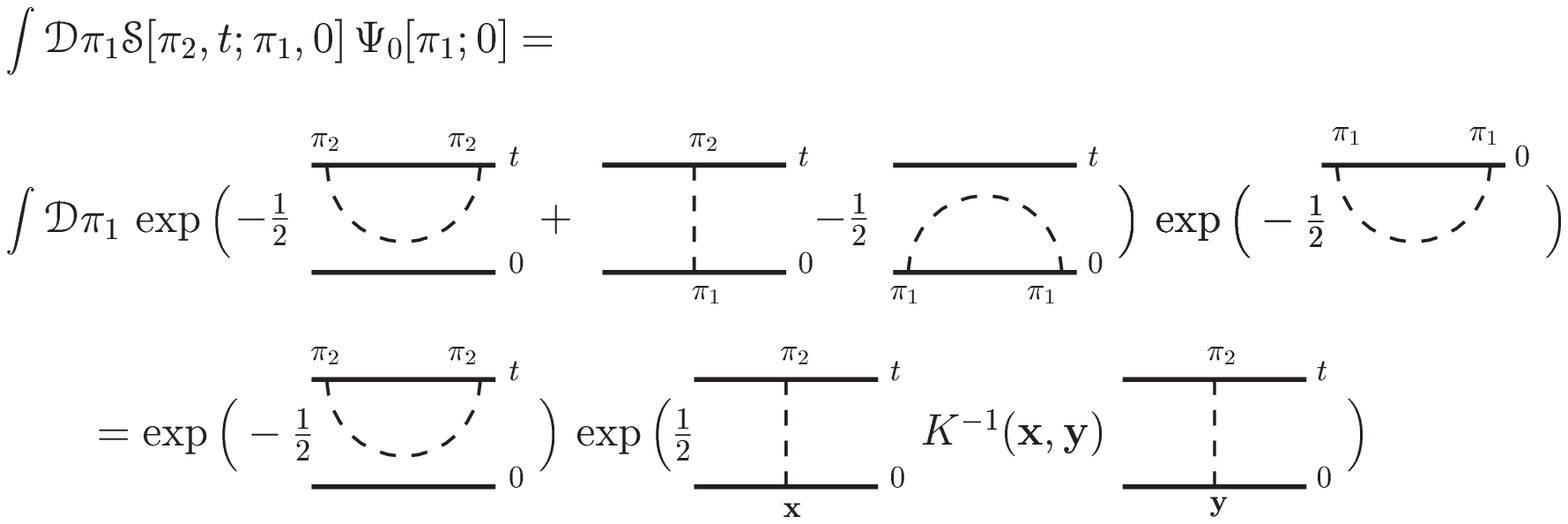}
\end{equation}
The first three terms are the Schr\"odinger functional, the final term is the vacuum. This time the operator $K$ and its inverse are given by
\begin{equation}\label{sh-evol2.eps}
  \includegraphics[width=0.7\textwidth]{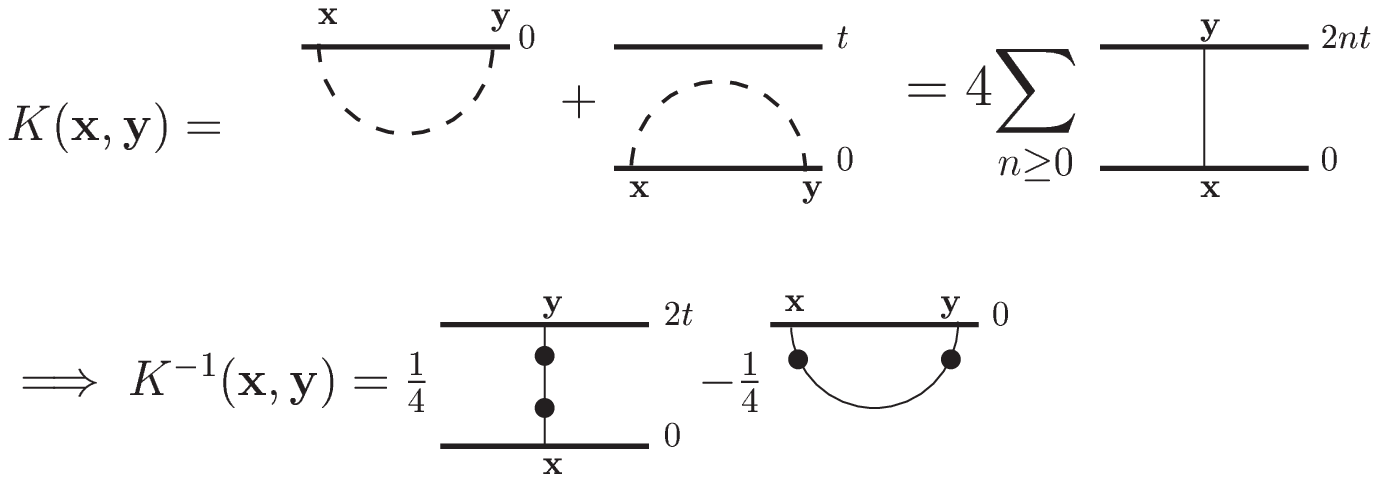}  
\end{equation}
Again we can check that the inverse is correct,
\begin{equation}\label{sh-evol3}
    \includegraphics[width=\textwidth]{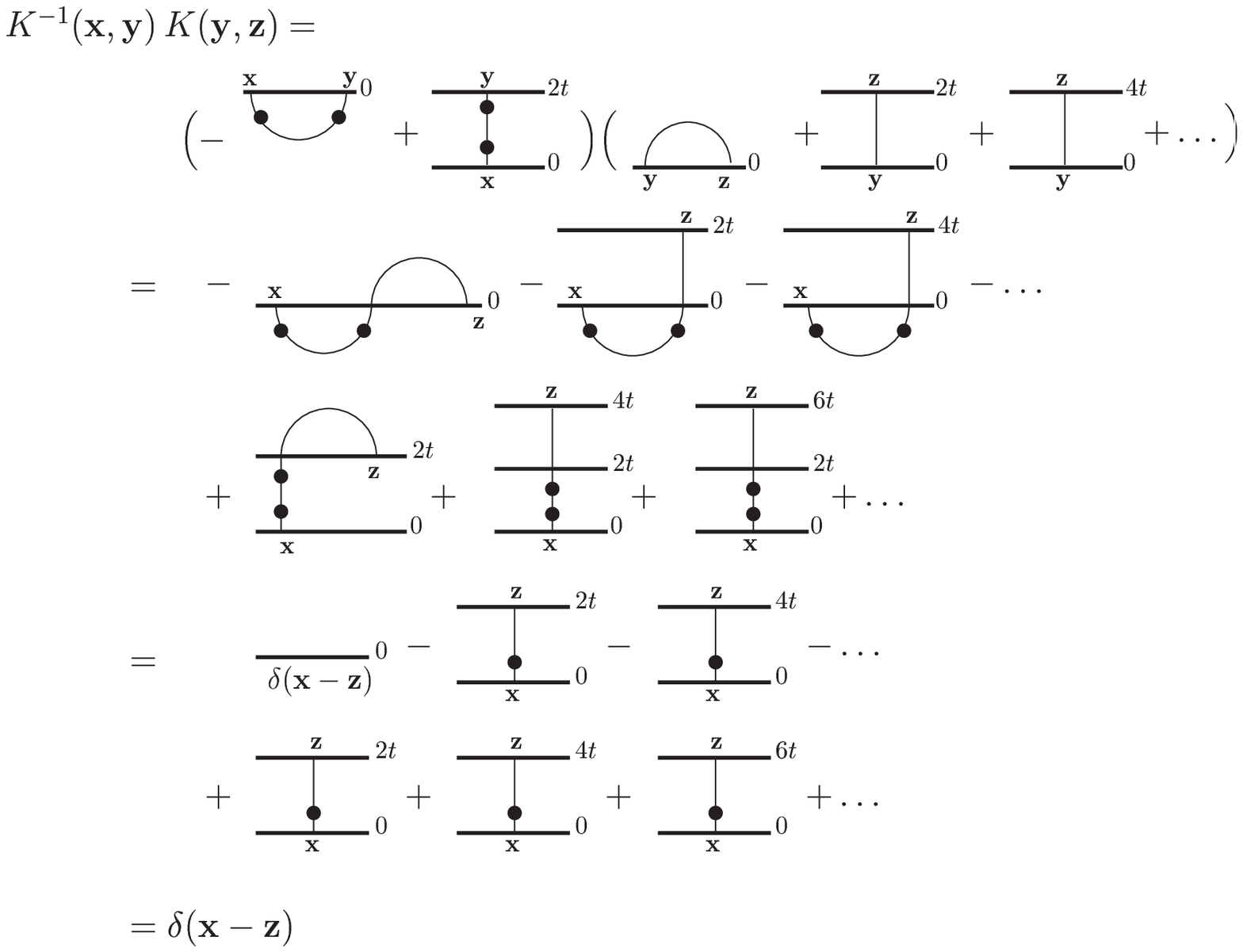}
\end{equation}
The Schr\"odinger functional term to be contracted with $K^{-1}$ is
\begin{equation}
    \includegraphics[width=0.85\textwidth]{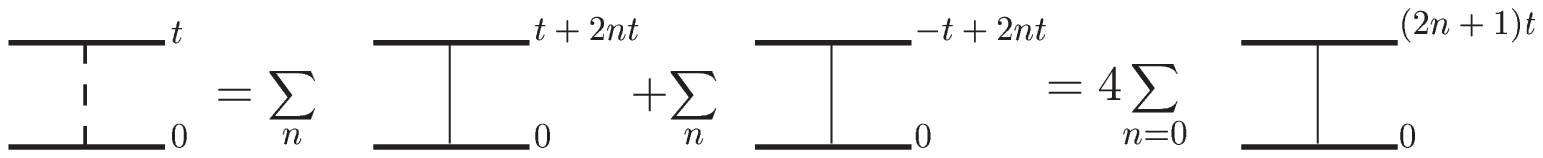}
\end{equation}
Finally, the Gaussian integral gives
\begin{equation}\label{sh-evol5.eps}
  \includegraphics[width=0.9\textwidth]{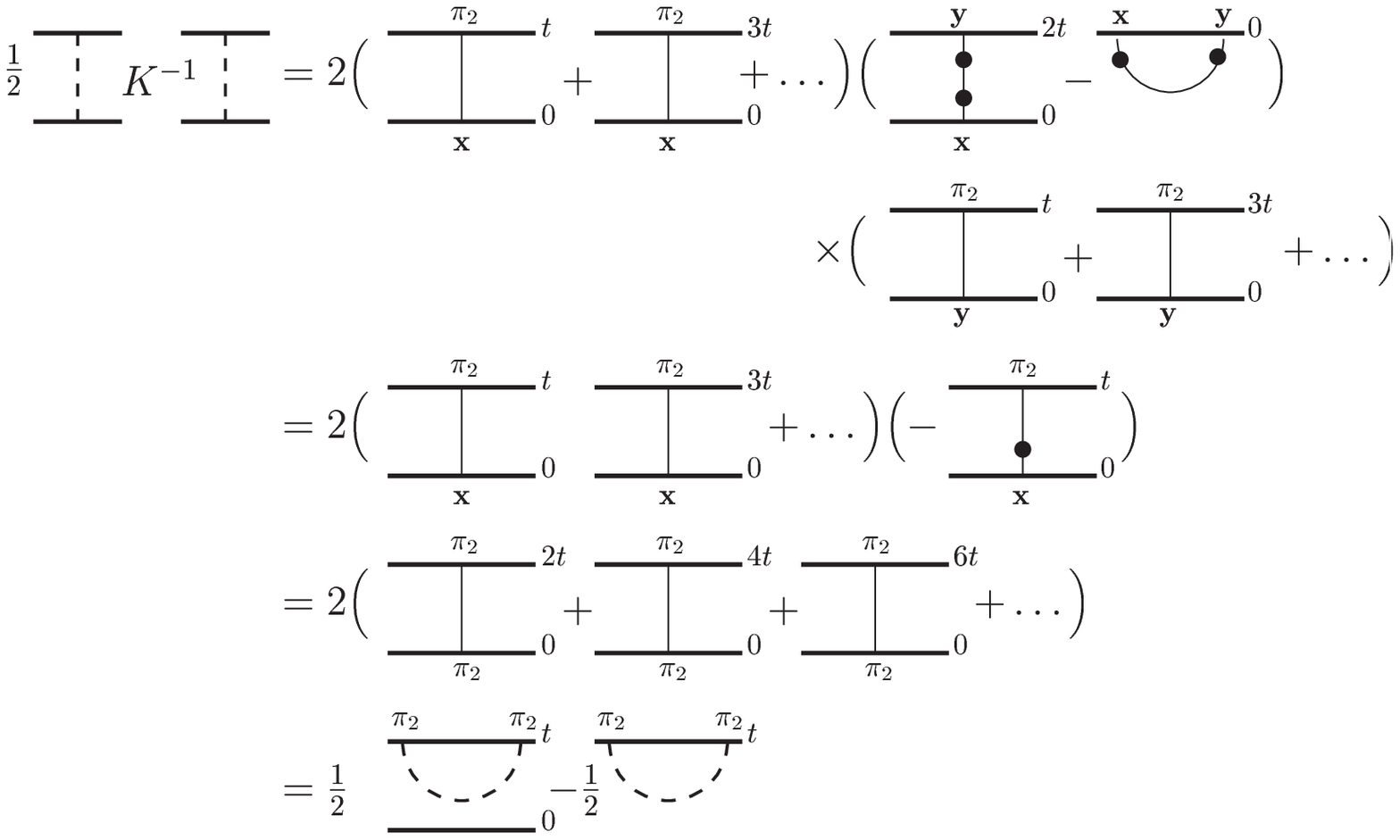}
\end{equation}
which again implies the correct result for time evolution of the vacuum,
\begin{equation}\label{sh-evol6}
    \includegraphics[width=0.7\textwidth]{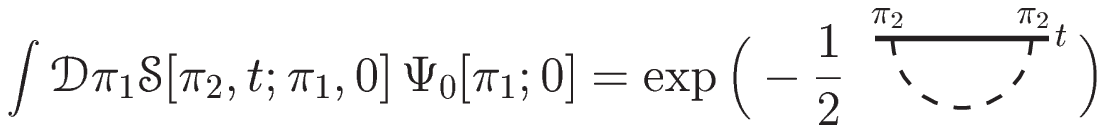}.
\end{equation}
Finally, the two point function in the field momentum representation is
\begin{equation}
  \langle \pi(\x , t)\pi(\y, 0)\rangle = \int\pathD (\pi_2, \pi_1)\Psi_0[\pi_2] \pi_2(\x) \mathscr{S}[\pi_2, t; \pi_1, 0]\pi_1(\y)\Psi_0[\pi_1].
\end{equation}
The $\pi_1$ integral is
\begin{equation}\label{2point1}
    \includegraphics[width=\textwidth]{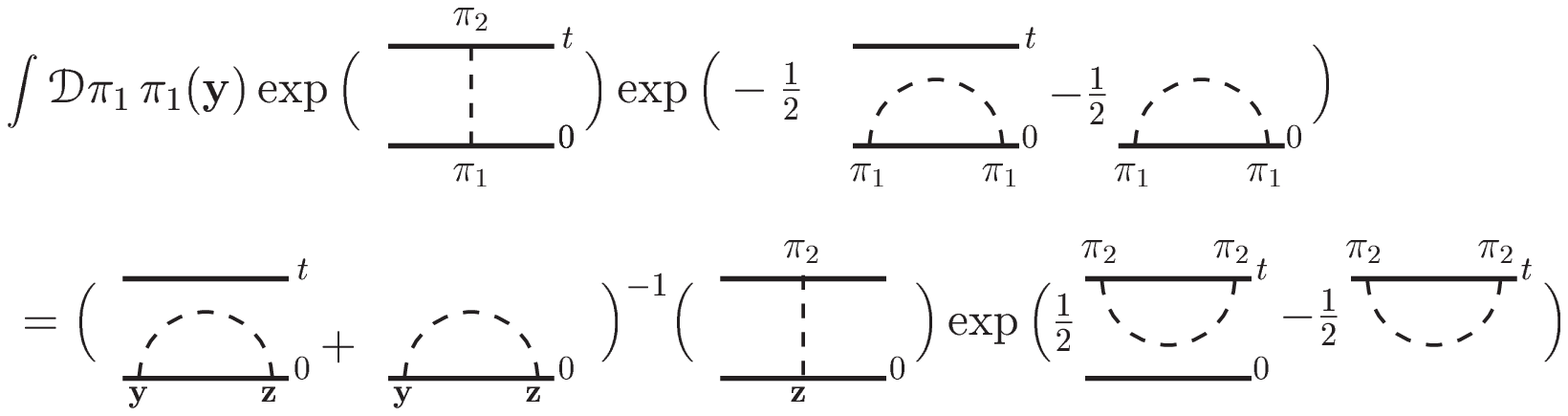}
\end{equation}
and the final integral is
\begin{equation}\label{2point2}
    \includegraphics[width=\textwidth]{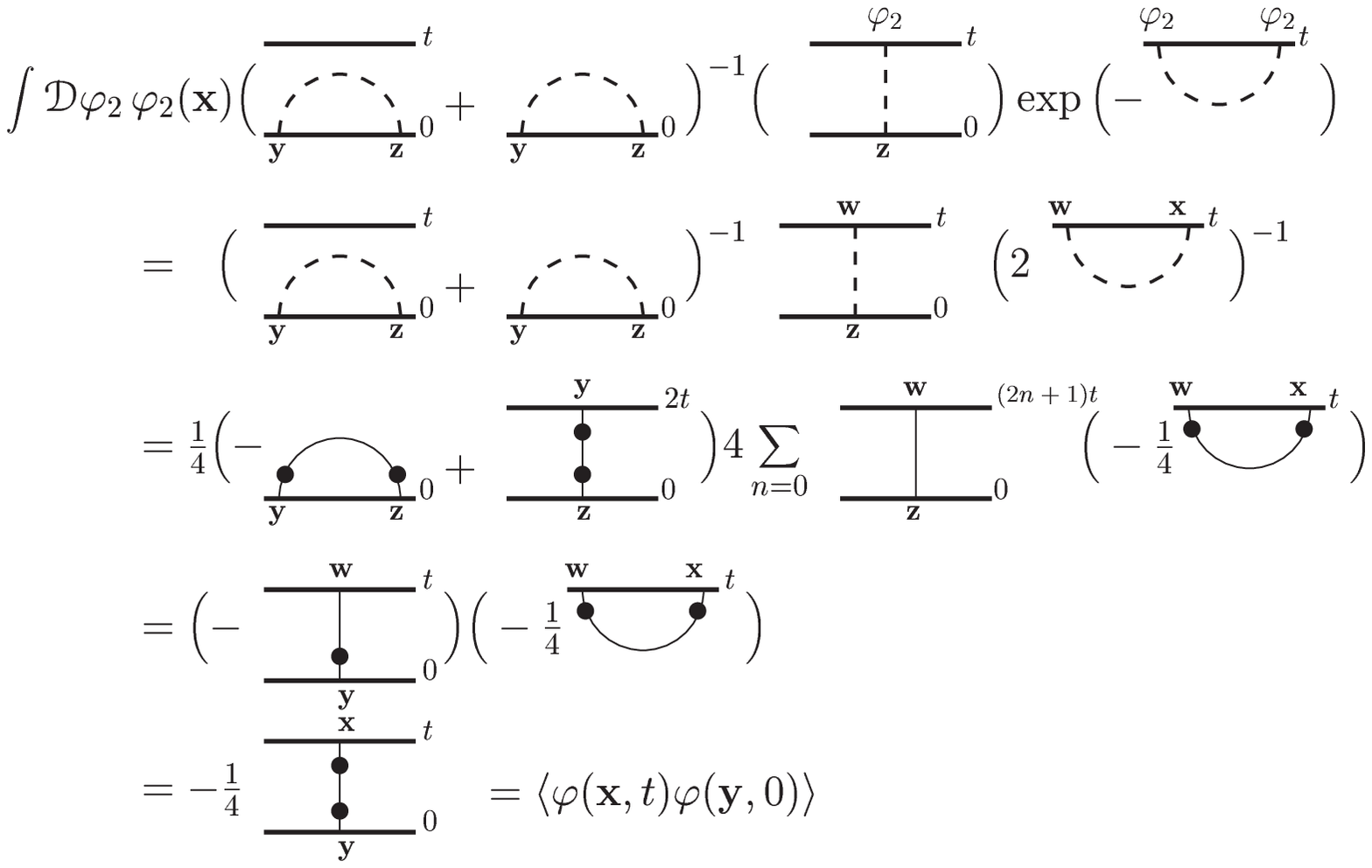}
\end{equation}
In the final line, the dots on the propagator appear since we are correlating the field momenta $\pi = \dot \varphi$. The factor is a result of our conventions; the minus sign is an artifact of Euclideanisation, as in (\ref{inversions-E}), and the quarter cancels the four coming from the time derivatives, giving the correct two point function. This can be verified by using the field momentum vacuum functional to generate the leading term in the equal time two point function,
\begin{equation}\label{2pt2eq-mom}
    \includegraphics[width=0.5\textwidth]{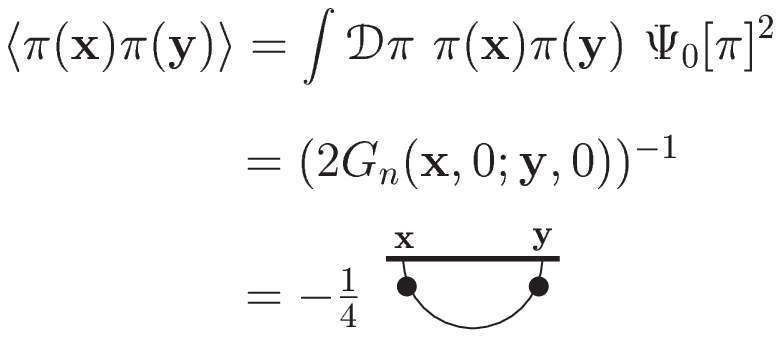}.
\end{equation}

We have shown that our arguments hold in Euclidean space and in the field momentum representation. In the next section we generalise the gluing property to string theory so that we may repeat our diagrammatic arguments and construct the second quantised string Schr\"odinger functional.

\section{Schr\"odinger representation of string field theory}

The string field propagator can be constructed in much the same way as in QFT \cite{Brink}, \cite{PolyakovBook} as the transition amplitude $G(X_f;X_i)$ between arbitrary spacetime curves $X_i(\sigma)$ and $X_f(\sigma)$. The boundary curves of interest to us have $X^0(\sigma)=$ constant, for which we denote the propagator by
\begin{equation}\label{prop-trans}
  G_{t_f-t_i}(\X_f; \X_i) := \int\pathD(X,g)\,\, e^{-\frac{1}{4\pi\alpha'}\int\!\ud^2\sigma\sqrt{g}\,  g^{ab} \partial_a X^\mu \partial_b X_\mu}\bigg|_{\X=\X_i(\sigma),\,\, X^0=t_i.}^{\X=\X_f(\sigma),\,\, X^0=t_f}
\end{equation}
At tree level the worldsheet is a finite strip (cylinder) for open (closed) strings. An arbitrary metric can be written as a diff$\times$Weyl transformation (orthogonal to the CKV) of a reference metric $\hat g_{ab}(T)$ for some value of the Teichm\"uller parameter $T$. The propagator is
\begin{equation}\label{prop-def}
  G_{t_f-t_i}(\X_f; \X_i) = \int\limits_0^\infty\!\ud T\, \text{Jac}(T)(\Det' \widehat{P}^\dagger \widehat{P})^{\frac{1}{2}}(\Det\widehat{\Delta})^{-13}\int\pathD\xi\, e^{-S_\text{cl}[X_\text{cl},\hat{g}(T)]}.
\end{equation}
The measure on Teichm\"uller space is Jac$(T)$ given by
\begin{equation*}
\text{Jac}(T)_\text{open}=\frac{(h_{ab}|\chi_{ab})}{(h_{ab}|h_{ab})^{1/2}}, \quad \text{Jac}(T)_\text{closed}=\frac{(h_{ab}|\chi_{ab})}{(V^a|V^a)^{1/2}(h_{ab}|h_{ab})^{1/2}}
\end{equation*}
where $h_{ab}$ is the zero mode of $\hat{P}^\dagger$, $\chi_{ab}$ is the symmetric traceless part of $\hat{g}_{ab,T}$ and $V =\partial/\partial\sigma$ is the CKV on the cylinder. $X_\text{cl}$ satisfies the wave equation in metric $\hat{g}$ with boundary conditions ${X^{\xi}_\text{cl}}|_{\tau=0} = X_i(\sigma)$, ${X^{\xi}_\text{cl}}|_{\tau=1} = X_f(\sigma)$. The remaining $\xi$ integral is over reparametrisations of the boundary data. If we attach reparametrisation invariant functionals $\Pi_i[\X_i]$, $\Pi_f[\X_f]$ to the boundaries of the worldsheet then this integral can be done trivially to give an (infinite) constant factor, for then
\begin{equation}
\begin{split}
  \int\pathD(X_f, X_i)&\int\pathD\xi\, e^{-S_\text{cl}[X_\text{cl},\hat{g}]}\,\,\Pi_i[X_f]\Pi_f[X_i] \\
  &= \int\pathD\big({X_\text{cl}}|_{\tau=1},{X_\text{cl}}|_{\tau=0}\big)\,\,e^{-S_\text{cl}[X_\text{cl},\hat{g}]}\,\,\Pi_f[{X_\text{cl}}|_{\tau=1}]\Pi_i[{X_\text{cl}}|_{\tau=0}]\int\pathD\xi.
\end{split}
\end{equation}
The same applies when we sew two worldsheets together, since $G$ itself is reparametrisation invariant. 

Rather than work with reparametrisation invariant functionals it is customary to consider the BRST approach. The metric degrees of freedom are represented by the ghosts as
\begin{equation}
  (\Det P^\dagger P)^{\frac{1}{2}} = \int\pathD(c,b)\,\exp\bigg(\int\!\ud^2\sigma\sqrt{g}\,\, b_{ab}(P c)^{ab} + \int\!\ud^2\sigma\sqrt{g}\,\, b_{ab}\chi^{ab}\bigg).
\end{equation}
There is an additional term to the usual ghost action which must be included due to the presence of a zero mode of $P^\dagger$ to make the functional integral non zero. A full FP treatment fixing the Weyl invariant quantity $\sqrt{g}g^{ab}$ includes this term automatically, see \cite{Paul-BRST}. Expanding  
\begin{equation*}
  b_{ab} = b_0 \frac{h_{ab}}{(h_{ab}|h_{ab})^{1/2}} + b'_{ab}
\end{equation*}
where $b_0$ is Grassmann and $b'$ is constructed from the non-zero eigenvectors of $P^\dagger$ the integral over $b_0$ is now saturated and reproduces the Teichm\"uller Jacobian,
\begin{eqnarray*}
\int\pathD(c,b)\,\exp\big(-S_{gh}[b,c]\big)= \frac{(h_{ab}|\hat{g}_{ab,T})}{(h_{ab}|h_{ab})^{1/2}}\,\,  \int\pathD(c,b')\, e^{-S_{gh}[b',c]}.
\end{eqnarray*}
The ghosts inherit the Alvarez boundary conditions \cite{Alvarez}, \cite{Beer}
\begin{equation}
  n^a c_a =0\quad\text{on }\partial\mathcal{M}, \qquad n^a t^b (Pc)_{ab} =0\quad\text{on }\partial\mathcal{M}.
\end{equation}
In the conformal gauge, $\hat{g}_{ab} =\text{diag}(1,T^2$, the contributions to the propagator integrand are
\begin{eqnarray}
  \label{class act} S_\text{cl}[X_{cl},\hat g] &=& \frac{1}{4\pi}\int\limits_{\partial\mathcal{M}}\ud s\, X_\text{bhd}n^a\partial_a X_\text{cl}, \\
  \label{det lap} (\Det\Delta)^{-1/2} &=&\begin{cases} T^{-1/2}\eta(T)^{-1/2} &\text{open} \\ T^{-1/2}\eta^(T){-1} &\text{closed},\end{cases} \\
  (\Det' P^\dagger P)^{1/2} &=& \begin{cases}T^{1/2}\eta(T) &\text{open} \\ T\,\eta^2(T) &\text{closed,}\end{cases}\\
  \text{Jac}(T) &=& \begin{cases}T^{-1/2} &\text{open}\\ T^{-1} &\text{closed,}\end{cases}
\end{eqnarray}
where the Dedekind eta function is
\begin{equation*}
  \eta(T) := e^{-T/12}\prod\limits_{m=1}^\infty (1-e^{-2mT}).
\end{equation*}
The determinants have been zeta-function regulated and any irrelevant (that is $T$ independent) factors have been neglected.

The propagator in the extended state space of the co-ordinates and ghosts should interpolate between boundary configurations of the ghosts also. To fix the values of $c$ and $b$ on the Dirichlet ($\tau=0,1$) boundaries delta-functionals can be inserted into the integration \cite{Ordonez}, giving
\begin{equation}\label{ghost integral done}
  (\Det' P^\dagger P)^{1/2}\prod\limits_{m=1}\exp\bigg( \frac{\cosh mT}{\sinh mT}(c^i_m b^i_m + c^f_m b^f_m) - \frac{1}{\sinh mT}(c^i_m b^f_m + c^f_m b^i_m)\bigg)
\end{equation}
where $b^j_m$, $c^j_m$ are the Fourier modes of the initial, final ghost configurations.

\subsection{Corners and sewing for the open string}

There is a caveat to these calculations in the case of the open string. We know that the integrand in (\ref{prop-trans}) is independent of the Liouville mode when $D+1=26$ \cite{Moore} at least for worldsheets without boundaries, giving a Weyl invariant string theory. When boundaries are present, as in our case, there are extra contributions to the Weyl dependence of the determinants in (\ref{prop-def}).

As an example consider the determinant of the Laplacian. Using the usual heat kernel regularisation,
\begin{equation}
	\delta\log{\Det \Delta} = -\int\!\ud^2\sigma \sqrt{g}\,\, \mathcal{K} (\sigma,\sigma, \tau,\tau;\epsilon)\delta\rho(\sigma),
\end{equation}
we know that the variation must have an expansion in powers of the cutoff of form
\begin{equation}
	\begin{split}
	\delta\log{\Det \Delta} = &-\frac{1}{24\pi}\int\!\ud^2\sigma\sqrt{g}\,\, R\,\delta\rho - \frac{1}{4\pi\epsilon}\int\!\ud^2\sigma\sqrt{g}\, \delta\rho \\
	&+\sum\limits_i A_i\frac{1}{\sqrt{\epsilon}}\int\!\ud s\, \delta\rho + B_i\int\!\ud s\, n^a\partial_a\delta\rho + C_i\int\!\ud s\,\, K_g\delta\rho \\
	&+ E\sum\limits_j \delta\rho(\sigma_j) + \mathcal{O}(\sqrt{\epsilon})
	\end{split}
\end{equation}
where $i$ runs over the boundaries and $j$ over distinguished points on the manifold where the boundary conditions change-- the corners. The divergent volume and surface terms can be removed by local counter-terms. What remains is removed by the metric integral when $D+1=26$ excluding the contribution from the corners \cite{Varughese}, \cite{Paul-ghosts}. So even in the critical dimension, the off-shell propagator is not truly Weyl invariant. 

We will need to compute corner anomalies for various determinants. Each corner contributes equally and independently (for the same change in boundary conditions). Since the anomaly is a local effect it is insensitive to the global topology of the worldsheet, so we can work on the simpler geometry of the upper right quadrant with appropriate conditions on the axes. Since the anomaly only depends on a single value of the Liouville field we can compute it using a constant $\delta\rho$.

For the Laplacian the quadrant has Dirichlet conditions on the $x$-axis and Neumann on the $y$-axis. The heat kernel for this geometry is given by the method of images,
\begin{equation}
\begin{split}
  \k(x,y,x',y';\epsilon) &= \k_0(x,y,x',y';\epsilon) - \k_0(x,y,x',-y';\epsilon) \\
  &\hspace{5pt} +\k_0(x,y,-x',y';\epsilon) - \k_0(x,y,-x',-y';\epsilon),
\end{split}
\end{equation}
where the free space heat kernel for the Laplacian is
\begin{equation}\label{free heat kernel}
  \k_0(x,y,x',y';\epsilon) = \frac{1}{4\pi\epsilon}\exp\bigg(-\frac{(x-x')^2 + (y-y')^2}{4\epsilon}\bigg).
\end{equation}
The variation in the determinant with a constant Weyl scaling is, writing down only the non-zero corner anomaly in the final line,
\begin{eqnarray}
\nonumber  \delta\log\Det\Delta &=& -\delta\rho\, \int\limits_0^\infty\!\ud x\,\ud y\,\, \k(x,y,x,y;\epsilon) \\
\nonumber	&=& -\delta\rho\int\limits_0^\infty\!\ud x\,\ud y\,\,  \frac{1}{4\pi\epsilon} - \frac{1}{4\pi\epsilon}e^{-y^2/\epsilon} + \frac{1}{4\pi\epsilon}e^{-x^2/\epsilon} - \frac{1}{4\pi\epsilon}e^{-(x^2 + y^2)/\epsilon} \\
	&=& \frac{1}{16}\,\delta\rho + \ldots
\label{Lap c.anom}
\end{eqnarray}
The corner anomaly is not a sickness of the off-shell theory. It is there to cancel anomalies generated by the process of sewing worldsheets together or sewing string functionals onto Dirichlet sections. When we sew two worldsheets $\mathcal{M}_1$ and $\mathcal{M}_2$ together by integrating over all possible boundary values of $X$ (the ghosts behave similarly), the classical actions combine to give the classical action on the sewn worldsheet. The Gaussian integration brings down the determinant of $n^a\partial_a$ to the power minus one half, computed with the harmonic extension of the boundary data into the bulk, which sews the determinants together;
\begin{equation}
\label{X-sewing}
\begin{split}
  (\Det\Delta)^{-1/2}_{\mathcal{M}_1}(\det\Delta)^{-1/2}_{\mathcal{M}_2} &\int\pathD X_\text{bhd} e^{-\int\limits_{\mathcal{M}_1}\ud s X_\text{bhd} n^a\partial_a {X_{cl}}}\,\, e^{-\int\limits_{\mathcal{M}_2}\ud s X_\text{bhd} n^a\partial_a {X_{cl}}} \\
  &=(\Det\Delta)^{-1/2}_{\mathcal{M}_1 +\mathcal{M}_2}e^{-\int_{\mathcal{M}_1 + \mathcal{M}_2}\ud s X_\text{bhd} n^a\partial_a {X_{cl}}}.
\end{split}
\end{equation}
The anomalous boundaries become a bulk piece of the sewn worldsheet, which has no privileged structure and so cannot carry the anomaly-- how does this come about? The answer is the determinant of $n^a\partial_a$ has its own corner anomaly. The calculation is given in reference \cite{Paul-ghosts} so we will just state
\begin{equation}\label{bhd-anom}
  \delta\log\Det n^a\partial_a = \mp\frac{1}{8}\delta\rho -\frac{1}{2\pi\epsilon}\int\!\ud s\, \delta\rho
\end{equation}
where the minus (plus) sign is for Neumann (Dirichlet) boundary conditions at $\sigma=0,\pi$. The co-ordinates have Neumann conditions so the anomaly is minus twice that coming from a single Laplacian, (\ref{Lap c.anom}), ensuring that two worldsheets are sewn without anomaly in the bulk.

\subsection{Sewing worldsheets}
The appearance of the corner anomaly has led us naturally to sewing. In this section we will prove the extension of the gluing property (\ref{factorisation}) to the string propagator as a method of sewing worldsheets. The details are a little involved, but in the next section we will give explicit examples.

If we represent the propagator as an amplitude
\begin{equation}\label{prop-canon def}
  G(X_f, b^f , c^f; X_i, b^i, c^i) =\int\limits_0^\infty\!\ud T\,\,\bra{X_f, b^f, c^f}e^{-H T}\ket{X_i, b^i, c^i}
\end{equation}
then the standard sewing prescription is to integrate over all boundary values of $X^\mu$ and the ghosts shared between two propagators and returns
\begin{equation}\label{bad-sew}
  \int\limits_0^\infty\!\ud U\int\limits_0^\infty\!\ud T\,\,\bra{X_f, b^f, c^f}e^{-H (T+U)}\ket{X_i, b^i, c^i},
\end{equation}
which is incorrect, since we have a redundant Teichm\"uller integral which gives an infinite factor. Carlip showed \cite{Carlip} that by inserting the Hamiltonian acting on the boundary of one worldsheet, the moduli spaces are correctly sewn. This can be seen from (\ref{prop-canon def}); the Hamiltonian is the derivative of the integrand with respect to the Teichm\"uller parameter, so this reduces one propagator to a delta functional\footnote{Strictly speaking this is only true for a positive definite Hamiltonian, so for the bosonic string there is the tachyon divergence to be taken care of.}. Sewing this onto the second propagator trivially produces the desired result.

However, this method is inappropriate for our means. In the Schr\"odinger representation we are interested in evolving states between successive times so the propagator is calculated with the particular boundary conditions $X^0=$ constant and we do not integrate over $X^0$ when sewing, just as in the field theory case. This would leave behind one Laplacian determinant from each worldsheet, so there is something more going on. However, we have not yet considered the metric. Taking the Alvarez boundary conditions in the conformal gauge, a basis of reparametrisations $\xi^a = (\xi^\sigma, \xi^\tau)$ on the strip (with metric as given earlier, so $\tau \in[0,1]$ and $\sigma \in(0,\pi)$ is given by
\begin{equation}\label {ghosty}
  \xi^1_{nm} = N_m\bigg( \begin{array}{c} \cos(m\pi\tau)\sin(n\sigma)\\ 0\end{array}\bigg), \quad \xi^2_{nm} = N_m\bigg(\begin{array}{c} 0 \\ \sin(n\pi\tau)\cos(m\sigma)\end{array}\bigg)
\end{equation}
with normalisation $N_m = 2^{1-\tfrac{1}{2}\delta_{m=0}}/\sqrt{T\pi}$. The reparametrisations split into orthogonal pieces, with the set $\{\xi^2\}$ obeying
\begin{equation*}
  n^a\xi^2_a = t^a\xi^2_a = \partial_\sigma \xi^2_\sigma = 0
\end{equation*}
on the Dirichlet boundaries -- only half of all reparametrisations couple to the boundary. If we were to compute the propagator with conditions on the metric fixing $g_{\sigma\sigma}$, much as we do for the co-ordinates, then to sew two propagators together we would integrate over all values of the boundary metric which would sew together only half of the determinant of $P^\dagger P$. Since $(\Det P^\dagger P)^{1/2}$ cancels two copies of $(\Det\Delta)^{-1/2}$, this would leave a determinant behind to cancel that from $X^0$.

That there should exist a method of sewing worldsheets appropriate to our needs may not be immediately obvious because of the extended nature of the string, but performing a similar trick to (\ref{prop-trick}) gives a strong indication; insert one into the functional integral defining the propagator,
\begin{equation}
  G_0(X_f;X_i) = \int\pathD(X,g)\,\bigg[1=\int\pathD C\,J(C,X^0)\dirac{X^0(\sigma,C(\sigma)) - t}\bigg]e^{-S_E(X,g)}
\end{equation}
where the new integral is over curves $C$ on the worldsheet. For $t_f>t>t_i$ the delta functional has support on the worldsheet, and the Jacobian is
\begin{equation}
  \prod\limits_\sigma \dot{X}^0(\sigma, C_X(\sigma))
\end{equation}
where $C_X$ is the curve on which $X^0(\sigma, C_X(\sigma))=t$. We will now describe how this can be implemented to generalise (\ref{E-glue}). The relation to Carlip's method will be apparent at the end of the section. The proof is in three stages.
\begin{enumerate}
\item We introduce a change of variables in the ghost sector allowing us to realise the discussion following (\ref{ghosty}).
\item We show that when integrating over the boundary data, the ghosts cancel the unwanted effects of not integrating over $X^0$. This combines the integrands of the propagators into the standard "sewn" form, as in (\ref{bad-sew}).
\item We insert a time derivative between the propagators being sewn and, similarly to Carlip's method, use it to remove the extra Teichm\"uller parameter and contract one propagator to a delta functional, integrating over which returns the sewn propagator.
\end{enumerate}

\subsubsection*{1. Ghosts}

The role of the ghosts in string theory is to cancel the undesirable effects of including the $X^0$ oscillators. They will do the same thing for us here, although to a different end. Starting with the standard $b-c$ system,
\begin{equation*}
  (\Det' P^\dagger P)^{1/2} = \int\pathD(c^a,b'_{ab})\,\, \exp\bigg(-\int\!\ud^2\sigma\sqrt{g}\,\, {b'}_{ab} P(c)^{ab}\bigg)
\end{equation*}
we change variables $b'=P\gamma$ and pick up a Jacobian $(\Det P^\dagger P)^{-1/2}$, which we can represent as a bosonic vector integral. Our new ghost system is therefore
\begin{equation}\label{det-new-2}
\begin{split}
  (\Det' P^\dagger P)^{1/2} = \int\pathD(c^a,\gamma^a, f^a)&\exp\left(-\frac{1}{2}\int\!\ud^2\sigma\sqrt{g}\,\, (Pf)_{ab}(Pf)^{ab}\right) \\
  &\times\exp\bigg(-\frac{1}{2}\int\!\ud^2\sigma\sqrt{g}\,\, (P\gamma)_{ab}(P c)^{ab}\bigg).
\end{split}
\end{equation}
We have included a factor of one half in front of the new action purely for convenience later. On the strip (cylinder) the operator $P$ is not invertible, but this should not worry us here since we are merely changing the representation of $\Det'P^\dagger P$, and the eigenvectors and eigenvalues of $b_{ab}$ and $P\gamma_{ab}$ are in one to one correspondence \cite{Alvarez}. However, this change of variable will have an interesting effect on the BRST transformation, as we will see.

The propagator in the extended state space is the transition amplitude between arbitrary values of $X^\mu, c^\sigma$, $\gamma^\sigma$, $f^\sigma$. The ghost integrals are done by expanding around a classical solution satisfying $P^\dagger P =0$. The $\tau$--components of the fields, corresponding to the set of reparametrisations $\{\xi^2\}$ discussed earlier, are integrated out. For the open string, the classical ghost fields obey
\begin{align}
  \nonumber J^\sigma_\text{cl}(\sigma, 0) &= \sum\limits_{m=1} J^i_m\sin(m\sigma)\sqrt{\frac{2}{\pi}}, \qquad J^\sigma_\text{cl}(\sigma, 1) = \sum\limits_{m=1} J^f_m\sin(m\sigma)\sqrt{\frac{2}{\pi}} \\
  \nonumber J_\text{cl}^\tau &\equiv 0
\end{align}
where the field $J\in\{f, c, \gamma\}$. The quantum fields obey
\begin{align}\label{new ghost BCs}
  \nonumber n^a \delta J_a &= t^a \delta J_a =0 \quad\text{on Dirichlet boundaries,} \\
  n^a \delta J_a &= n^a t^b P(\delta J)_{ab} =0\quad\text{on Neumann boundaries.}
\end{align}
For the closed string the change of variables is only well defined up to shifts proportional to the CKV, $J^a\rightarrow J^a+\lambda V^a$ for $J^a\in\{c^a,\gamma^a,f^a\}$, so we choose $(J|V)=0$ which removes the c.o.m. from the classical pieces. 

With these boundary conditions the actions separate into a quantum piece giving the determinants,
\begin{equation}
  \int\!\ud^2\sigma\sqrt{g}\, \delta \gamma_a\, (P^\dagger P \delta c)^a + \int\!\ud^2\sigma\sqrt{g}\, \delta f_a\, (P^\dagger P \delta f)^a,
\end{equation}
and a classical action
\begin{equation}
  \int\!\ud s\, \gamma_{cl}^\sigma\, (Pc_\text{cl})_{\sigma\tau} +\int\!\ud s\, f_\text{cl}^\sigma \,(Pf_\text{cl})_{\sigma\tau}
\end{equation}
evaluated at $\tau=0,1$. With the given boundary conditions the classical action reduces to that for a single free boson and a pair of free Grassmann fields,
\begin{equation}
  \int\limits_0^\pi\!\ud s\, \gamma_{cl}^\sigma\,\partial_\tau {c_\sigma}_\text{cl} + \int\limits_0^\pi\!\ud s\, f_\text{cl}^\sigma \,\partial_\tau {f_\sigma}_\text{cl}
\end{equation}
The exponent in the Grassmann part is easily verified to give (\ref{ghost integral done}) with $b_m \rightarrow \gamma_m$.

\subsubsection*{2. Integrating over boundary data}

We can write the propagator as the amplitude
\begin{equation}
  G_{t_f-t_i}(\BB_f;\BB_i) = \int\limits_0^\infty\!\ud T\,\,\bra{t_f}e^{-H_0 T}\ket{t_i}\bra{\BB_f}e^{-\mathbf{H}T}\ket{\BB_i}
\end{equation}
where the states $\bra{t}$ are ordinary quantum mechanical states for the time variable and $\bra{\BB}$ represents boundary data for $\X$, $c^\sigma,\gamma^\sigma,f^\sigma$, the $X^0$ oscillators (equal to zero), and the conditions on the ghost $\tau$ components. The worldsheet Hamiltonian is
\begin{equation}\label{H-split}
  H=H_0 + \mathbf{H}
\end{equation}
with $H_0 =-\frac{\p^2}{\p t^2}$ and $\mathbf{H}$ the Hamiltonian for all remaining degrees of freedom. Explicitly,
\begin{equation}\begin{split}
  \bra{t_f}e^{-HT}\ket{t_i} &= \frac{1}{\sqrt{T}}e^{-(t_f-t_i)^2/4\pi\alpha' T}, \\
  \bra{\BB_f}e^{-HT}\ket{\BB_i} &= \frac{1}{T^{12}}e^{-S_\text{cl}[\BB]}f(T)^{-24}
\end{split}\end{equation}
where $S_\text{cl}$ is the classical action for $\X$ and the ghost action (\ref{ghost integral done}) with $b_m\rightarrow \gamma_m$. The contribution from $X^0$ not included in the first inner product above ($f(T)^{-1/2}$ open, $f(T)^{-1}$ closed) precisely cancels against the $\tau$ contribution from the ghosts and the Teichm\"uller Jacobians. It is a simple matter to check that when we integrate over the remaining boundary data shared by two propagators with Teichm\"uller parameters $T$ and $U$ we get back the amplitude for the remaining boundary data with Teichm\"uller parameter $T+U$-- the new ghosts have made this integration into a resolution of the identity,
\begin{equation}
  \int\pathD\BB\,\,\bra{\BB_f}e^{-HT}\ket{\BB}\bra{\BB}e^{-HU}\ket{\BB_i} = \bra{\BB_f}e^{-H(T+U)}\ket{\BB_i}.
\end{equation}

\subsubsection*{3. Sewing with the time derivative}

Now insert the time derivative, just as in the particle case, between the two propagators, and integrate over the boundary data as we have described to obtain
\begin{equation}
\int\limits_0^\infty\!\ud T\ud U\,\,\bra{t_f}e^{-H_0 T}\ket{t}\overleftrightarrow{\frac{\p}{\p t}}\bra{t}e^{-H_0 U}\ket{t_i}\bra{\BB_f}e^{-\mathbf{H}(T+U)}\ket{\BB_i}.
\end{equation}
Write the time derivative by a double derivative and an integral,
\begin{equation*}
    -\int\limits_{t}^\infty\!\ud\tilde{t}\int\limits_0^\infty\!\ud T\int\limits_0^\infty\!\ud U\, \bra{t_f}e^{-H_0 T}\ket{\tilde{t}}\overleftrightarrow{\frac{\partial^2}{\partial {\tilde{t}}^2}} \bra{\tilde{t}}e^{-H_0 U}\ket{t_i} \bra{\BB_f}e^{-\mathbf{H}(T+U)}\ket{\BB_i}.
\end{equation*}
Using the definition (\ref{H-split}) this is
\begin{align*}
  \int\limits_{t}^\infty\!\ud\tilde{t}\int\limits_0^\infty\!\ud T\!\int\limits_0^\infty\!\ud U & \left(\frac{\partial}{\partial U} - \frac{\partial}{\partial T}\right)\!\bigg\{ \bra{t_f}e^{-H_0 T}\ket{\tilde{t}}\bra{\tilde{t}}e^{-H_0 U}\ket{t_i} \bra{\BB_f}e^{-\mathbf{H}(T+U)}\ket{\BB_i}\bigg\} \\
  +\int\limits_{t}^\infty\!\ud\tilde{t}\int\limits_0^\infty\!\ud T\!\int\limits_0^\infty\!\ud U & \bra{t_f}e^{-H_0 T}\ket{\tilde{t}} \bra{\tilde{t}}e^{-H_0 U}\ket{t_i} \bra{\BB_f}\mathbf{H}e^{-\mathbf{H}(T+U)}\ket{\BB_i} \\
  -\int\limits_{t}^\infty\!\ud\tilde{t}\int\limits_0^\infty\!\ud T\!\int\limits_0^\infty\!\ud U & \bra{t_f}e^{-H_0 T}\ket{\tilde{t}} \bra{\tilde{t}}e^{-H_0 U}\ket{t_i} \bra{\BB_f}\mathbf{H}e^{-\mathbf{H}(T+U)}\ket{\BB_i}.
\end{align*}
The final two terms cancel. In the above we see the similarity to Carlip's result. We can do the integral over one Teichm\"uller parameter,
\begin{equation*}
\begin{split}
&\int\limits_{t}^\infty\!\ud\tilde{t}\int\limits_0^\infty\!\ud U\,\dirac{t_f -\tilde{t}} \bra{\tilde{t}}e^{-H_0 U}\ket{t_i}\bra{\BB_f}e^{-\mathbf{H}U}\ket{\BB_i} \\
& - \int\limits_{t}^\infty\!\ud\tilde{t}\int\limits_0^\infty\!\ud T\,\dirac{\tilde{t}-t_i} \bra{t_f}e^{-H_0 T}\ket{\tilde{t}}\bra{\BB_f}e^{-\mathbf{H}T}\ket{\BB_i},
\end{split}
\end{equation*}
and for $t_f > t > t_i$ the $\tilde{t}$ integral has support on only one delta function, giving
\begin{equation*}
  \int\limits_0^\infty\!\ud U\,\bra{t_f}e^{-H_0 U}\ket{t_i}\bra{\BB_f}e^{-\mathbf{H}U}\ket{\BB_i} \equiv\int\limits_0^\infty\!\ud U\,\bra{\BB_f, t_f}e^{-HU}\ket{\BB_i, t_i},
\end{equation*}
recovering the propagator. Overall, we have shown that the Euclidean generalisation of (\ref{factorisation}) holds in string theory as 
\begin{equation}\label{string-sew}
  \int\!\pathD \BB \,\, G_{t_2-t}(\BB_2;\BB)\,\overleftrightarrow{\frac{\partial}{\partial t}}\,G_{t-t_1}(\BB;\BB_1)= G_{t_2-t_1}(\BB_2;\BB_1),\quad t_2> t>t_1.
\end{equation}
We close this section with some explicit examples of sewing worldsheets.

\subsection{Pointlike states and light cone gauge}

The simplest example is for pointlike initial and final co-ordinate states and vanishing ghost states $c_m=\gamma_m=0$. In this case the integral over boundary data as we have described returns
\begin{equation*}
\begin{split}
  \int\limits_0^\infty\ud (T,U) &\frac{1}{\sqrt{T}}e^{ -(t_f-t)^2/4T}\bigg(\overleftrightarrow{\frac{\partial}{\partial t}}\bigg)\frac{1}{\sqrt{U}}e^{-(t-t_i)^2/4U} \\
  &\times \frac{1}{(T+U)^{25/2}}e^{-(\x_f-\x_i)^2/4(T+U)}e^{T+U}\prod\limits_{m=1}(1-e^{-2m(T+U)})^{-12}.
\end{split}
\end{equation*}
The final product is the correct determinant appearing in the propagator with Teichm\"uller parameter $T+U$. Expanding this product out gives
\begin{equation*}
\begin{split}
  \sum_m \eta_m \int\limits_0^\infty\ud (T,U) &\frac{1}{\sqrt{T}}e^{ -(t_f-t)^2/4T}\bigg(\overleftrightarrow{\frac{\partial}{\partial t}}\bigg)\frac{1}{\sqrt{U}}e^{-(t-t_i)^2/4U} \\
  &\times \frac{1}{(T+U)^{25/2}}e^{-(\x_f-\x_i)^2/4(T+U)}e^{-(2m-1)T+U}
\end{split}
\end{equation*}
where the $\eta_m$ are constants. Comparing with (\ref{E-prop}) this is a sum over masses of the particle gluing result, which we have already proven.

Two propagators can be explicitly glued together in the light cone gauge (see for example \cite{Martinec}, \cite{Lowe}). The propagator for positive, Wick-rotated $x^+$ is
\begin{equation}\begin{split}
  &G(\delta x^+, \delta x^-, \delta\X) = \int\limits_0^\infty \frac{\ud p}{2p} e^{-ip\delta x^-}\bigg(\frac{p}{2\pi\delta x^+}\bigg)^{12} e^{-p\delta\x^2/2\delta x^+ + \delta x^+/p} \\
  &\prod\limits_{m=1}\big(1-e^{-2m\delta x^+/p}\big)^{-12}\exp\bigg(\frac{-2m}{\sinh(\delta x^+/p)}\big[({\X^f_m}^2+{\X^f_m}^2)\cosh(\delta x^+/p)-2\X^f_m\cdot\X^i_m\big]\bigg)
\end{split}\end{equation}
Despite the lengthy expression it is a simple matter to check that
\begin{align}
  \nonumber \int\ud^{24}\X\ud x^-\,\,G(x^\pm_f , \X^f ; x^\pm , \X&)\bigg(-2i\frac{\partial}{\partial x^-}\bigg)G(x^\pm , \X ; x^\pm_i , \X_i) \\
  &= G(x^\pm_f , \X_f ; x^\pm_i , \X_i)\quad\text{if $x^+_f>x^+>x^+_i$} \\
  &= G_I(x^\pm_f , \X_f ; x^\pm_i , \X_i)\quad\text{if $x^+_f,x^+_i>x^+$}
\end{align}
and $G_I$ is the propagator from one point to the reflection of another in the plane at time $\delta x^+$. The insertion required is different to the conformal gauge case but should be expected since in the light cone gauge $-i\p/\p x^- = \pi_- = \dot{x}^+$.

\subsection{Discussion}

This factorisation of the ghosts may seem ad-hoc but in fact it follows from the gauge choice $P^\dagger_{\hat{g}}(g)^a =0$ which is equivalent to the usual gauge fixing $g_{ab}\propto \hat g_{ab}$. We discuss this and the BRST symmetry in section 7. 

The sewing we have given may seem surprising given the extended nature of the string. Fortunately, the open string affords us a check on our method- we can verify that the corner anomalies correctly cancel between the determinants of $(\Det' P^\dagger P)^{1/4}$ and $(\Det \Delta)^{-1/2}$ when we sew. The variation under an infinitesimal Weyl transformation of $(\Det' P^\dagger P)$ is \cite{Alvarez}
\begin{align*}
  \delta\log\Det P^\dagger P &= -\int\limits_{\epsilon}^\infty \!\frac{\ud s}{s}\text{Tr} \big[\,\delta \,e^{-sP^\dagger P}\,\big] \\
  &= \int\limits_{\epsilon}^\infty \!\ud s\, \text{Tr}\big[\, (-2P^\dagger \delta\rho P + P^\dagger P\delta\rho)e^{-sP^\dagger P}\,\big] \\
  &= \int\limits_{\epsilon}^\infty \!\ud s\, \text{Tr}\big[\, (-2\delta\rho PP^\dagger e^{-sPP^\dagger }+ \delta\rho P^\dagger P)e^{-sP^\dagger P}\,\big] \\
  &= -2\text{Tr}\big[\,\delta\rho e^{-\epsilon PP^\dagger}\,\big] + \text{Tr}\big[\,\delta\rho e^{-\epsilon P^\dagger P}\,\big].
\end{align*}
We have used the cyclicity of the trace in going from the second to the third line. However we know that we can calculate the corner anomaly with a constant $\delta\rho$ in which case the above becomes
\begin{equation}
  \delta\log\Det P^\dagger P = -\delta\rho\,\text{Tr}\big[\, e^{-\epsilon P^\dagger P}\,\big].
\end{equation}
It is for this reason that the corner anomalies calculated in \cite{Varughese} appear to be incorrect. For the ghost $\tau$--components the non-zero contribution to the heat kernel for $P\dagger P$ on the upper right quadrant is, by the method of images,
\begin{equation*}
  \k^{\tau\tau} = \frac{1}{8\pi\epsilon} - \frac{1}{8\pi\epsilon}e^{-\frac{1}{2\epsilon}(\sigma^2 + \tau^2)} -\frac{1}{8\pi\epsilon}e^{-\frac{1}{2\epsilon}\sigma^2}+\frac{1}{8\pi\epsilon}e^{-\frac{1}{2\epsilon}\tau^2}
\end{equation*}
contributing
\begin{equation}
  \delta\log (\Det' P^\dagger P)^{1/2} = \frac{1}{2}\frac{1}{16}\,\delta\rho +\ldots
\end{equation}
This part of the determinant, along with $(\Det\Delta)^{-1/2}$, is not sewn, and the corner anomaly is minus that from a single boson, so this piece cancels the corner anomaly from the $X^0$ determinant. For the $\sigma$ components the heat kernel is
\begin{equation*}
  \k^{\sigma\sigma} = \frac{1}{8\pi\epsilon} + \frac{1}{8\pi\epsilon}e^{-\frac{1}{2\epsilon}(\sigma^2 + \tau^2)} -\frac{1}{8\pi\epsilon}e^{-\frac{1}{2\epsilon}\sigma^2}-\frac{1}{8\pi\epsilon}e^{-\frac{1}{2\epsilon}\tau^2}.
\end{equation*}
The corner contribution is therefore
\begin{equation}
  \delta\log (\Det' P^\dagger P)^{1/2} = -\frac{1}{2}\frac{1}{16}\,\delta\rho +\ldots
\end{equation}
As we have seen the classical action for the $\sigma$ components involves $n^a\partial_a$, so we know from the Dirichlet case in (\ref{bhd-anom}) that this part of the determinant sews without anomaly in the bulk.

\subsection{Reconstructing the vacuum functional}

The sewing rule (\ref{string-sew}) confirms that despite the extended nature of the string a Schr\"odinger representation makes sense for string field theory, and we can carry over the (Euclidean continuation of the) diagrammatic arguments of Section 3 to free string field theory. 

The next step is to include interactions. Although our free theory has a local time co-ordinate, there is no reason to expect that the interaction should be local in this time \cite{Eliezer}. In this case the functional description of, for example, the vacuum as a large time integral fails, as it attempts to describe the vacuum by a sum over field histories on the half space $X^0<0$ but through a non-local interaction the field couples to itself at arbitrary times.

The reconstructive method used in Section 4 may be adapted, however, since first quantised string theory generates its own loop expansion through the sum over topologies. Rather than pick an interaction vertex as we did in Section 4 we will again use the first quantised theory to construct the a second quantised objects: the interacting string field vacuum functional. We will find that it differs only slightly from the `expected' functional description.

To begin, the free field Euclidean vacuum functional can be written (suppressing the $\hbar$ dependence)

\begin{equation}
  \includegraphics[height=1.3cm]{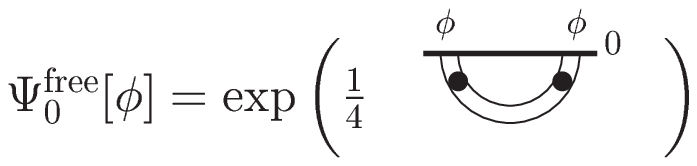},
\end{equation}
using the string equivalent of (\ref{inversions-E}). $\phi$ here is the string field. The double line is the string field propagator. Our arguments are based in the Schr\"odinger representation which treats all spatial arguments in the same way as covariant methods. Therefore we may choose the interaction to be local or non-local in the spacial or ghost co-ordinates without affecting the results. Accordingly we will suppress the spatial dependencies of functionals whenever possible from here on, and abbreviate our representation of the free string field propagator to
\begin{align*}
  G_\BB(0;t_j) &:= G(t_j,\BB_j;0,\BB), \\
  G_\BB(t_j;0)\phi_\BB &:= \int\pathD\BB\,\,G(t_j,\BB_j;0,\BB)\phi[\BB].
\end{align*}
Now suppose the three field expectation value is given by some functional of three strings, $T_3[\BB_1,t_1;\BB_2,t_2;\BB_3,t_3]$, evaluated at $t_i=0$ as in (\ref{3vev}), which would be represented by the diagram
\begin{equation}\label{3vev}
  \includegraphics[height=2.2cm]{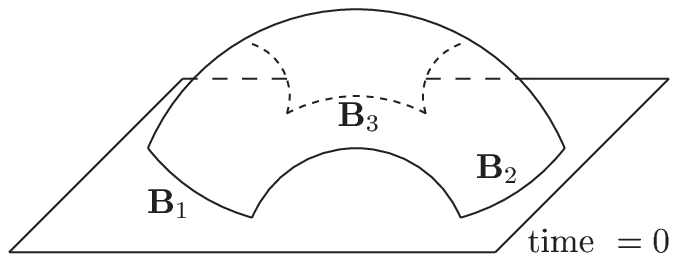}.
\end{equation}
We abbreviate this to $T_3[0,0,0]$. This functional could be computed in first quantisation with the Polyakov integral on a disk with marked sections on the boundary. We propose the lowest order expansion of the vacuum functional,
\begin{equation}\label{Gamma}
  \includegraphics[height=2.2cm]{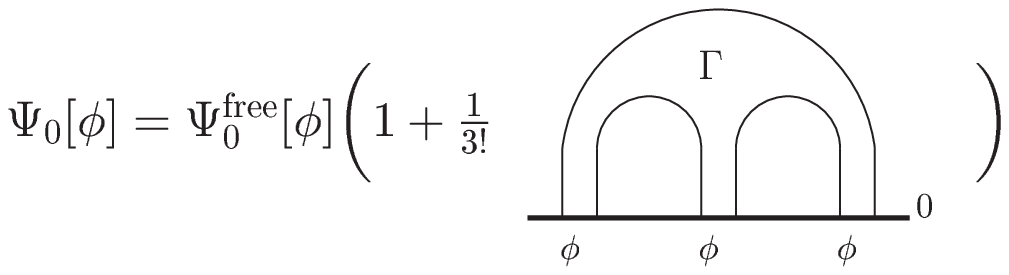}.
\end{equation}
where $\Gamma$ is an unknown. The three field vacuum expectation is
\begin{equation*}
  \langle \hat\phi[\BB_1,0]\hat\phi[\BB_2,0]\hat\phi[\BB_3,0] \rangle = \int\pathD\phi\,\,\phi[\BB_1,0]\phi[\BB_2,0]\phi[\BB_3,0]\Psi_0^2[\phi]
\end{equation*}
which implies

\begin{equation}\label{svwf10}
  \includegraphics[height=2.2cm]{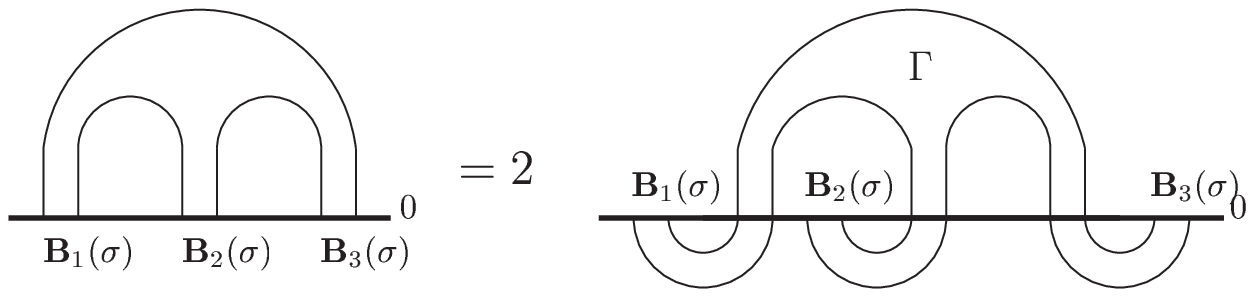}
\end{equation}
where the diagram on the left will represent $T_3[0,0,0]$ for simplicity. As in (\ref{4point2}) we invert the equal time propagators on the right hand side of (\ref{svwf10}). We find the first order cubic term in the vacuum state functional is
\begin{equation}\label{odd}
  \includegraphics[height=2.2cm]{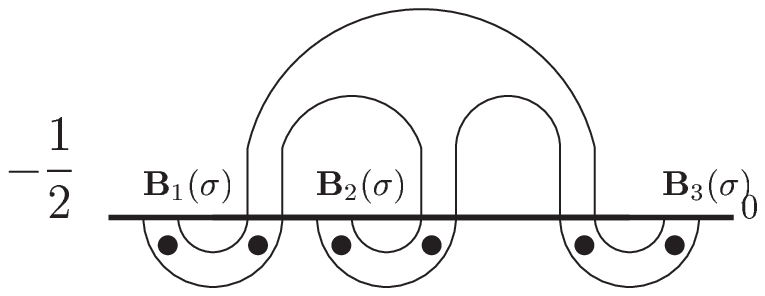}
\end{equation}
Our gluing rules give us the result of joining propagators when each end lies on a constant time surface, but this does not suffice to sew more general surfaces together. In \cite{Carlip} it was shown that to correctly sew the moduli spaces of two worldsheets the inverse propagator (first quantised Hamiltonian) should be attached to one of the boundaries being sewn, and then the field arguments should be integrated over. This removes a divergent integral over a redundant moduli parameter and gives the correct moduli space for the sewn worldsheets.

Using the simple identity
\begin{equation}
  f(0) = \int\!\ud s\,\, f(s) \delta(s) = \int\!\ud s\ud q\,\, f(s)G^{-1}(s,q)G(q,0).
\end{equation}
we can write the expectation value as
\begin{equation}
  T_3[0,0,0] = \int\!\prod\limits_{j=1}^3\ud t_j'\ud t_j\,\, T_3\big[t_1',t_2',t_3'\big]\prod\limits_{i=1}^3G^{-1}(t_i',t_i)G(t_i,0)
\end{equation}
Attaching the inverse of the equal time propagator to the left hand side of (\ref{svwf10}) has the effect of differentiating the propagators at the boundary $t=0$,
\begin{equation}
  \int\pathD\BB\,\, G(t,\BB_2;0,\BB)G(\BB,\overset{\bullet}{0};\BB_1,\overset{\bullet}{0}) = \text{sign}(t)G(t,\BB_2;\overset{\bullet}{0},\BB_1) = G(|t|,\BB_2;\overset{\bullet}{0},\BB_1).
\end{equation}
where the bullet is the time derivative with factor $-2$ as before, and we find that the vacuum functional is
\begin{equation}
  \Psi_0[\phi] = \Psi_0^\text{free}[\phi]\bigg(1- \frac{1}{2.3!}\int\!\prod\limits_{j=1}^3\ud t_j'\ud t_j \,\,T_3\big[t_1',t_2',t_3'\big]\prod\limits_{i=1}^3G^{-1}(t_i',t_i)\phi_\BB\cdot G_\BB(|t_k|;\overset{\bullet}{0})\bigg).
\end{equation}
In our case the inverse propagator is that part of the first quantised string theory Hamiltonian $H$ which depends on $\BB$ and $t$. We can represent the vacuum functional by

\begin{equation}
  \includegraphics[height=2.8cm]{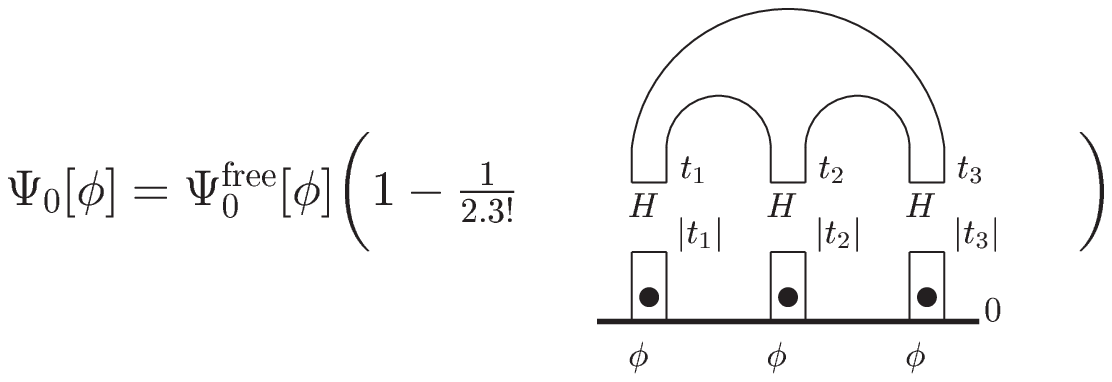}
\end{equation}
where we have written $H$ in place of the inverse propagators for clarity. The tree level $n$--field terms in the vacuum functional can be constructed from the tree level $n$--field expectation value in the same way. The first loop term in the vacuum functional is the tadpole. We introduce another unknown $\mu$ into the vacuum functional,
\begin{equation*}
  \Psi_0[\phi] = \Psi^\text{free}_0[\phi]\bigg(1+ \frac{\lambda}{3!\hbar}\Gamma\phi\phi\phi+ \mu\phi\bigg)
\end{equation*}
and calculate
\begin{equation}\label{tad-int}
  \langle\hat\phi[\BB,0]\rangle = \int\pathD\phi\,\,\phi[\BB]\Psi_0^2[\phi].
\end{equation}
If we put the factors of $\hbar$ back in we find that the tadpole receives contributions from both $\mu$ and $\Gamma$. The new unknown $\mu$ must cancel the unwanted contributions from $\Gamma$ and leave only the tadpole.  We perform the functional integral and invert the equal time propagators as before to identify $\mu$ as
\begin{equation}
  \includegraphics[height=2.9cm]{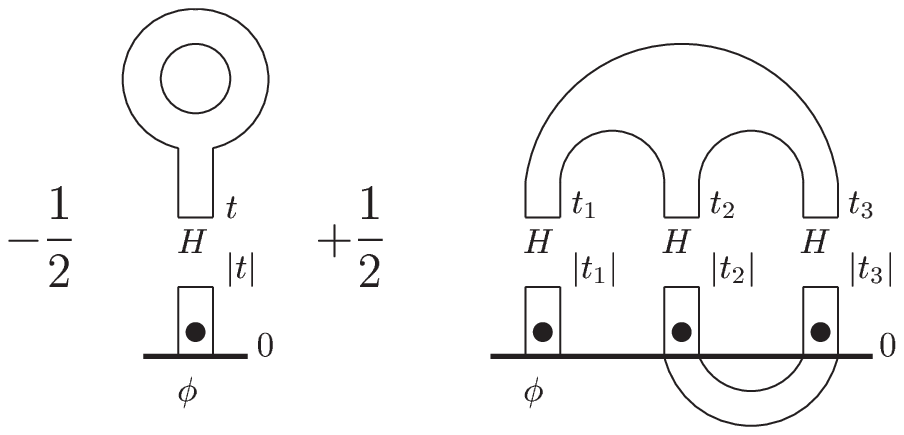},
\end{equation}
where the first term generates the tadpole and the second cancels the contribution from $\Gamma$ when we perform the integration in (\ref{tad-int}). Explicitly, if the three field expectation value is $T_3$ and the tadpole is $R_1$ then we have found that the fist terms of the vacuum state functional are
\begin{align}
    \nonumber \Psi_0[\phi] = \Psi_0^\text{free}[\phi]\bigg(1 &-\frac{1}{2.3!}\int\!\prod\limits_{j=1}^3\ud t_j'\ud t_j \,\,T_3\big[t_1',t_2',t_3'\big]\prod\limits_{i=1}^3G^{-1}(t_i',t_i)\phi_\BB\cdot G_\BB(|t_k|;\overset{\bullet}{0}) \\
\nonumber  &-\frac{1}{2}\int\!\ud t' R_1[t']G^{-1}(t';t)G(|t|;\overset{\bullet}{0};)\phi \\
+\frac{1}{2}\int\!\prod\limits_{j=1}^3\ud t_j'\ud t_j \,\,T_3\big[t_1',&t_2',t_3'\big] G^{-1}(t_1',t_1)   G_\BB(|t_1|;\overset{\bullet}{0})\phi_\BB G(t_2;t_3)\prod\limits_{i=2}^3 G^{-1}(t_i',t_i)G(|t_i|;\overset{\bullet}{0})\bigg)
\end{align}
The series is very similar to that found for local field theory, and we can continue these arguments to higher order in perturbation theory. We point out that this construction applies to both open and closed strings, whereas other approaches to closed string field theory are usually more complex than their open counterparts.

Rather than start with an action and try to derive amplitudes we began with the first quantised theory and used it to construct a second quantised object. We used Carlip's sewing methods together with our own to give the correct measures on moduli space when sewing worldsheets. Unlike in other string field theories our time co-ordinate is a time at which the whole spatially extended string (with ghosts) exists. This seems to work around the problems of, for example, quantising Witten's theory. There, time is normally taken to be the midpoint of $X^0(\sigma)$ but the string remains extended in time and there are difficulties with quantisation without going to the light cone \cite{Maeno}.

\section{The string field Schr\"odinger functional and T-duality}
In this section we return to the field momentum representation, where the free Schr\"odinger functional is
\begin{equation}\label{string-schro}
  \includegraphics[width=0.85\textwidth]{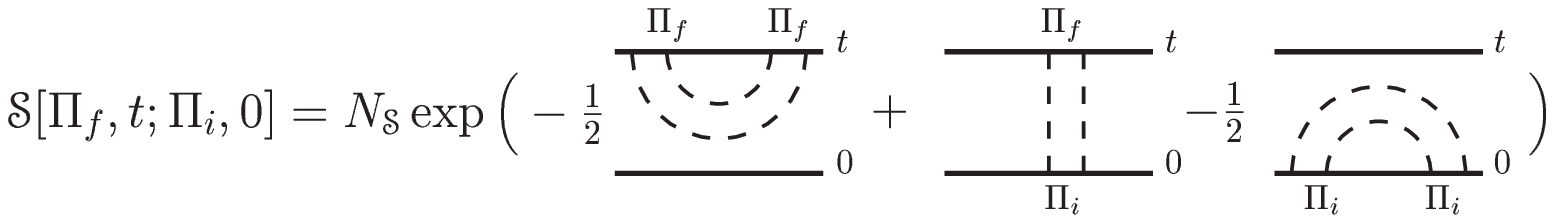}
\end{equation}
for momentum string fields $\Pi[\X, {J^\sigma}]$, and the double dashed represents the orbifolded propagator for either the open or closed string. The normalisation constant will be discussed shortly. The orbifold leads naturally to the question of what r\^ole T-duality plays. We now show that T-duality exchanges the states attached to the Schr\"odinger functional with backgrounds in the dual picture, and vice versa.

\subsection{Closed strings}

The closed string Schr\"odinger functional is T-dual to the loop diagrams appearing in the normalisation of the open string Schr\"odinger functional. We set $t=\pi R$, making the orbifold radius explicit, and use Poisson resummation and a change in modular parameter to convert the closed propagators into open loops so that, in an obvious notation, the Schr\"odinger functional becomes
\begin{equation}\label{logs}\begin{split}
  \log \mathscr{S}_\text{closed} &= -\frac{1}{2}\sum_\text{$n$ even}\!\Pi_f G_{\pi Rn}\Pi_f +\sum_\text{$n$ odd}\!\Pi_f G_{\pi Rn}\Pi_i-\frac{1}{2}\sum_\text{$n$ even}\!\Pi_i G_{\pi Rn}\Pi_i \\
  &= -\frac{1}{2}\sum_\text{$n$ even}\Pi_f G_{\pi\til{R}n}\Pi_f +\!\!\sum_\text{$n$ even}\Pi_i G_{\pi\til{R}n}e^{in\pi/2}\Pi_f-\frac{1}{2}\sum_\text{$n$ even}\Pi_i G_{\pi\til{R}n}\Pi_i
\end{split}\end{equation}
where the sums impose the Neumann boundary conditions on the propagators, $\til{R}=\alpha'/R$ and $G$ in the second line is an open string contribution. The new exponent in the second line comes from the Poisson resummation,
\begin{equation}\begin{split}
  \sum\limits_{n\in\mathbb{Z}} e^{-\pi an^2 +2\pi inb} = \frac{1}{\sqrt{a}}\sum\limits_{m\in\mathbb{Z}}e^{-\frac{\pi}{a}(m+b)^2}
\end{split}\end{equation}
The fields originally glued onto the Dirichlet sections of the closed string propagator become an averaging over backgrounds, characterised by $\Pi_i$ and $\Pi_f$, coupling to the ends of the open string. These backgrounds, as they must be, are the same at each end of the string, for we can write the above as
\begin{equation}
  \log\mathscr{S}_\text{closed} = -\frac{1}{2}\sum\limits_\text{$n$ even}\big(\Pi_i-e^{iA\int\!\ud X^0}\Pi_f\big) G_{\pi\til{R}n} \big(\Pi_i-e^{iA\int\!\ud X^0}\Pi_f\big)
\end{equation}
with Wilson line value $A=(2\til{R})^{-1}$. Let us give an explicit example. We will focus on the co-ordinates. Consider the reparametrisation invariant boundary states
\begin{equation}\label{states}
  \Pi_{\,i,f}[\X]=\delta^p(\X(\sigma)-\q_{\,i,f})
\end{equation}
for $\q_{\,i,f}$ constant $p$-vectors. These are pointlike states in $p$ directions and Neumann states in $25-p$ directions, $0\leq p\leq 25$. The closed string Schr\"odinger functional is
\begin{equation}\label{example1}\begin{split}
\log \mathscr{S}_\text{closed}= &-\text{Vol}^{25-p}\int\limits_0^\infty\!\frac{\ud T}{T^\frac{p+1}{2}}\,\,e^{2T}\prod\limits_{m=1}\big(1-e^{-2mT}\big)^{-24}\sum\limits_\text{$n$ even} e^{-\frac{\pi^2 R^2}{2\alpha'T}n^2} \\
&+\text{Vol}^{25-p}\int\limits_0^\infty\!\frac{\ud T}{T^\frac{p+1}{2}}\,\,e^{-\frac{\delta\q^2}{2\alpha' T} + 2T}\prod\limits_{m=1}\big(1-e^{-2mT}\big)^{-24}\sum\limits_\text{$n$ odd} e^{-\frac{\pi^2 R^2}{2\alpha'T}n^2}
\end{split}\end{equation}
with $\delta\q=\q_f-\q_i$. Since the open string runs from $\sigma=0\ldots\pi$ and the closed string from $\sigma=0\ldots 2\pi$ we must scale the closed string worldsheet to interpret (\ref{example1}) as an open loop. We include this in a change of modular parameter $U:=2\pi^2/T$. After this and the Poisson resummation we find
\begin{equation}\label{example2}\begin{split}
  \log\mathscr{S}_\text{closed}=-\text{Vol}^{25-p}\int\limits_0^\infty\frac{\ud U}{U}\frac{1}{U^\frac{26-p}{2}}e^U&\prod\limits_{m=1}(1-e^{-mU})^{-24} \\
  &\times\sum\limits_\text{$n$ even}e^{-\frac{\pi^2\til{R}^2}{4\alpha'U}n^2}\bigg(1-e^{\frac{in\pi}{2}} e^{-\frac{\delta\q^2 U}{4\pi\alpha'}}\bigg).
\end{split}\end{equation}
Now consider an open string loop. The measure on Teichm\"uller space is $\ud U/U$ (this can be viewed as giving the logarithm of the trace of the worldsheet propagator). If the string has Neumann conditions at its endpoints in $26-p$ directions (including $X^0$) and Dirichlet conditions in $p$ directions, as for a string on a D$(25-p)$-brane then the trace over $\X$ gives the eta function and the factor $(U^{-1/2}\text{Vol})^{(25-p)}$ from the $25-p$ zero modes. The sum and remaining factor of $U^{-1/2}$ come from the trace over $X^0$ in the co-ordinate representation. We arrive at (\ref{example2}), if the term in large brackets represents an averaging over backgrounds of Wilson lines and D$(25-p)$-branes of separation $\delta\q$.

\subsection{Open strings}

We interpret the open string duality as taking us from one Schr\"odinger functional to another with an exchange of boundary states and backgrounds. Poisson resummation implies
\begin{equation}\label{new-R-sums}
\begin{split}
  \bigg(\frac{\pi R^2}{\alpha' T}\bigg)^{1/2}\sum\limits_\text{$n$ even}e^{-\frac{\pi^2 R^2}{4\alpha' T}n^2} &= \sum\limits_\text{$n$ odd}e^{-\frac{\tilde{R}^2 T}{4\alpha'}n^2} + \sum\limits_\text{$n$ even}e^{-\frac{\tilde{R}^2 T}{4\alpha'}n^2} \\
  \bigg(\frac{\pi R^2}{\alpha' T}\bigg)^{1/2}\sum\limits_\text{$n$ odd}e^{-\frac{\pi^2 R^2}{4\alpha' T}n^2} &= -\sum\limits_\text{$n$ odd}e^{-\frac{\tilde{R}^2 T}{4\alpha'}n^2} + \sum\limits_\text{$n$ even}e^{-\frac{\tilde{R}^2 T}{4\alpha'}n^2}
\end{split}
\end{equation}
where the dual radius is now $\overline{R}=2\alpha'/R$. Following a modular transformation $S:=\pi^2/T$ these are the sums in the open string Schr\"odinger functional. Again the states now represent an averaging over backgrounds. The new momentum states are characterised by the original Neumann condition on the open string ends. The open string Schr\"odinger functional becomes
\begin{equation}
  \log\mathscr{S}_\text{open} = -\frac{1}{2}\sum_\text{$n$ even}(\Pi_i - \Pi_f) G_{n\pi\overline{R}}(\Pi_i-\Pi_f) - \frac{1}{2}\sum_\text{$n$ odd}(\Pi_i + \Pi_f) G_{n\pi\overline{R}}(\Pi_i + \Pi_f)
\end{equation}
To interpret this as strings moving in a single background we can introduce a Wilson line, $\Pi_i - e^{iA\int\!\ud X^0}\Pi_f$, with value $A=(\overline{R})^{-1}$.

Explicit examples are difficult to construct for two reasons: potential difficulties in finding reparametrisation invariant or BRST invariant states, and keeping track of the corner anomaly in the states and backgrounds. A simple example of a state independent of parameterisation would seem to be a string collapsed to a point, as we used for the closed string previously. The closed string invariant states are more commonly known as boundary states \cite{Ishy} arising via closed string channel descriptions of open string loop amplitudes \cite{Callan1}. They are characterised by the condition
\begin{equation}\label{cl-cond}
	L_n - \bar{L}_{-n} = 0\qquad \forall\, n\in\mathbb{Z}
\end{equation}
(at worldsheet $\tau=0$ for simplicity). The pointlike, or localised, state is given by
\begin{equation*}	
	\ket{D;q} = \exp\left(\sum\limits_{n=1}\frac{1}{n} \alpha_{-n}\cdot\tilde\alpha_{-n}\right)\ket{q},\qquad \hat{X}^\mu(\sigma,0)\ket{D;q}=q^\mu\ket{D;q}
\end{equation*}
and the state $\ket{q}$ is an eigenstate of the c.o.m. of the co-ordinate $X^\mu$. By writing the Virasoro generators in terms of the oscillator modes $\alpha_n$ it is easy to check that this state satisfies (\ref{cl-cond}). The corresponding generators for the open string are usually taken to be $L_n - L_{-n}$. This was derived in \cite{Ramond} by inserting the mode expansion for $X^\mu$ into the functional expression for the generators $M_n$ of reparametrisations,
\begin{equation}\label{rep-gen}
	M_n := \sqrt{\frac{2}{\pi}}\int\limits_0^\pi\!\ud\sigma\,\sin(n\sigma) X'(\sigma)\frac{\delta}{\delta X(\sigma)}.
\end{equation}
Since this is an operator expression the authors chose to normal--order the result. It is however already well defined -- normal ordering unreasonably removes a finite constant. The generators are in fact
\begin{equation}\label{anomalous}
	M_n = \frac{1}{\sqrt{2\pi}} \left( L_n - L_{-n} + \frac{nd}{4}\delta_{n/2 \in \mathbb{Z}}\right)
\end{equation}
in $d$ dimensions. Consistency can be checked in that the Neumann state $\ket{N}$ obeying $\hat{\pi}(\sigma)\ket{N}=0$ is killed by $M_n\,\forall\, n$, as is appropriate from (\ref{rep-gen}). The pointlike open string state $\ket{D}$ obeys
\begin{equation}
  \begin{split}
    \ket{D;q}_\text{open} &= \exp\bigg(\sum\limits_{n=1}\frac{1}{2n}\alpha_{-n}\alpha_{-n}\bigg)\ket{q}, \\
	\bigg(L_n - L_{-n} &- \frac{nd}{4}\delta_{n/2\in\mathbb{Z}}\bigg)\ket{D;q} = 0.
	\end{split}
\end{equation}
The eigenvalue has the wrong sign to be a solution. Similar observations have been made before in the consideration of pointlike states, for example \cite{Green1}. However, the state is indeed independent of the co-ordinate $\sigma$,
\begin{equation*}
	\partial_\sigma\hat{X}(\sigma,0)\ket{D;q}=0,
\end{equation*}
and the state wave functional $\bracket{X}{D;q}$ is reparametrisation invariant -- it is a delta functional,
\begin{equation}
	\bracket{X}{D;q} = \dirac{X(\sigma)-q} = \int\pathD\lambda^\mu\, \exp\left(\int\!\ud\sigma\sqrt{g}\, \lambda^\mu (X_\mu-q_\mu)\right),
\end{equation}
with reparametrisation invariant measure $(\lambda,\lambda) = \int\!\ud^2\xi \sqrt{g}\lambda_\mu\lambda^\mu$. There is a mismatch between the functional and Hilbert space approaches. In the latter the integral over the metric has been performed and so the meaning of a reparametrisation is unclear, to which we attribute the discrepancy. There are further contributions to the reparametrisation generators from quantum effects, which we discuss in the next section.

Let us give an example of the open string duality based on the pointlike and Neumann states. Take the states attached to the Schr\"odinger functional to be the open string equivalent of (\ref{states}). The logarithm of the Schr\"odinger functional is given by
\begin{equation}\label{old}
\begin{split}
  \frac{\log \mathscr{S}}{\text{Vol}^{25-p}} = -&\int\limits_0^\infty\!\frac{\ud T}{T^\frac{p+1}{2}}\sum\limits_\text{$n$ even} e^{-\frac{\pi^2 R^2}{4\alpha' T}n^2+T}\prod\limits_{m=1}\big(1-e^{-2mT}\big)^{-12} \\
  &+\int\limits_0^\infty\!\frac{\ud T}{T^\frac{p+1}{2}}e^{-\frac{(\x_f-\x_i)^2}{4\alpha' T}}\sum\limits_\text{$n$ odd} e^{-\frac{\pi^2 R^2}{4\alpha' T}n^2+T}\prod\limits_{m=1}(1-e^{-2mT})^{-12}
\end{split}
\end{equation}
where Vol is the volume of space generated from attaching the $25-p$ Neumann states. Following the Poisson resummations in (\ref{new-R-sums}) the logarithm of the Schr\"odinger functional becomes
\begin{equation}\label{new}
\begin{split}
  \frac{\log \mathscr{S}}{Vol^{25-p}} = -&\int\limits_0^\infty\!\frac{\ud T}{T^{p/2}}\sum\limits_\text{$n$ even} e^{-\frac{\tilde{R}^2 T}{4\alpha'}n^2+T}\bigg(1 + e^{-\frac{(\x_f-\x_i)^2}{4\alpha' T}}\bigg)\prod\limits_{m=1}\big(1-e^{-2mT}\big)^{-12} \\
  &+\int\limits_0^\infty\!\frac{\ud T}{T^{p/2}}\sum\limits_\text{$n$ odd} e^{-\frac{\tilde{R}^2 T}{4\alpha'}n^2+T}\bigg(-1 - e^{-\frac{(\x_f-\x_i)^2}{4\alpha' T}}\bigg)\prod\limits_{m=1}\big(1-e^{-2mT}\big)^{-12} 
\end{split}
\end{equation}
The terms inside the large brackets represent the new background for the string. As with the closed string, we could now perform a change of variable $U=\pi^2/T$ which would give us back the sums and determinants of (\ref{old}). First, we can interpret the above expression in terms of a Schr\"odinger functional constructed in a different gauge. Choosing our gauge fixed worldsheet metric to be $\tilde{g}=\text{diag}(T^2,1)$ amounts to rotating the worldsheet by $90^\circ$ so that $T$ represents the length of the string (the co-ordinate ranges are reversed $\sigma\in[0,1]$, $\tau\in[0,\pi]$). The $X^0$ contributions to the Schr\"odinger functional in this gauge are
\begin{equation*}
  e^{T/24}\prod\limits_{m=1}\big(1-e^{-2mT}\big)^{-1/2}\,\sum e^{-\frac{\tilde{R}^2 T}{4\alpha'}n^2}.
\end{equation*}
A co-ordinate with Neumann conditions on its endpoints attached to Neumann states gives
\begin{equation*}
  \text{Vol}\,\,e^{T/24}\prod\limits_{m=1}\big(1-e^{-2mT}\big)^{-1/2}
\end{equation*}
and a co-ordinate with Dirichlet conditions on its endpoints attached to a Neumann state gives
\begin{equation*}
  \frac{1}{T^{1/2}}\,e^{T/24}\prod\limits_{m=1}\big(1-e^{-2mT}\big)^{-1/2}
\end{equation*}
in addition to any contribution from the background it couples to. We conclude that (\ref{new}) is the Schr\"odinger functional for a string in this gauge, with initial and final field momentum Neumann states $\Pi=1$, with an averaging over backgrounds of D(25-p)-branes and Wilson lines.

We might expect to see a description of the same system if we perform the change of variables on the Teichm\"uller parameter. Doing this gives us
\begin{equation}\label{new2}
\begin{split}
  \frac{\log \mathscr{S}}{Vol^{25-p}} = -&\int\limits_0^\infty\!\frac{\ud U}{U^\frac{26-p}{2}}\,U^5\,\sum\limits_\text{$n$ even} e^{-\frac{\tilde{R}^2 \pi^2}{4\alpha'U}n^2+U}\bigg(1 + e^{-\frac{U(\x_f-\x_i)^2}{4\alpha'\pi^2}}\bigg)\prod\limits_{m=1}\big(1-e^{-2mU}\big)^{-12} \\
  &+\int\limits_0^\infty\!\frac{\ud U}{U^\frac{26-p}{2}}\,U^5\,\sum\limits_\text{$n$ odd} e^{-\frac{\tilde{R}^2 \pi^2}{4\alpha'U}n^2+U}\bigg(-1 - e^{-\frac{U(\x_f-\x_i)^2}{4\alpha' \pi^2}}\bigg)\prod\limits_{m=1}\big(1-e^{-2mU}\big)^{-12} 
\end{split}
\end{equation}
There is an additional factor of $U^5$ in both terms to the `expected' result. However, we have not been careful with the corner anomaly. The change of Teichm\"uller parameter $T\rightarrow \pi^2/T$ corresponds to a scaling of the metric. If we put the parameter dependence into the metric (so $\hat{g}_{\sigma\sigma}\rightarrow \pi^2$ then the change of variable corresponds to the scaling
\begin{equation}
  g_{ab}=\bigg(\begin{array}{cc}\pi^2 & 0 \\ 0 & T^2\end{array}\bigg)\rightarrow\bigg(\begin{array}{cc}\pi^4/T^2 & 0 \\ 0 & \pi^2\end{array}\bigg) = \frac{\pi^2}{T^2}\bigg(\begin{array}{cc}\pi^2 & 0 \\ 0 & T^2\end{array}\bigg).
\end{equation}
This is a constant Weyl scaling with Liouville mode $\rho = 2\log(\pi^2/T)$. The logarithms of the various determinants in the Polyakov integral depend linearly on $\rho$ at the corners, so a scaling must produce \emph{powers} of $T$ in the propagator. Joining states and background contributions produces the same effect, thus the additional factors in (\ref{new2}).

\section{Gauge fixing and BRST}
The factorisation of the ghosts we have used may seem ad-hoc but follows from a choice of gauge equivalent to the usual choice $\sqrt{g}g^{ab} = \sqrt{\hat{g}}\hat{g}^{ab}(T)$. We begin this section by showing this with Faddeev Popov gauge fixing and then investigate the BRST symmetry.

Our gauge fixing conditions are
\begin{eqnarray}\label{GF-conds}
  \int\!\ud^2\sigma\sqrt{g}\,g^{ab}\hat{h}_{ab}\equiv(\hat{h}_{ab}|g_{ab}) &=0, \\
  \hat{P}^\dagger\bigg(\frac{\sqrt{g}g^{rs}}{\sqrt{\hat{g}}}\bigg)^a &=0.
\end{eqnarray}
These conditions are equivalent to the usual choice $\sqrt{g}g^{ab} = \sqrt{\hat{g}}\hat{g}^{ab}(T)$ as we now show. A variation of the metric can be decomposed into a Weyl scaling, a reparametrisation and a modular transformation as
\begin{eqnarray}\label{metric-var1}
  \delta g_{ab} &=& \delta\rho\,g_{ab} + \nabla_{(a}\delta\xi_{b)} + \delta T\, g_{ab,T}, \\
  \label{metric-var2}
  \implies \delta (\sqrt{g}g^{ab}) &=& -\sqrt{g}P(\delta\xi)^{ab} + \sqrt{g}\,\delta T\,\chi^{ab}
\end{eqnarray}
where $\chi^{ab}$ is the traceless symmetric part of ${g^{ab}}_{,T}$. The first constraint above implies
\begin{equation}
  \delta T (\hat{h}_{ab}|\hat{\chi}^{ab}) =0 \implies \delta T =0
\end{equation}
since the inner product is not zero. The second constraint gives
\begin{equation}
  \hat{P}^\dagger \hat{P}{\delta\xi}^a =0 \implies {\delta\xi}^a =0
\end{equation}
by the boundary conditions on $\delta\xi^a$ (for the closed string we take $\delta\xi^a$ to be orthogonal to the CKV to get a good co-ordinate system). Inserting into (\ref{metric-var2}) we have
\begin{equation}
  \delta(\sqrt{\hat{g}}\hat{g}^{ab}) = 0 \implies \sqrt{g}g^{ab} = \sqrt{\hat{g}}\hat{g}^{ab}(T)
\end{equation}
as claimed. We now compute the Faddeev-Popov determinant. Define the determinant $\Delta_\text{FP}$ by
\begin{equation}\label{FP-def}
  1 = \int\pathD\zeta\int\!\ud T\,\, \Delta_\text{FP}(g,T)\delta\big[(\hat{h}_{ab}\,|\, g^{ab})^\zeta\big]\prod\limits_{a=1}^2 \delta\bigg[\hat{P}^\dagger\left(\frac{\sqrt{g}g^{rs}}{\sqrt{\hat{g}}} \right)^a\bigg]^\zeta
\end{equation}
where $\pathD\zeta$ is the measure on the diff$\,\times$Weyl group. This and $\Delta_\text{FP}$ are invariant under the action of $\zeta$. Suppose that given $g$ the constraints have a solution $\zeta={\rho\xi}$ and $T$. Expanding about this solution gives
\begin{equation}\label{FP-vary}
  1 = \int\pathD(\delta\rho,\delta\xi)\int\!\ud \delta T\,\, \Delta_\text{FP}\,\,\delta\big[\delta T\, (\hat{h}_{ab}\,|\, \hat\chi^{ab})\big]\,\, \delta\big[\hat{P}^\dagger\hat{P}\delta\xi^a\big].
\end{equation}
We can integrate out the Teichm\"uller parameter and represent the delta functional by an integral over a vector Lagrange multiplier $\lambda^a$, 
\begin{equation}
  1 = \Delta_\text{FP}(\hat{g})(\hat{h}\,|\,\hat{g}(T)_{ab,T})^{-1}\int\pathD(\delta\rho, \delta\xi,\lambda)\exp\bigg(\int\!\ud^2\sigma\sqrt{\hat{g}}\,\,\lambda^{a}\hat{P}^\dagger \hat{P}\delta\xi^a\bigg).
\end{equation}
In the critical dimension the integral over the Liouville mode $\delta\rho$ contributes a volume (if we cancel the corner anomalies). As usual we invert the expression for $\Delta_\text{FP}$ by replacing bosonic with Grassmann variables. Following this and an integration by parts we have
\begin{equation}
  \Delta_{FP}(\hat{g}) = \big(\hat{h}\,|\,\hat{g}_{ab,T}(T)\big)\int\pathD(\gamma,\, c)\exp\bigg(\int\!\ud^2\sigma\sqrt{\hat{g}}\,\,P(\gamma)^{ab} \hat{P}(c)_{ab}\bigg).
\end{equation}
where $\gamma^a$, $c^a$ are Grassmann vectors obeying the Alvarez boundary conditions. We now wish to insert our expression for ``1" into the Polyakov integral and carry out the integration over the metric. Since we known our gauge choice is equivalent to the usual choice we have two expressions for ``1",
\begin{equation}
  1 = \int\pathD\zeta\int\!\ud T\,\Delta^0_\text{FP}\delta(\sqrt{g}g - \sqrt{\hat{g}}\hat{g}) = \int\pathD\zeta\int\!\ud T\,\, \Delta_\text{FP}(g,T)\delta\big[(\hat{h}_{ab}\,|\, g^{ab})\big]\delta\bigg[\hat{P}^\dagger\left(\frac{\sqrt{g}g^{rs}}{\sqrt{\hat{g}}} \right)\bigg]
\end{equation}
where $\Delta_{FP}^0$ is the standard $b$-$c$ determinant. By changing variables $b^{\perp}=P\gamma$ in the expression for the FP determinants where $b^{\perp}$ is orthogonal to the zero mode of $\hat{P}^\dagger$ we find
\begin{equation}
  \Delta^0_\text{FP} = \Det'(\hat{P}^\dagger \hat{P})^{-1/2}\Delta_\text{FP}
\end{equation}
and we can carry out the integration over $g$ in our functional integral using the usual, single, delta function,
\begin{equation}
  \int\pathD g\frac{1}{\text{Vol diff$\,\times$Weyl}} = \int\!\ud T\, (\hat{h}_{ab}|\hat{\chi}^{ab})\big(\Det' \hat{P}^\dagger\hat{P}\big)^{-1/2}\int\pathD(\gamma,c)e^{ \int\hat{P}\gamma\cdot\hat{P}c_{ab}}.
\end{equation}

\subsection{BRST invariance}
The BRST action for our constraints is, setting $\alpha' = 1/2\pi$,
\begin{equation}\begin{split}
  S_\text{BRST} = \frac{1}{2}\int\!\ud^2\sigma\sqrt{g}g^{ab}\p_a X\p_b X &+ \frac{1}{2}\delta_Q\bigg[ \int\!\ud^2\sigma\sqrt{\hat{g}}\,\,\hat{P}^\dagger\bigg(\frac{\sqrt{g}}{\sqrt{\hat{g}}}g^{rs}\bigg)^a\hat{g}_{ab}\gamma^b \bigg] \\
  &+ \frac{1}{2}\delta_Q\bigg[\Gamma\int\!\ud^2\sigma\sqrt{g}g^{ab}\hat{h}_{ab} \bigg]
\end{split}\end{equation}
where $\gamma^a(\sigma,\tau)$ and $\Gamma$ are antighosts and the nilpotent BRST transformations are
\begin{equation}\begin{split}\label{BRST-trans1}
  \delta_Q X &= c^a\partial_a, X \\
  \delta_Q \sqrt{g}g^{ab} &= -\sqrt{g}P(c)^{ab}, \\
  \delta_Q \gamma^a &= B^a, \quad \delta_Q B^a =0, \\
  \delta_Q \Gamma &= B, \quad \delta_Q B =0. \\
\end{split}\end{equation}
Inserting these transformations into the action above we find, after an integration by parts,
\begin{equation}\begin{split}
  S_\text{BRST} = \frac{1}{2}\int\!\ud^2\sigma\sqrt{g}g^{ab}\p_a X\p_b X &+ \frac{1}{2}\int\!\ud^2\sigma\sqrt{g}\,\,\big(\hat{P}(\gamma)_{ab} + \Gamma\hat{h}_{ab}\big)P(c)^{ab} \\
  &+\frac{1}{2}\int\!\ud^2\sigma\sqrt{g}\,\,g^{ab}\big(\hat{P}(B)_{ab} + B\hat{h}_{ab}\big).
\end{split}\end{equation}
Integrating over the Lagrange multipliers in the final term imposes the gauge fixing constraints ($\ref{GF-conds}$). To find the BRST transformations which result after this integration we should solve the equations of motion of $g^{ab}$. Notice that the first two terms in $S_\text{BRST}$ have exactly the $g$ dependence of the standard string action,
\begin{equation*}
  \frac{1}{2}\int\!\ud^2\sigma\sqrt{g}g^{ab}\p_a X\p_b X + \frac{1}{2}\int\!\ud^2\sigma\sqrt{g}\,\,b_{ab}P(c)^{ab} \\
\end{equation*}
with $b_{ab}$ replaced by $\hat{P}\gamma + \Gamma\hat{h}$. The equations of motion for $g$ therefore imply
\begin{equation}
  2T_{ab} = -\hat{P}(B)_a -\hat{h}_{ab}B = -\delta_Q\big( \hat{P}(\gamma)_a +\hat{h}_{ab}\Gamma\big)
\end{equation}
where $T$ is the usual energy momentum tensor of the string with $b_{ab}=\hat{P}(\gamma)_{ab} + \Gamma\hat{h}_{ab}$. The variations of $\gamma$ and $\Gamma$ can be disentangled by multiplying both sides by $\hat{h}^{ab}$, since $\hat{P}(\gamma)$ is orthogonal to $\hat{h}$. However, after imposing the constraints any dependence on $\hat{h}_{ab}$ vanishes from the action, which becomes
\begin{equation}\label{theact}
  S_\text{BRST} = \frac{1}{2}\int\!\ud^2\sigma\sqrt{\hat{g}}\hat{g}^{ab}\p_a X\p_b X + \frac{1}{2}\int\!\ud^2\sigma\sqrt{\hat{g}}\,\,\hat{P}(\gamma)_{ab}\hat{P}(c)^{ab}.
\end{equation}
The BRST transformations reduce to
\begin{equation}\begin{split}\label{BRST-trans2}
  \delta_Q X &= c^a\partial_a\, X \\
  \delta_Q c^a &= c^b\partial_b\, c^a \\
  \delta_Q \hat{P}(\gamma)_{ab} &= -2\hat{T}_{ab}.
\end{split}\end{equation}
The energy momentum tensor is $\hat{T} = \hat{T}^X + \hat{T}^\text{gh}$,
\begin{equation}\begin{split}
  2\hat{T}^X_{ab} &= \p_a X\p_b X -\frac{1}{2}\hat{g}_{ab}\p^r X\p_r X, \\
  2\hat{T}^\text{gh}_{ab} &= \hat{P}(\gamma)_{p(a}\hat\nabla_{b)}c^p + (\hat\nabla_p \hat{P}(\gamma)_{ab})c^p -\hat{g}_{ab}\hat{P}(\gamma)_{rs}\hat\nabla^r c^s.
\end{split}\end{equation}
We have a non-local transformation for the antighost. Just as is found in the standard case the transformation for the antighost is not nilpotent,
\begin{equation*}
  \delta^2 \hat{P}(\gamma)_{ab} = -\mathcal{L}\hat{P}(c)_{ab}
\end{equation*}
where $\mathcal{L}$ is the Lagrangian density in (\ref{theact}). On a manifold without boundary, or a manifold with boundary where the Alvarez boundary conditions hold, we recover nilpotency when the ghost $c^a$ is on shell, for then $P^\dagger P c =0\iff P(c)=0$.

Although the BRST transformation is now non-local, it has the natural interpretation of generating reparametrisations of the boundary as we now describe. Consider the string field propagator written as
\begin{equation}
  G(\X_f;\X_i) = \int\pathD(X,\gamma,c)(\Det\hat{P}^\dagger \hat{P})^{-1/2}e^{-S_\text{BRST} -S_J}\bigg|_{X=X_i}^{X=X_f}
\end{equation}
where $S_\text{BRST}$ is as in (\ref{theact}) and $S_J$ is a source term which generates boundary values of the ghosts,
\begin{equation*}
  S_J = \frac{1}{2}\int\!\ud^2\sigma\sqrt{\hat{g}}\,\,  (\hat{P}^\dagger\hat{P}\gamma)_ac^a_\text{cl} + \gamma^a_\text{cl}(\hat{P}^\dagger\hat{P}c)_a.
\end{equation*}
In the above, $c^a_\text{cl}$ obeys $\hat{P}^\dagger\hat{P}=0$ and equals the boundary values of the ghosts on the Dirichlet sections of the worldsheet,
\begin{equation*}
  c^a(\sigma,0)=c^a_i(\sigma), \qquad c^a(\sigma,1)=c^a_f(\sigma), 
\end{equation*}
and similarly for $\gamma^a$. The integration variables obey the boundary conditions (\ref{new ghost BCs}) and $c^\tau_\text{cl}=\gamma^\tau_\text{cl}\equiv 0$. After repeated integration by parts we can write the source term as
\begin{equation}\label{source}
  S_J = \int\limits_\text{bhd}\!\ud\Sigma^b\,\,  (\hat{P}\gamma)_{ab}c^a_\text{bhd} + \gamma^a_\text{bhd}(\hat{P}c)_{ab}.
\end{equation}
The bulk action $S_\text{BRST}$ is invariant for arbitrary boundary values of $X,c,\gamma$,
\begin{equation*}
    \delta_Q S_\text{BRST} = -\int\!\ud^2\sigma\frac{\p}{\p\sigma^a} \bigg(c^a\mathcal{L}\bigg) =0
\end{equation*}
using the ghost boundary conditions $n^ac_a=0$. The source term $S_J$ does not respect this symmetry, so we are led to expect a Ward identity resulting from a shift in integration variables corresponding to (\ref{BRST-trans2}),
\begin{equation}\label{Ward}
  \langle\,\delta_Q S_J\,\rangle = 0
\end{equation}
where the expectation value is defined as 
\begin{equation*}
  \langle\Omega\rangle := \int\pathD(X,\gamma,c)\,\,\Omega\,\,(\Det\hat{P}^\dagger \hat{P})^{-1/2}e^{-S_\text{BRST} -S_J}\bigg|_{X=X_i}^{X=X_f}
\end{equation*}
so the propagator is $\langle1\rangle$. To simplify the remaining presentation we now fix the gauge $\hat{g}_{ab}=\delta_{ab}$ and absorb dependency on the Teichm\"uller parameter into the co-ordinates, so $\sigma\in[0,\pi]$ and $\tau\in[0,T]$. 
Consider calculating the expectation value of the ghosts away from the boundary. Each $c^a$ ($\gamma^a)$ is contracted with $\gamma^a$ ($c^a$) in the source term and brings down the classical field,
\begin{equation}
  \langle\,c^a(\sigma,\tau)\,\rangle = c^a_{cl}(\sigma,\tau),\qquad \langle\,\gamma^a(\sigma,\tau)\,\rangle = \gamma^a_{cl}(\sigma,\tau).
\end{equation}
As $\tau\rightarrow 0,1$ this gives the boundary values of the ghosts. In general an expectation value separates into quantum and classical pieces. In addition to the usual short distance divergences the quantum pieces will contain finite, non-zero contributions from the image charges- the corner anomaly contributes to the Ward identity. To illustrate, the contribution to (\ref{Ward}) from the first term in (\ref{source}), at $\tau=\epsilon>0$ is
\begin{equation}\label{source2}
  \langle\delta_Q \int\!\ud\sigma\, (P\gamma)_{\sigma\tau}c^\sigma_\text{b}\rangle = -\int\!\ud\sigma\, \big\langle\, X'\dot{X} + (P\gamma)_{p(\sigma}\partial_{\tau)}c^p + (\partial_p (P\gamma)_{\sigma\tau})c^p \,\big\rangle\,c^\sigma_\text{cl}(\sigma,\epsilon).
\end{equation}
The two boundaries contribute similarly so we will focus on $\tau=0$, and let a subscript ``$\text{b}$" denote boundary value. We will deal with the co-ordinate term first. The $X$--functional integrals are carried out by splitting $X$ into classical and quantum parts, $X=X_\text{cl} + X_q$. The expectation value becomes
\begin{equation*}
  -\int\!\ud\sigma\,\, c^\sigma_\text{cl}(\sigma,\epsilon)X'_\text{cl}(\sigma,\epsilon)\dot{X}_\text{cl}(\sigma,\epsilon) + c^\sigma_\text{cl}(\sigma,\epsilon)\big\langle\, X_q'(\sigma,\epsilon)\dot{X}_q(\sigma',\epsilon)\big\rangle.
\end{equation*}
To calculate the quantum contribution from the corners, we again go to the upper right quadrant with Dirichlet conditions at $\tau=0$ and Neumann conditions at $\sigma=0$. The Green's function, $F$, for $X_q$ on this geometry is given by the method of images,
\begin{equation}\label{F-images}
  F(\sigma,\tau;\sigma',\tau') = F_0(\sigma,\tau;\sigma',\tau') + F_0(\sigma,\tau;-\sigma',\tau') - F_0(\sigma,\tau;\sigma'\,-\tau') - F_0(\sigma,\tau;-\sigma',-\tau')
\end{equation}
in terms of the free space Green's function
\begin{equation}
  F_0(\sigma,\tau;\sigma'\tau') = -\frac{1}{4\pi}\log\big((\sigma-\sigma')^2 + (\tau-\tau')^2\big).
\end{equation}
The contribution we are interested in comes from the final term in (\ref{F-images}), as this is the only term which sees both reflections and hence the corner. Taking $\sigma=\sigma'$ and $\tau=\tau'=\epsilon$ this term contributes
\begin{equation}
  -\frac{1}{4\pi}\frac{\sigma}{\sigma^2 + \epsilon^2}\frac{\epsilon}{\sigma^2 + \epsilon^2} \rightarrow -\frac{1}{4}\frac{\delta(\sigma)}{\sigma}\quad\text{as $\epsilon\rightarrow 0$}.
\end{equation}
Since each corner contributes equally and independently we know that the expectation value on the strip will be given by
\begin{equation}
  \big\langle\, X_q'(\sigma,0)\dot{X}_q(\sigma',0)\big\rangle =-\frac{1}{4}\frac{\delta(\sigma)}{\sigma} -\frac{1}{4}\frac{\delta(\sigma-\pi)}{\sigma-\pi}
\end{equation}
per spacetime dimension. This leads to a quantum contribution to the Ward identity,
\begin{equation}
-\int\limits_0^\pi\!\ud\sigma\,\, c^\sigma_\text{b}(\sigma)\bigg(-\frac{1}{4}\frac{\delta(\sigma)}{\sigma} -\frac{1}{4}\frac{\delta(\sigma-\pi)}{\sigma-\pi}\bigg) = \frac{1}{4}\bigg({{c^\sigma}_{\!\text{b}}}'(0) + {{c^\sigma}_{\!\text{b}}}'(\pi)\bigg),
\end{equation}
where we have applied L'H\^opital's rule since $c^\sigma(0)=c^\sigma(\pi)=0$. As we take $\epsilon\rightarrow0$ the total contribution from the co-ordinates to the Ward identity is
\begin{equation}
  -\int\!\ud\sigma\, {{c^\sigma}_{\!\text{b}}}(\sigma) X'_\text{b}(\sigma)\dot{X}_\text{cl}(\sigma,0) + \frac{1}{4}\bigg({{c^\sigma}_{\!\text{b}}}'(0) + {c^\sigma}_{\!\text{b}}'(\pi)\bigg).
\end{equation}
The remaining classical field can be written as a derivative acting on the propagator $\langle 1\rangle$ with respect to the $X$ boundary data, so we have the operator expression
\begin{equation}\label{7.33}
2\int\!\ud\sigma\, c^\sigma_\text{b}(\sigma) X'_\text{b}(\sigma)\frac{\delta}{\delta X_\text{b}(\sigma)} + \frac{1}{4}\bigg({c^\sigma}_{\!\text{b}}'(0) + {c^\sigma}_{\!\text{b}}'(\pi)\bigg)
\end{equation}
per co-ordinate. This looks like a reparametrisation apart from the terms coming from the corners. These are of precisely the form of the constant terms found in (\ref{anomalous}) as we can see by using a Fourier representation for the fields. The ghost $c^\sigma$ is expanded as
\begin{equation*}\begin{split}
  &c^\sigma_\text{b}(\sigma) = \sqrt{\frac{2}{\pi}}\sum\limits_{m=1}^\infty c_m \sin m\sigma \\  
  \implies &{c^\sigma}_{\!\text{b}}'(0) + {c^\sigma}_{\!\text{b}}'(\pi) = \sqrt{\frac{2}{\pi}}\sum\limits_{m=1}^\infty 2m c_m\, \delta_{m/2\in\mathbb{Z}},
\end{split}\end{equation*}
and each mode multiplies a generator of reparametrisations in (\ref{7.33}) and a constant term as found in (\ref{anomalous}).

All that remains is to calculate the ghost contributions from (\ref{source}) to the Ward identity. Recall that we found the total contribution to the corner anomaly in $\Det P^\dagger P$ vanished using our ghosts, cancelling between the $\sigma$ and $\tau$ components. The classical ghost terms in $(\ref{source2})$ are
\begin{equation}
  -\int\!\ud\sigma\,\, c^\sigma_\text{b}(\sigma){\gamma^\sigma}'_\text{b}(\sigma)\dot{c}^\sigma_\text{cl}(\sigma,0) - c^\sigma_\text{b}(\sigma){c^\sigma}'_\text{b}(\sigma)\dot{\gamma}^\sigma_\text{cl}(\sigma,0).
\end{equation}
The quantum pieces immediately simplify since $P^\dagger P$ and its inverse are diagonal,
\begin{equation}\begin{split}
-&\int\!\ud\sigma\, \big\langle\, P\gamma_{p(\sigma}\partial_{\tau)}c^p + (\partial_p (P\gamma)_{\sigma\tau})c^p \,\big\rangle\,c^\sigma_\text{cl}(\sigma,\epsilon)  \\
= -&\int\!\ud\sigma\, \big\langle\, {\gamma^\sigma}'\dot{c}^\sigma + {\gamma^\tau}'\dot{c}^\tau + \dot{\gamma}^\sigma {c^\sigma}' + \dot{\gamma}^\tau {c^\tau}' + {{\dot{\gamma'}}^\sigma} c^\sigma + {{\dot{\gamma'}^\tau}} c^\tau \,\big\rangle\,c^\sigma_\text{cl}(\sigma,\epsilon).
\end{split}\end{equation}
The Green's function for $P^\dagger P$ on the upper right quadrant is given by
\begin{equation}
  F_{ab} = \bigg(\begin{array}{cc} F_{\sigma\sigma} & 0 \\ 0 & F_{\tau\tau}\end{array}\bigg)
\end{equation}
where
\begin{equation}\begin{split}
  F_{\sigma\sigma}(\sigma,\tau;\sigma',\tau') = \frac{1}{2}F_0(\sigma,\tau;\sigma',\tau') &- \frac{1}{2}F_0(\sigma,\tau;-\sigma',\tau') \\
  &- \frac{1}{2}F_0(\sigma,\tau;\sigma'\,-\tau') + \frac{1}{2}F_0(\sigma,\tau;-\sigma',-\tau'), \\
  F_{\tau\tau}(\sigma,\tau;\sigma',\tau') = \frac{1}{2}F_0(\sigma,\tau;\sigma',\tau') &+ \frac{1}{2}F_0(\sigma,\tau;-\sigma',\tau') \\
  &- \frac{1}{2}F_0(\sigma,\tau;\sigma'\,-\tau') - \frac{1}{2}F_0(\sigma,\tau;-\sigma',-\tau').
\end{split}\end{equation}
Following the same argument as we used for the co-ordinates it is a simple matter to check that the corner contributions once again cancel between the $\sigma-$ and $\tau-$components.

Finally, the second term in (\ref{source}) contributes only a classical piece,
\begin{equation}
  \langle\delta_Q \int\!\ud\sigma\, \gamma^\sigma_\text{b}(Pc)_{\sigma\tau}\rangle = -\int\!\ud\sigma\,\gamma^\sigma\langle \partial_\tau(c^\sigma {c^\sigma}' + c^\tau{\dot{c}}^\sigma)\rangle = -\int\!\ud\sigma\, 2{c^\sigma_\text{b}}'\gamma^\sigma_\text{b} \dot{c}^\sigma_\text{cl} + c^\sigma_\text{b} {\gamma^\sigma_\text{b}}'\dot{c}^\sigma_\text{cl},
\end{equation}
since for the quantum pieces $\langle cc\rangle=0$ as the ghosts anti commute.
We can turn this into an operator acting on the propagator using
\begin{equation}
  \dot{c}^\sigma_\text{cl} = -\frac{\delta}{\delta \gamma^\sigma_\text{b}}, \qquad \dot{\gamma}^\sigma_\text{cl} = \frac{\delta}{\delta c^\sigma_\text{b}}, \qquad \dot{X}_\text{cl} = -2\frac{\delta}{\delta X_\text{b}},
\end{equation}
where the factors come from the conventions in the action. Collecting all the terms we find the Ward identity
\begin{equation}
\bigg[\int\limits_0^\pi\!\ud\sigma\,\, c_\text{b}^\sigma \X_\text{b}'\frac{\delta}{\delta \X_\text{b}} + \big(c^\sigma_\text{b}\gamma^\sigma_\text{b}\big)'\frac{\delta}{\delta \gamma^\sigma_\text{b}} +\frac{1}{2} c^\sigma_\text{b} {{c^\sigma}'_\text{b}}\frac{\delta}{\delta c^\sigma_\text{b}} +\frac{26}{8}\big({c^\sigma}'(0)+{c^\sigma}'(\pi)\big)\bigg]G=0.
\end{equation}
This operator describes the transformation of 25 scalars, $\X$ ($X^0$ drops out since its tangential derivative is zero on the boundary) and the tangential component of a vector $\gamma^\sigma$, under a reparametrisation of the boundary generated by the ghost $c^\sigma$, and quantum corrections. The nonlocal BRST transformations correspond to local reparametrisations of the boundary as claimed. Demanding BRST invariance here does not put the string field on shell, as the reparametrisations are only a subset of those described by BRST in the usual formalism.

\section{Conclusions}
We have shown that the scalar field Shr\"odinger functional can be written in terms of particles moving on $\mathbb{R}^D\times\mathbb{S}^1/\mathbb{Z}_2$, and that the action of the Schr\"odinger functional, describing time evolution, reduces to a Feynman diagram expansion and the gluing property.

The bosonic string field propagator, for both the open and closed string, obeys a generalisation of the gluing property which sews together the propagator worldsheets between definite times. The string field Schr\"odinger functional can therefore be written down using the diagram expansion found in the field theory case.

Timelike T-duality of the string theory becomes a large/small time duality of the string field theory under which boundary states are exchanged with backgrounds coupling to the ends of the dual string. In the examples given we have seen the usual features of T-duality appearing, namely Wilson lines and D-branes. In the open string case, the presence of the corner Weyl anomaly is necessary in any string field theory for building more complicated surfaces by sewing propagators and vertices together and care must be taken to include its effects in calculations.

As a very first step to performing a similar construction of string field objects in non-flat spacetimes, we generalise some of our field theory results to anti-de Sitter spacetimes or arbitrary dimension in \cite{Anton}.

\acknowledgments
This work was partly supported by the EC network ``EUCLID", contract number HPRN-CT-2002-00325. All Feynman diagrams were created using JaxoDraw \cite{Jax}.


\begin{thebibliography}{99}

\bibitem{Witten} E. Witten, `{\em Noncommutative geometry and string field theory}', \Journal{Nucl. Phys.}{B268}{1986}{253}.

\bibitem{Siopsis} G. Siopsis, `{\em Hamiltonian formulation of string field theory}', \Journal{Phys. Lett.}{B195}{1987}{541}

\bibitem{Maeno} Maeno `{\em Canonical quantisation of Witten's SFT using midpoint light cone time}', \Journal{Phys. Rev.}{D43}{1991}{}

\bibitem{Symanzik} K. Symanzik, `{\em Schr\"odinger representation and Casimir effect in renormalisable quantum field theory}', \Journal{Nucl. Phys. }{B190}{1983}{1}

\bibitem{us1} A.~Ilderton and P.~Mansfield, `{\em Time evolution in string field theory and T-duality}', Phys.\ Lett.\ B {\bf 607} (2005) 294 [arXiv:hep-th/0410267].


\bibitem{Brink} L. Brink, P. Di Vecchia, P. Howe `{\em A locally supersymmetric and reparametrisation invariant action for the spinning string}', \Journal{Phys.Lett.}{B65}{1976}{471}

\bibitem{PolyakovBook}
A.M. Polyakov, `{\em Gauge fields and strings}', Harwood Academic Publishers, 1987, ISBN 3-7186-0492-6

\bibitem{Paul-Vevs} A. Jaramillo and P. Mansfield, `{\it Finite VEVs from a Large Distance Vacuum Wave Functional}', hep-th/9808067

\bibitem{Paul-BRST} P. Mansfield `{\em The consistency of topological expansions in field theory: `BRST anomalies' in strings and Yang-Mills}', \Journal{Nucl.Phys.}{B416}{1994}{205-226}, hep-th/9308117

\bibitem{Alvarez} O. Alvarez, `{\it Theory of strings with boundaries: Fluctuation, topology and quantum geometry}', \Journal{Nucl. Phys. }{B126}{1983}{125}.

\bibitem{Beer} W. de Beer, `{\it Conformal counterterms and boundary conditions for open strings}', \Journal{Phys. Rev. D}{37}{1988}{1696}

\bibitem{Ordonez} C.R. Ordonez et al, `{\it Path integral with ghosts for the bosonic string propagator}', \Journal{J. Phys A:}{22}{1989}{3399}; \\
C.R. Ordonez et al, `{\it Polyakov path integral with ghosts: Closed string and one-loop amplitudes}', \Journal{Phys. Lett. B}{215}{1988}{103}.

\bibitem{Moore} G. Moore and P. Nelson, `{\it Absence of nonlocal anomalies in the Polyakov string}', \Journal{Nucl.Phys.}{B266}{1986}{58}.

\bibitem{Varughese} C. Varughese and W.I. Weisberger, `{\it Off-shell amplitudes for open bosonic strings}', \Journal{Phys. Rev. D.}{37}{1988}{981}.

\bibitem{Paul-ghosts} P. Mansfield, `{\it Bosonic ghosts and Witten's non-commutative geometry}', \Journal{Nucl. Phys. }{B306}{1988}{630}.

\bibitem{Carlip} S. Carlip, `{\it Bordered surfaces, off-shell amplitudes, sewing and string field theory}', published in `{\it Functional integration, geometry and strings}', Birkh\"auser Verlag Basel, 1989, ISBN 3-7643-2387-6.

\bibitem{Martinec} E.Martinec, `{\em The light cone in string theory}', \Journal{Class.Quant.Grav.}{10}{1993}{L187-L192}, hep-th/9304037

\bibitem{Lowe} D.A.Lowe, `{\em Causal properties of string field theory}', \Journal{Phys.Lett.}{B326}{1994}{223-330}, hep-th/9312107

\bibitem{Eliezer} D.~A.~Eliezer and R.~P.~Woodard, ``The Problem Of Nonlocality In String Theory,''
Nucl.\ Phys.\ B {\bf 325}, 389 (1989)

\bibitem{Ishy} N. Ishibashi,`{\it The boundary and crosscap states in conformal field theories}', \Journal{Mod. Phys. Lett A}{4}{1989}{251}

\bibitem{Callan1} C.G. Callan et al, `{\it Adding holes and cross-caps to the superstring}, \Journal{Nucl. Phys. B}{293}{1987}{83}

\bibitem{Ramond} G. Kleppe, P. Ramond, R. Viswanathan, `{\em A reparametrisation-invariant approach to string field theory}', \Journal{Phys.Lett.B}{206}{1988}{466}

\bibitem{Green1} M. B. Green `{\em Modifying the Bosonic String Vacuum}', \Journal{Phys. Lett.}{B21}{1988}{42}

\bibitem{Anton} A. Ilderton, `{\em Radial evolution in AdS space}', hep-th/0501218

\bibitem{Jax} D. Binosi and L. Theu\ss l, `{\it JaxoDraw: A graphical user interface for drawing Feynman diagrams}' \Journal{Comp.Phys.Comm}{161}{2004}{76-86}, hep-ph/0309015

\end{thebibliography}
\end{document}